\documentclass[12pt]{book}
\usepackage[english]{babel}
\usepackage[utf8]{inputenc}
\usepackage[T1]{fontenc}
\usepackage{csquotes} 
\usepackage[
	backend=bibtex,
    citestyle=authoryear, 
    natbib=true,
]{biblatex}
\usepackage[colorlinks=true,linkcolor=blue,filecolor=black, urlcolor=blue, citecolor=blue, unicode, pdfencoding=auto, psdextra]{hyperref}
\newcommand\pdfmath[1]{\texorpdfstring{$#1$}{#1}}

\DeclareCiteCommand{\textcite}
  {\boolfalse{cbx:parens}}
  {\usebibmacro{citeindex}%
   \printtext[bibhyperref]{\usebibmacro{textcite}}}
  {\ifbool{cbx:parens}
     {\bibcloseparen\global\boolfalse{cbx:parens}}
     {}%
   \multicitedelim}
  {\usebibmacro{textcite:postnote}}

\DeclareCiteCommand{\parencite}[\mkbibparens]
  {\usebibmacro{prenote}}
  {\usebibmacro{citeindex}%
    \printtext[bibhyperref]{\usebibmacro{cite}}}
  {\multicitedelim}
  {\usebibmacro{postnote}}
  
\usepackage[a4paper,left=3.5cm,right=2.5cm,top=3cm,bottom=3cm, twoside]{geometry}
\usepackage{xcolor}
\usepackage{stmaryrd}
\usepackage{amssymb}
\usepackage{amsmath}
\usepackage{libertine}
\usepackage[pdftex]{graphicx}
\usepackage{caption}
\usepackage{subcaption}
\usepackage{float}
\usepackage{multirow}
\usepackage{apalike}
\usepackage{minitoc}
\usepackage{tikz}
\usepackage{enumitem}
\usepackage{epigraph}
\usepackage{lscape}
\usepackage{titletoc}
\usepackage{titlesec}
\usepackage{xspace}
\usepackage{adjustbox}
\usepackage{booktabs,caption}
\usepackage{booktabs}
\usepackage{arydshln}
\usepackage[nonumberlist]{glossaries}
\usepackage{calc}
\usepackage[]{titlesec} 
\definecolor{linkColor}{HTML}{32a852}
\usepackage{fancyhdr}
\usepackage{lipsum}
\usepackage{wasysym}
\usepackage{breqn}
\DeclareUnicodeCharacter{2009}{\^a}

\newcommand\Msun{\text{M}_{\astrosun}} 
\newcommand\Zsun{\text{Z}_{\astrosun}} 
\newcommand\thesan{\mbox{\textsc{thesan}}\xspace}
\newcommand\thesanhigh{\mbox{\textsc{thesan-high}}\xspace}
\newcommand\thesanlow{\mbox{\textsc{thesan-low}}\xspace}

\definecolor{yourcolor}{HTML}{8a0e19}

\titleformat{\chapter}[display]
{\normalfont\color{yourcolor}}
{\filleft\huge\color{black}\textsc\chaptertitlename\hspace*{2mm}%
	\begin{tikzpicture}[baseline={([yshift=-.6ex]current bounding box.center)}]
	\node[fill=yourcolor,circle,text=white] {\thechapter};
	\end{tikzpicture}
}
{1ex}
{\titlerule[1.5pt]\vspace*{1.5ex}\Huge\color{black}\textsc}
[]

\titleformat{name=\chapter,numberless}[display]
{\normalfont\color{black}}
{}
{1ex}
{\vspace*{-5cm}\Huge\textsc}
[]

\newcommand{\printmyminitoc}[1]{%
	\noindent\hspace{1cm}%
	\colorlet{chpnumbercolor}{black}%
	\begin{tikzpicture}
	\node(s){
		\begin{minipage}{.9\linewidth}
		\printcontents[chapters]{}{1}{}
		\end{minipage}
	};
	{
		\color{yourcolor}
		\draw(s.north west)--(s.north east) (s.south west)--(s.south east);
	}
	\end{tikzpicture}
	\vspace*{3ex}
	
	#1
	\vfill
	\pagebreak
}

\begin{document}

\begin{titlepage}
\begin{center}
{\huge\bf UNVEILING THE COSMIC DAWN AND EPOCH OF REIONIZATION USING COSMIC 21-CM SIGNAL}\\
\end{center}

\vspace*{2cm}

\begin{center}
\vspace*{1cm}
{\Large Thesis submitted for the partial fulfillment of the requirements for the degree of Doctor of Philosophy in Science} \\
\vspace*{1cm}
{\Large by} \\
\vspace*{1cm}
{\Large\bf Ankita Bera}\\
\end{center}

\vspace*{5cm}
\begin{center}
{\Large School of Astrophysics}\\
{\Large Presidency University }\\
{\Large Kolkata, India}\\
{\Large 2023}\\
\end{center}
\end{titlepage}
\pagenumbering{roman}

\pagestyle{fancy}
\cfoot{\thepage}
\fancyhead{}
 \frontmatter
 
\addcontentsline{toc}{chapter}{Title page}

\begin{titlepage}
\begin{center}
{\huge\bf UNVEILING THE COSMIC DAWN AND EPOCH OF REIONIZATION USING COSMIC 21-CM SIGNAL}\\
\end{center}

\vspace*{2cm}

\begin{center}
\vspace*{0.7cm}
{\Large Thesis submitted for the partial fulfillment of the requirements for the degree of Doctor of Philosophy in Science} \\
\vspace*{0.7cm}
{\Large by} \\
\vspace*{0.7cm}
{\Large\bf Ankita Bera}\\
\vspace*{1.5cm}
{\Large Under the supervision of}\\
{\Large\bf Dr. Saumyadip Samui \\ \& \\ Dr. Kanan Kumar Datta}\\
\end{center}

\vspace*{3.5cm}
\begin{center}
{\Large School of Astrophysics}\\
{\Large Presidency University }\\
{\Large Kolkata, India}\\
{\Large 2023}\\
\end{center}
\end{titlepage}

\chapter*{}
\addcontentsline{toc}{chapter}{Registration details}
\noindent
{\Large\bf UNVEILING THE COSMIC DAWN AND EPOCH OF REIONIZATION USING COSMIC 21-CM SIGNAL}\\
\vspace*{3cm}\\
{\bf Name of the Candidate : Ankita Bera \\
Registration Number : R-18RS09160150 \\
Date of Registration : 28 August, 2019 \\
Department : School of Astrophysics\\
Institution : Presidency University\\
\vspace{5.5cm}
\begin{flushleft}
Signature :\\
Date : 
\end{flushleft} }

\chapter*{}
\vspace*{5cm}
\begin{center}
{\centering  Then even nothingness was not, nor existence.
There was no air then, nor the heavens beyond it.
Who covered it? Where was it? In whose keeping?
Was there then cosmic water, in depths unfathomed?
But, after all, who knows, and who can say,
Whence it all came, and how creation happened?
The gods themselves are later than creation,
so who knows truly whence it has arisen?} \\
{\centering \bf -- Rig Veda, X, 129}
\end{center}

\chapter*{\hfill{\centering Acknowledgements}\hfill}
\addcontentsline{toc}{chapter}{Acknowledgements}

{
This PhD thesis marks the culmination of my extensive journey in obtaining the PhD degree.
I am deeply grateful for the support, guidance and encouragement provided by numerous individuals throughout my doctoral studies. 
I would like to take this opportunity to express my sincere gratitude to all those people who have played a significant role in my academic and personal development.

First and foremost, I would like to express my sincere gratitude to my supervisor, Dr. Saumyadip Samui, and my co-supervisor, Dr. Kanan Kumar Datta for their invaluable guidance, support, and mentorship throughout my PhD journey. Their expertise, insights, and constructive feedback in various stages of my research have been instrumental in shaping my research and enhancing the quality of this thesis. I am grateful for their patience, understanding, and encouragement during the ups and downs of my research journey. Their unwavering belief in me and my abilities has been a source of inspiration and motivation. I could not have accomplished this without their guidance and support. I am honored to have had the opportunity to learn from them. I would also like to extend my heartfelt thanks to the faculty members of the School of Astrophysics and Department of Physics, at Presidency University who have provided a supportive and stimulating environment for my academic growth. Their constant commitment to academic excellence has been a source of constant motivation for me.

I would like to thank my mother, Pratima Bera, my father, Asis Bera, and my sister, Ahana Bera for their unwavering love, support, and encouragement throughout my academic journey. I am grateful to my parents for their sacrifices, which have enabled me to pursue my academic goals. My sincere appreciation goes to my sister who has shown a lot of patience and tried to help in all possible ways during the challenging times of my research. I would like to express my heartfelt thanks to the most significant person in my life, my husband, Dr. Soham Mukhopadhyay who has played a vital role in ensuring that this journey has been seamless. 
For the past thirteen years, I have been fortunate to receive consistent support, academic and non-academic advice, and unyielding enthusiasm from him, which have helped me to maintain my motivation and paved the way for my success. His patience, support, and understanding during my demanding work schedule have been invaluable and his constant presence by my side ensured that I never felt alone or fearful.
This thesis is a testament to their unwavering belief in me, and I am forever grateful for everything they have done for me. I am also thankful to all other family members for their support and encouragement during my academic journey.

I am grateful to my collaborator Dr. Sultan Hassan for guiding me throughout my participation in the pre-doctoral program at the Center for Computational Astrophysics, Flatiron Institute in New York, USA. It was a great leap for me and I have been introduced to a completely new place and culture for the first time but he helped me a lot to overcome all my fears. I am privileged to get the opportunity to interact with all the great minds at the Flatiron Institute and I never felt unwelcome.
I am indebted to Simons Foundation for providing all the financial support during my visit to Flatiron Institute. I would like to thank my other collaborators, Dr. Rahul Kannan, Dr. Aaron Smith, Dr. Renyue Cen, Dr. Enrico Garaldi and Prof. Mark Vogelsberger who shared their valuable time and insights with me on which Chapter 5 of this thesis is based. Their contributions to this research have been immensely valuable, and I am grateful for their willingness to participate. I would like to express my appreciation to Prof. Lars Hernquist for granting me the opportunity to visit one of the world's leading institutions, Harvard University, where I got the opportunity to be a part of the THESAN collaboration.

I am indebted to the University Grants Commission (UGC) for their financial support during my PhD studies. Their support has enabled me to complete this research.

I am grateful to all my colleagues at the School of Astrophysics and Department of Physics, in particular Rhombik Roy, Avinanda Chakraborty, Rudrani Kar Chowdhury, Sangita Bera, Shibsankar Dutta, Susmita Das, Arijit Sar and Anirban Chowdhary for their support and assistance during my research. Their professionalism and friendship have been crucial in making this journey a smooth one.

I would like to thank all the authors whose work I have cited in this thesis. Their research has been instrumental in shaping my thinking and has contributed significantly to my research. Lastly, I would like to acknowledge the anonymous referee of my thesis for their time and effort in reviewing and providing feedback. Their input has certainly improved the quality of this thesis. 

In conclusion, writing a PhD thesis is a long and challenging journey, but with the support and encouragement of the people around me, I have been able to complete this research successfully. I am grateful to all those who have contributed to my academic journey in one way or another. I will cherish this experience for the rest of my life, and I hope to contribute to the field of astrophysics in the future.

\vspace{6cm}
\noindent Date:\\
Place: Kolkata 
\hspace*{9 cm} (Name)
}


\chapter*{\hfill{\centering Declaration}\hfill}
\addcontentsline{toc}{chapter}{Declaration}
\noindent
I hereby declare that this thesis  contains original research work carried out by me under the guidance of Dr. Saumyadip Samui, Coordinator and Assistant Professor, School of Astrophysics, Presidency University, Kolkata, India and Dr. Kanan Kumar Datta, Assistant Professor, Jadavpur University, Kolkata, India as part of the Ph.D. programme. \vspace*{1cm}   \\   
All information in this document have been obtained and presented in accordance with academic rules and ethical conduct.
\vspace*{1cm} \\
 I also declare that, as required by these rules and conduct, I have fully cited and referenced all materials and results that are not original to this work.
\vspace*{1cm}\\
 I also declare that, this work has not been submitted for any degree either in part or in full to any other Institute or University before.
\vspace*{6cm} \\
\hspace*{7cm} Signature of the candidate with Date

\chapter*{\hfill{\centering Certificate}\hfill}
\addcontentsline{toc}{chapter}{Certificate}
This is to certify that the thesis entitled {\bf {``UNVEILING THE COSMIC DAWN AND EPOCH OF REIONIZATION USING COSMIC 21-CM SIGNAL"}} submitted by \\
{\underline{ \bf {Ankita Bera}}} who got her name registered for Ph.D. programme under our supervision (Registration Number : R-18RS09160150, Date of Registration : 28 August, 2019) and that neither her thesis nor any part of the thesis has been submitted for any degree/diploma or any other academic award anywhere before.
\vspace*{3cm} \\
 Signature of the supervisor with date and official stamp\\
 \vspace*{2cm} \\
 Signature of the co-supervisor with date and official stamp\\

\renewcommand{\chaptermark}[1]{\markboth{\textsc{#1}}{}}

\chapter*{Abstract}
\addcontentsline{toc}{chapter}{Abstract} 
The cosmic dawn (CD) and epoch of reionization (EoR) is one of the least understood episodes of universe. The cosmological 21-cm signal from neutral hydrogen, which is considered as a promising tool, is being used to observe and study the epoch. A significant part of this thesis focuses on the semi-analytical modeling of the global HI 21-cm signal from CD considering several physical processes. Further, it investigates the nature of galaxies that dominate during CD and EoR using  current available observations.

Understanding different physical processes through which the thermal and ionization states of the intergalactic medium (IGM) during CD and EoR evolved is the key to unlocking the mysteries of the early universe. The study of the 21-cm signal is a powerful tool that can be used to investigate different physical processes in the early universe. It provides us with a window into the time before the formation of the first galaxies. 
Our study is partly motivated by the first-ever detection of the global 21-cm signal  reported by the EDGES low-band antenna and ongoing observations by several global 21-cm experiments. 
One of the promising avenues to interpret the EDGES signal is to consider a `colder IGM' background. 

In our first work, we study the redshift evolution of the primordial magnetic field (PMF) during the dark ages and cosmic dawn and prospects of constraining it in light of EDGES 21-cm signal in the `colder IGM' background. Our analysis has been carried out by considering the dark matter-baryon interactions for the excess cooling mechanism. We find that the colder IGM suppresses both the residual free electron fraction and the coupling coefficient between the ionized and neutral components. The Compton heating also gets affected in colder IGM backgrounds. Consequently, the IGM heating rate due to the PMF enhances compared to the standard scenario.
Thus, a significant fraction of the magnetic energy, for $B_0 \lesssim 0.5 \, {\rm nG}$, get transferred to the IGM, and the magnetic field decays at a much faster rate compared to the simple $(1+z)^2$ scaling during the dark ages and cosmic dawn. 
We also find that the upper limit on the PMF depends on the underlying DM-baryon interaction. Higher PMF can be allowed when the interaction cross-section is higher and/or the DM particle mass is lower. Our study shows that the PMF with $B_0$ up to  $\sim 0.4 \, {\rm nG}$, which is ruled out in the standard model, can be allowed if  DM-baryon interaction with suitable cross-section and DM mass is considered. 
 However, this low PMF is an unlikely candidate for explaining the rise of the EDGES absorption signal at lower redshift.

We further consider, in detail, the heating of the IGM owing to cosmic ray protons generated by the supernovae from both early Pop~III and Pop~II stars. The low-energy cosmic ray protons from Pop~III supernovae can escape from minihalos and heat the IGM via collision and ionization of hydrogen. Moreover, the high-energy protons generated in Pop~II supernovae can escape the hosting halos and heat the IGM via magnetosonic Alfv\'en waves. We show that the heating due to these cosmic ray particles can significantly impact the IGM temperature and hence the global 21-cm signal at $z\sim 14-18$. The depth, location, and duration of the 21-cm absorption profile are highly dependent on the efficiencies of cosmic ray heating and star formation rate. In particular, the EDGES signal can be well fitted by the cosmic ray heating along with the Lyman-$\alpha$ coupling and the dark matter-baryon interaction that we consider to achieve a `colder IGM background'. Further, we argue that the properties of cosmic rays and the nature of the first generation of stars could be constrained by accurately measuring the global 21-cm absorption signal during the cosmic dawn.

We, further, explore the conditions by which the EDGES detection is consistent with current reionization and post-reionization observations, including the volume-averaged neutral hydrogen fraction of the intergalactic medium at $z\sim6$--$8$, the optical depth to the cosmic microwave background, and the integrated ionizing emissivity at $z\sim5$. By coupling a physically motivated source model derived from radiative transfer hydrodynamic simulations of reionization to a Markov Chain Monte Carlo sampler, we find that high contribution from low-mass halos along with high photon escape fractions are required to simultaneously reproduce the high-redshift (cosmic dawn) and low-redshift (reionization) existing constraints. Low-mass faint-galaxies dominated models produce a flatter emissivity evolution that results in an earlier onset of reionization with gradual and longer duration and higher optical depth. Our results provide insights into the role of faint and bright galaxies during cosmic reionization, which can be tested by upcoming surveys with the James Webb Space Telescope (JWST).

With the extreme effort in building more advanced and sophisticated telescopes, such as the Square Kilometre Array (SKA) and several global 21-cm telescopes the future 21-cm signal detection would be able to provide better constraints on the amplitude of PMF and the efficiencies on cosmic ray protons, and consequently on early star formation rates. We would be able to get a more complete picture of the cosmic dawn and the epoch of reionization by combining the future detection of the 21-cm signal with other observational constraints at high redshifts.

\chapter*{List of publications}
\addcontentsline{toc}{chapter}{List of Publications}

\begin{itemize}
 \item {\textbf {Published manuscripts:}}
\begin{enumerate}
\item { Primordial magnetic fields during the cosmic dawn in light of EDGES 21-cm signal; {\textbf{Ankita Bera}}, Kanan Kumar Datta, Saumyadip Samui, {\textbf{MNRAS}} 498, 1, (2020).}

\item { Cosmic recombination history in light of EDGES measurements of the cosmic dawn 21-cm signal; Kanan K. Datta, Aritra Kundu, Ankit Paul, {\textbf{Ankita Bera}}, {\textbf{Physical Review D}}, 102, 8, 083502, (2020).}

\item{ Studying Cosmic Dawn using redshifted HI 21-cm signal: A brief review; \textbf{Ankita Bera}, Raghunath Ghara, Atrideb Chatterjee, Kanan K. Datta, Saumyadip Samui, {\textbf{Journal of Astrophysics and Astronomy}}, 44, 1, (2023).}

\item{ Impact of cosmic rays on the global 21-cm signal during cosmic dawn; \textbf{Ankita Bera}, Saumyadip Samui, and Kanan K. Datta, {\textbf{MNRAS}}, 519, 4, (2023). }
\end{enumerate}

\item {\textbf {Communicated manuscripts:}}
\begin{enumerate}
\item { Bridging the gap between Cosmic Dawn and Reionization favors Faint Galaxies-dominated Models; \textbf{Ankita Bera}, Sultan Hassan, Aaron Smith, Renyue Cen, Enrico Garaldi, Rahul Kannan, and Mark Vogelsberger, Under review in {\bf {ApJ}}, (2022), {[https://ui.adsabs.harvard.edu/abs/2022arXiv220914312B]}.}

\item { JWST constraints on the UV luminosity density at cosmic dawn: implications for 21-cm cosmology; Sultan Hassan, Christopher C. Lovell, Piero Madau, Marc Huertas-Company, Rachel S. Somerville, Blakesley Burkhart, Keri L. Dixon, Robert Feldmann, Tjitske K. Starkenburg, John F. Wu, Christian Kragh Jespersen, Joseph D. Gelfand, \textbf{Ankita Bera}, Submitted to {\bf ApJL}, (2023), \\{[https://ui.adsabs.harvard.edu/abs/2023arXiv230502703H/abstract]}. }
\end{enumerate}

\end{itemize}

\tableofcontents
\addcontentsline{toc}{chapter}{Table of contents}

\listoffigures
\addcontentsline{toc}{chapter}{List of figures}

\listoftables
\addcontentsline{toc}{chapter}{List of tables}

\mainmatter

\setlength{\parskip}{.7em}

\titlespacing*{\section}{0pt}{.9em}{.8em}
\renewcommand{\baselinestretch}{1.1}

\chapter{Introduction}
\epigraph{\itshape  Now entertain conjecture of a time, when creeping murmur and the poring dark fills the wide vessel of the universe.}{-- William Shakespeare}

\startcontents[chapters]

\section{Cosmic Dawn and the Epoch of Reionization: \texorpdfstring{\\}{} an overview} \label{chap:intro}
According to current understanding, the universe came into existence with a Big Bang approximately $13.7$ billion years ago. Universe got cooled as it expanded adiabatically, and over time, different ingredients of our universe such as quarks, protons, neutrons, and then electrons froze out. Finally, about $380,000$ years after the Big Bang, neutral hydrogen atoms started to form by recombination of protons with free electrons. With the formation of these first hydrogen atoms, the universe entered a period called the `Dark Ages' \citep{2003Sci...300.1904M, 2004PhRvL..92u1301L}. During this time, hydrogen was mostly neutral until the first stars, quasars, and/or the first generation of galaxies appeared. The ignition of the first stars marked the end of the Dark Ages and the beginning of `Cosmic Dawn' (CD). It happened approximately $100$ million years after the Big Bang. When the first stars and galaxies were beginning to form, they emitted the ionizing photons that eventually reionized the neutral hydrogen atoms, and the universe made a transition to the `Epoch of Reionization' (EoR) \citep[see][for reviews]{barkana01, 2005SSRv..116..625C, 2006astro.ph..3360L, 2016ASSL..423....1H, 2022arXiv220307864L}. This period occurred roughly between $100$ million and one billion years after the Big Bang and marked a major transition in the evolution of the universe.

Over the past two decades, significant progress has been made in understanding the evolution of universe, from the infancy of the universe to the present day. However, the first billion years of the universe, particularly the period during which the first stars and galaxies were formed, remain largely unobserved. Several probes such as observations of the cosmic microwave background radiation, high redshift quasars, and distant galaxies are being used to unveil the epoch of reionization and cosmic dawn. We give a brief description on the probes of reionization and cosmic dawn.


\section{Probes of studying the Cosmic Dawn and the Epoch of Reionization}
\label{sec:probes}
During cosmic dawn, universe was in a very different state than it is today, and so studying this era can provide insights into the origin of structure and formation of the first galaxies.
Probes of cosmic dawn include observations of first stars, galaxies, the cosmic microwave background radiation and the intergalactic medium \citep{1998ASPC..133...73L, 2019PhRvD..99d3524D}.
The Epoch of Reionization is a critical era in the history of universe. It happened when the neutral hydrogen gas that pervaded the cosmos was reionized by the first stars and galaxies \citep[see][for review]{2001ARA&A..39...19L}.
A variety of probes such as the cosmic microwave background radiation CMBR) \citep{1996A&A...311....1A, 2008arXiv0811.3918Z}, high redshift galaxies including the Lyman-alpha emitters \citep{2007MNRAS.377.1175D, 2014PASA...31...40D, 2019MNRAS.486.2197B, 2023arXiv230316225W}, high redshift quasars  \citep{Mortlock2015npr, 2022MNRAS.517.1264L} etc are being used to study the epoch of reionization and cosmic dawn. 


The Cosmic Microwave Background Radiation (CMBR) is an important probe of the epoch of reionization \citep{1999ASPC..181..227H, zahn05, zahn11a, 2021ApJ...914...44A}. 
During the epoch of reionization, neutral hydrogen gas in the intergalactic medium was ionized by the ultraviolet radiation emitted by the first galaxies. This process imprints signatures on the CMBR radiation that can be detected through observations \citep{2015IJMPD..2430004B}. Specifically, as the CMBR photons pass through the ionized intergalactic medium they are scattered off by free electrons and become partially polarized. Measurements of the CMBR polarization  have been used to probe the timing and during of reionization epoch \citep{2008ApJ...672..737M, 2022ApJ...930..140S}.

High redshift quasars are also important probes of the cosmic dawn and the epoch of reionization \citep{2015ApJ...813L...8M, 2022arXiv221206907F, 2023arXiv230308918L, 2023arXiv230405378A}.
Quasars are among the brightest and most energetic objects in universe, powered by accretion onto supermassive black holes. Because of their brightness, quasars can be observed out to very high redshifts, corresponding to cosmic epochs as early as the epoch of reionization \citep{Mortlock2015npr}. Quasars emit radiation across a broad range of wavelengths. 
During the final phase of the epoch of reionization, the residual neutral hydrogen in the IGM left a measurable signature on the spectra of high redshift ($\gtrsim 5.5)$ quasars, known as the Gunn-Peterson trough \citep{1965ApJ...142.1633G}. The Gunn-Peterson trough is a region in the quasar spectrum where the transmission drops sharply to zero due to the absorption of Ly-$\alpha$ photons by neutral hydrogen atoms. Observations of Gunn-Peterson troughs have been used to provide insights into the final stage of the reionization \citep{1999ApJ...519..479H, 2001AJ....122.2850B, 2002AJ....123.1247F}.

High-redshift galaxies can also be used to study the reionization. 
The growing population of high redshift galaxies provides valuable insights into the sources that fueled reionization, as many of these galaxies, especially those at $z>7$, were formed during the EoR. These galaxies serve as observational probes for tracking the progress of reionization. In particular, 
quantities such as the galaxy luminosity functions \citep{2011ApJ...734..119K},  galaxy clustering \citep{2007MNRAS.381...75M, 2023A&A...671A...5H}, and line intensity mapping \citep{Suginohara_1999, Lidz_2011, 2019BAAS...51c.101K} can be used to constrain the ionization state, source properties during the reionization.

The discoveries of high-redshift Ly-$\alpha$ emitters (LAEs) have allowed for the study of reionization through their suppression by a neutral intergalactic medium (e.g. \citealt*{2008MNRAS.386.1990M}, and see \citealt*{2014PASA...31...40D} for a review). Constraints on reionization can also be obtained through various methods, including cross-correlating the distribution of LAEs with intergalactic 21-cm emission or adding up the ionized volume surrounding individual LAEs \citep{2007MNRAS.377.1175D}. Additionally, Ly-$\alpha$ line emissions are expected to be produced by the first primeval galaxies. The Ly-$\alpha$ photons emitted by a source typically scatter over a characteristic angular radius of about $\sim 15''$ around the source and are broadened and redshifted by about $10^3$ km/s relative to the source. Detection of the diffuse Ly-$\alpha$ halos around high redshift sources would provide a unique tool for probing the neutral intergalactic medium before the epoch of reionization \citep{1999ApJ...524..527L}.

Above probes provide some important information about the sources, ionization state, timing and duration of the epoch of reionization and cosmic dawn. However, in order to get the full picture of the epoch we have to depend on the cosmological 21-cm signal from neutral hydrogen \citep[see][for reviews]{furlanetto06, pritchard12, 2022arXiv220307864L}. Observations of  cosmological HI 21-cm signal will provide 3D picture of our universe over a significant period of cosmic history ranging from the dark ages to the post reionization epoch \citep{1997ApJ...475..429M, 2003ApJ...596....1C, mellema06, 2013ASSL..396...45Z}.


However, detection of the cosmological HI 21 cm radiation is challenging because the signal is extremely weak compared to other sources of radio emission in the universe. Additionally, the 21 cm signal is affected by a number of factors, including foreground emission from our own galaxy and other extragalactic sources, instrumental noise, and the complex nature of the astrophysical processes that affect the 21 cm signal \citep{2003MNRAS.346..871O, fan2006observational}.
Despite these challenges, the current and upcoming telescopes are trying to bring a new era of studying the high redshift Universe, which will partly bridge the gap between the very high redshift universe probed by the CMB (at $z \sim 1100$) and the low redshift universe (at $z \sim< 6$).  By studying the 21 cm signal, we hope to gain insights into the formation of the first stars and galaxies, the evolution of the intergalactic medium, and a complete picture of the epoch of reionization \citep[e.g.][for a review]{2016ASSL..423..247F}. The following section presents a brief description of some basics of 21-cm cosmology with a special emphasis on the global 21-cm signal.

\section{Cosmological HI 21-cm signal}
\label{sec:21cmth}
\begin{figure}
    \centering
    \includegraphics[width=\columnwidth]{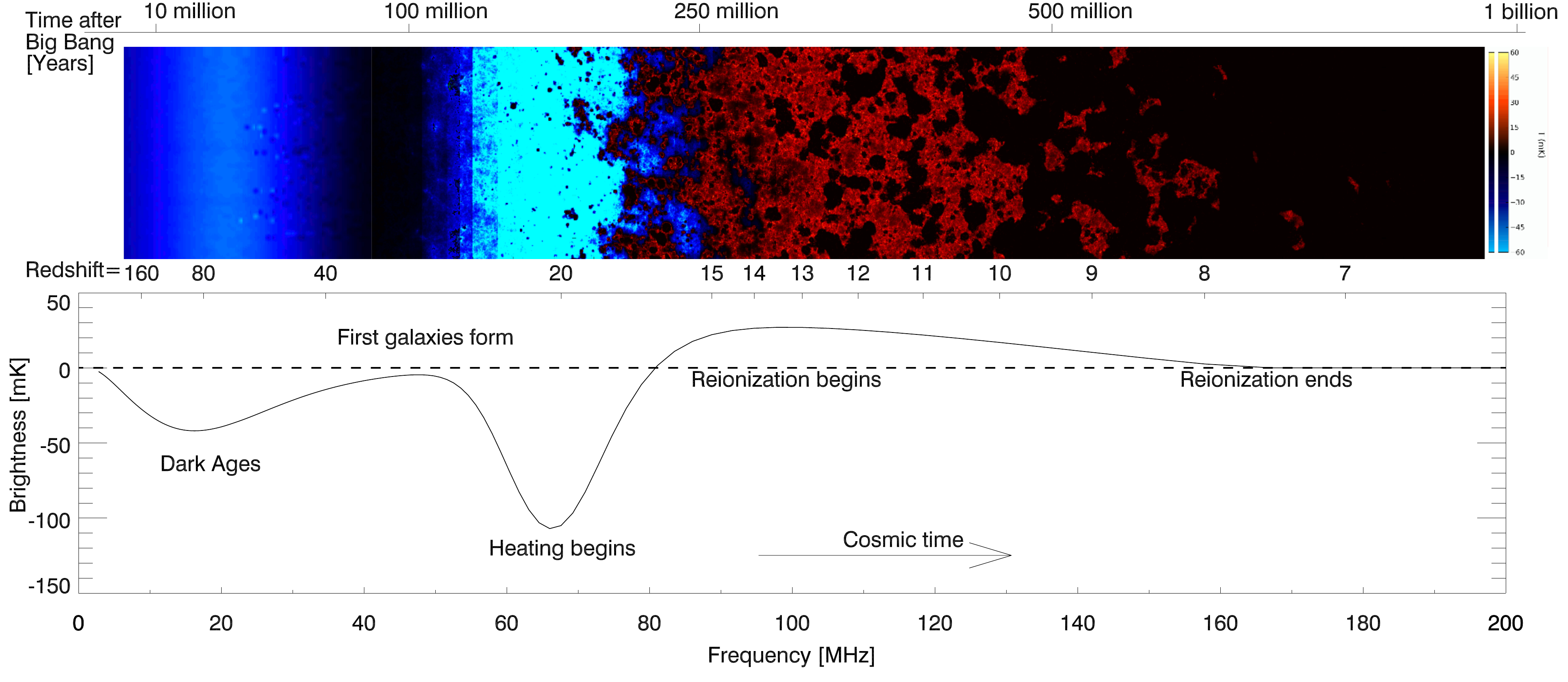}
    \caption[The time evolution of the fluctuations in the 21-cm brightness and the expected sky-averaged 21-cm signal.]{Figure is adapted from \citet{pritchard12}. The upper panel shows the time evolution of the fluctuations in the 21-cm brightness starting from the Dark Ages i.e. before the formation of first stars to the end of the reionization epoch. The lower panel depicts the expected sky-averaged 21-cm signal and its evolution from Dark Ages to the end of reionization. The solid curve is the 21-cm signal and the dashed line indicates the situation if $T_{ 21} = 0$.}
    \label{fig:21cm}
\end{figure}
The hyperfine splitting in the ground state of the hydrogen atom allows for a promising observational probe, emitting a photon with a wavelength of $\lambda = 21$ cm \citep{1958PIRE...46..240F}. Despite the extremely low transition rate, the amount of neutral hydrogen in universe makes it a good probe of the EoR, mapping the distribution of neutral hydrogen across the sky and probing the 3-dimensional distribution of matter at high redshifts. 
In Fig.~\ref{fig:21cm}, the main features of the 21-cm signal are depicted with relevant cosmic time, frequency, and redshift. The signal originates during Dark Ages when the cosmic gas decouples from the CMBR and cools adiabatically with the expansion of universe, before the formation of the first stars. During this period, the 21-cm absorption signal is observed, as shown in the bottom panel of Fig.~\ref{fig:21cm}, and fluctuations caused by density variations, shown in the top panel of Fig.~\ref{fig:21cm}. The emergence of the first stars and galaxies changes the properties of the gas, in particular, the scattering of Ly-$\alpha$ photons creates a strong coupling between the excitation of the 21-cm line spin states and the gas temperature, leading to a spatially varying absorption signal. Subsequently, with the emergence of the heating sources, the ambient gas gets heated producing a 21-cm emission signal. Finally, ultraviolet photons ionize the gas, creating dark regions in the 21-cm signal within the ionized bubbles surrounding groups of galaxies. Eventually, all the hydrogen gas except for that in a few dense pockets gets ionized.

The quantity of interest in the context of the cosmological HI 21-cm signal is the differential HI brightness temperature \citep[see, eg.][for brief reviews on cosmic 21-cm signal]{pritchard12, 2006PhR...433..181F, 2023JApA...44...10B}. It is defined as the excess brightness temperature relative to a background radio temperature, redshifted to the present observer, and is given by \citep{Bharadwaj_2005},
\begin{equation}
    T_{21}(\textbf{n},z) = T(z) \times \eta_{\rm HI}(\textbf{n},z) , 
    \label{eq:T21_full}
\end{equation}
where, 
\begin{equation}
    T(z) = 4.0 \,{\rm mK} (1+z)^2 \left( \frac{\Omega_b h^2}{0.02} \right) \left(\frac{0.7}{h} \right) \left(\frac{H_0}{H(z)} \right) 
    \label{eq:T21_z}
\end{equation}
that depends only on the redshift $z$, and the cosmological parameters. $\eta_{\rm HI}$ is the 21-cm radiation efficiency in redshift space and given by, 
\begin{equation}
    \eta_{\rm HI} = \left(\frac{\rho_{\rm HI}}{\Bar{\rho_H}} \right) \left( 1-\frac{T_{\gamma}}{T_s} \right)
    \times \left[1- \frac{(1+z)}{H(z)} \frac{\partial v}{\partial r} \right] .
    \label{eq:T21_n}
\end{equation}
Here, $\rho_{\rm HI}$ is the density of the neutral hydrogen whereas $\Bar{\rho}_{\rm H}$ is the mean hydrogen density, and $\textbf{n}$ is the direction of light propagation. 
Further, $({\rho_{\rm HI}}/{\Bar{\rho}_H})$ arises due to the non-uniform distribution of hydrogen, and the term inside the square brackets arises due to the redshift space distortion in which $\partial v/ \partial r$ is the divergence of the peculiar velocity along the line of sight \citep{Bharadwaj_2004}. Moreover, $T_{\gamma}$ is the background temperature of radio photons, mostly dominated by cosmic microwave background radiation (CMBR).
$T_s$ is the hydrogen spin temperature which is determined by the relative population of the singlet and triplet states of neutral hydrogen atom.
It is clear from Eq.~\ref{eq:T21_full} that 21-cm signal will be in the absorption or emission depending on whether $T_s<T_{\gamma}$ or $T_s>T_{\gamma}$ respectively.
The spin temperature ($T_s$) in Eq.~\ref{eq:T21_full} is governed by three coupling mechanisms: (i) radiative transition due to the absorption and stimulated emission of CMB photons (couples $T_s$ and CMBR temperature $T_{\gamma}$, (ii) spin-flip transition due to atomic collisions (couples $T_s$ and gas kinetic temperature $T_g$) and (iii) the Wouthuysen-Field effect \citep{Wouthuysen_1952, Field_1959} which also couples $T_s$ and $T_g$.
$T_s$ is related to $T_g$ and $T_{\gamma}$ as (\citealt*{Field_1958}; also see \citealt*{furlanetto06} for a detailed review), 
\begin{equation}\label{spin_temp}
    T^{-1}_s = \frac{T^{-1}_{\gamma} + x_{\alpha} T^{-1}_{\alpha} + x_c T^{-1}_g}{1 + x_{\alpha} + x_c} ,
\end{equation}
where, $T_{\alpha}$ is the color temperature corresponds to the Lyman-$\alpha$ radiation field. As the Lyman-$\alpha$ photons get absorbed and emitted repeatedly by hydrogen atoms, they are in equilibrium with H-atom, so $T_{\alpha} = T_{\rm g}$ during the cosmic dawn period \citep{Field_1959, Hirata2006}. The coupling coefficients, $x_c$ and $x_{\alpha}$ depend on the different processes such as Ly-$\alpha$ coupling \citep[due to Wouthuysen-Field mechanism][]{Wouthuysen_1952, Field_1958}, and collisional coupling due to the collisions between two hydrogen atoms, hydrogen atom and an electron or the H-atom and a proton.

The Wouthuysen-Field coupling coefficient is given by \citep{Pritchard_2006},
\begin{equation}
    x_{\alpha} = \frac{16 \pi^2 T_{*} e^2 f_{\alpha}}{27 A_{10} T_{\gamma} m_e c} S_{\alpha} J_{\alpha} ,
    \label{eq:x_a}
\end{equation}
where $f_{\alpha} = 0.4162$ is the oscillator strength for the Ly-$\alpha$ transition. Further, $J_{\alpha}$ is the Ly-$\alpha$ photon intensity which will be discussed in Sec.~\ref{sec:ly_alpha}. Moreover, $S_{\alpha}$ in Eq.~\ref{eq:x_a} is a correction factor that takes care of the redistribution of photon energies due to the repeated scattering off the thermal distribution of atoms. It can be expressed as, $S_{\alpha} = \exp{(-0.37 (1+z)^{1/2} \, T_g^{-2/3})}/{(1+4\,T_g^{-1})}$ \citep{Chuzhoy_2006, 2021ApJ...914...44A}. As this factor is of the order of unity \citep{Chen_2004}, we assume it to be $S_{\alpha} = 1$. Also, $T_* = h_p \nu_e/k_B = 0.068$~K is the characteristic temperature for the HI 21-cm transition.
The total collisional coupling coefficient can be written as a sum of coupling between H-H, H-p, H-$e^-$, ($x_c^{\rm HH}, x_c^{\rm pH}, x_c^{\rm eH}$ respectively), and is given by,
\begin{eqnarray}
    x_c & = & x_c^{\rm HH} + x_c^{\rm eH} + x_c^{\rm pH}   \\
    & = & \frac{T_*}{A_{10} T_{\gamma}} {\kappa^{\rm HH}_{10}(T_{\rm g}) n_{\rm H} + \kappa^{\rm eH}_{10}(T_{\rm g}) n_{\rm e} + \kappa^{\rm pH}_{10}(T_{\rm g}) n_{\rm p}} . \nonumber
    \label{eq:x_c}
\end{eqnarray}
All the specific rate coefficient values, $\kappa^{\rm HH}_{10}, \kappa^{\rm eH}_{10}$, and $\kappa^{\rm pH}_{10}$ are given in \citet{pritchard12}. As the universe was mostly filled with hydrogen during cosmic dawn, $\kappa^{\rm HH}_{10}$ dominate over $\kappa^{\rm eH}_{10}$, and $\kappa^{\rm pH}_{10}$ throughout this period.
Although, this collisional coupling is a dominant process during the dark ages, and the coupling between $T_s$ and $T_{\rm g}$ happens due to the Ly-${\alpha}$ coupling during cosmic dawn.

Note that, $\rho_{\rm HI}$ in eq.~\ref{eq:T21_n}, and the number densities of hydrogen ($n_{\rm H}$), electron ($n_{\rm e}$), and proton ($n_{\rm p}$) are determined by the ionization state of the intergalactic medium (IGM). This can be obtained by the evolution in the ionized fraction of hydrogen, $x_e$ ($=1-x_{\rm HI}$) which can be written as \citep{Peebles1968},
\begin{dmath}
    {\frac{dx_e}{dz}=} \left[ C_p \left( \beta_e\,(1-x_e)  \, e^{-\dfrac{h_p\nu_\alpha}{k_B T_{\rm g}}} - \alpha_e\,x_e^2 n_{\rm H}(z) \right) + \gamma_e\,n_{\rm H}(z) (1-x_e)x_e \\ \nonumber + c(1-x_e) I_{\rm CR} + \frac{\dot N_{\gamma}}{n_{\rm H}(z)} \right] \frac{dt}{dz}.
    \label{eq:ion_frac}
\end{dmath}
Here the evolution in ionization fraction is affected due to the photoionization by CMBR photons, recombination, collisional ionization, ionization by cosmic rays, and photoionization by UV photons respectively. The photoionization co-efficient, $\beta_e$ can be calculated using the relation, $\beta_e (T_{\gamma}) = \alpha_e (T_{\gamma}) \Big( \frac{2 \pi m_e k_B T_\gamma}{h^2_p} \Big)^{3/2} e^{-E_{2s}/k_B T_\gamma}$ \citep{Seager1999,Seager2000}. The recombination co-efficient, $\alpha_e (T_{\rm g}) = F \times 10^{-19} (\frac{a t^b}{1 + c t^d}) \hspace{0.1cm} {\rm m^3 s^{-1} } $, where $a = 4.309$, $b = -0.6166$, $c = 0.6703$, $d = 0.53$, $F = 1.14$ (the fudge factor)  and $t = \frac{T_{\rm g}}{10^4 \, {\rm K}} $. The Peebles factor is defined by $C_p = \frac{1+ K \Lambda (1-x) n_{H}}{1+K(\Lambda+\beta_{e})(1-x) n_{H}}$, where $\Lambda=8.3 \, {\rm s}^{-1}$ is the transition rate from (hydrogen ground state) $2s\rightarrow1s$ state through two photons decay, and $K=\frac{\lambda_{\alpha}^{3}}{8\pi H(z)}$. The collisional ionization coefficient, $\gamma_e(T_{\rm g}) =0.291 \times 10^{-7} \times U ^{0.39} \frac{\exp(-U)}{0.232+U}  \, {\rm cm^3/s} $ \citep{Minoda17} with $h_{p}  \nu_{\alpha} = 10.2 \, {\rm eV}$ and $U=\vert E_{1s}/k_B T_{\rm g} \vert$. We note that $\alpha_e$ and $\gamma_e$ depend on the IGM temperature $T_g$. In contrast, $\beta_e$ depends on the CMBR temperature \citep[see][for a detailed discussion]{Chluba15}. The second last term in eq.~\ref{eq:ion_frac} arises due to ionization by the cosmic rays where $c$ is the speed of light, and $I_{\rm CR}$ is described in section~\ref{sec:CR_th}. Further, $\dot N_{\gamma}$ is the rate of UV photons escaping into the IGM and $n_{\rm H}(z)$ is the proper number density of the hydrogen atoms \citep{barkana01}. 

Global signal experiments attempt to detect the sky-averaged $T_{21}$ where the average is taken over all directions of sky at a particular redshift, as shown in the lower panel of Fig.~\ref{fig:21cm}. Thus the globally averaged differential brightness temperature, $T_{21}(z)$ is
given by,
\begin{equation}
    T_{21}(z) =  4.0 \,{\rm mK} (1+z)^2 \left( \frac{\Omega_{\rm b} h^2}{0.02} \right) \left(\frac{0.7}{h} \right) \left(\frac{H_0}{H(z)} \right) \left(\frac{\rho_{\rm HI}}{\Bar{\rho_H}} \right) 
    \left( 1-\frac{T_{\gamma}}{T_s} \right).
    \label{eq:T21_global}
\end{equation}
To determine $T_{21}$, we also need to have the information on the evolution of the gas kinetic temperature, $T_g$ along with the ionization fraction. The evolution of $T_g$ as a function of redshift can be obtained by solving the equation  \citep{Kompaneets_1957,Peebles_1993, Seager1999}, 
\begin{equation}
    \frac{dT_g}{dz} = \frac{2T_g}{1+z} - \frac{8\sigma_T a_{\rm SB} T_{\gamma}^4}{3m_e c H(z) (1+z)}\left(T_\gamma-T_g\right) \frac{x_e}{1+x_e}  \\
    - \frac{2}{3k_B} \sum_i Q_i .
	\label{eq:Tg}
\end{equation}
The first two terms on the R.H.S. arise due to the adiabatic cooling of baryonic gas due to the expansion of universe and the Compton heating due to the interaction between CMBR and free electrons, respectively. Further, $ k_{\rm B}$, $ \sigma_{\rm T}$ are the Boltzmann constant, Thomson scattering cross-section respectively and $a_{\rm SB}=4 \sigma_{\rm SB}/c$, where $\sigma_{\rm SB}$ is the Stefan Boltzmann constant. The third term $Q_i$ includes other heating and cooling mechanisms such as heating due to X-rays, cosmic rays, primordial magnetic fields, cooling/heating due to the possible interaction between dark matter and baryons, etc. that impact the evolution in $T_g$. 

As the observations of redshifted 21-cm signal from the neutral hydrogen (HI) is sensitive to the heating and ionization of the IGM, it became one of the most promising tools for studying the cosmic dawn \citep{furlanetto06, pritchard12, 2019BAAS...51c..48C, 2021IJMPD..3030009P} and the subsequent Epoch of Reionization. 
There are mainly two separate approaches by which the cosmological HI 21-cm signal from the CD/EoR can be detected, namely, the measurements of (i) global HI 21-cm signal, and, (ii) statistical signal such as power spectrum, Bi-spectrum etc. Several ongoing and upcoming experiments such as the Experiment to Detect the Global reionization Signature~\citep[EDGES,][]{EDGES18}, Shaped Antenna measurement of the background RAdio Spectrum~\citep[SARAS,][]{Saurabh_2021}, Large-Aperture Experiment to Detect the Dark Ages~\citep[LEDA,][]{2018MNRAS.478.4193P}, the Radio Experiment for the Analysis of Cosmic Hydrogen~\citep[REACH,][]{8879199} etc. are dedicated to detecting the global HI 21-cm signal.

However, the global HI 21-cm signal can not retain information regarding the spatial distribution of HI field and sources, and provide only the global evolution of the HI differential brightness temperature. Radio interferometers, such as the Giant Metrewave Radio Telescope~\citep[GMRT,][]{Ali_2008, Pal21}, the Murchison Widefield Array~\citep[MWA,][]{2016ApJ...833..102B, patwa21},  Low Frequency Array~\citep[LOFAR,][]{2017ApJ...838...65P,martens20}, the Hydrogen Epoch of Reionization Array~\citep[HERA,][]{2017PASP..129d5001D, hera22}, etc. are trying to detect radio fluctuations in the redshifted 21-cm background arising from variations in the amount of neutral hydrogen. These instruments seek to make detailed maps of ionized regions during reionization and measure properties of hydrogen out to z = 30. Next-generation instruments, such as the Square Kilometre Array~\citep[SKA,][]{2015aska.confE..10M}, will provide a powerful tool for learning about the first stars and galaxies, and they also have the potential to inform us about fundamental physics, such as the density field, neutrino masses, and the initial conditions from the early epoch of cosmic inflation in the form of the power spectrum.

\section{Observations of global HI 21-cm signal}
\label{sec:EDGES}
\begin{figure}
    \centering
    \includegraphics[width=\columnwidth]{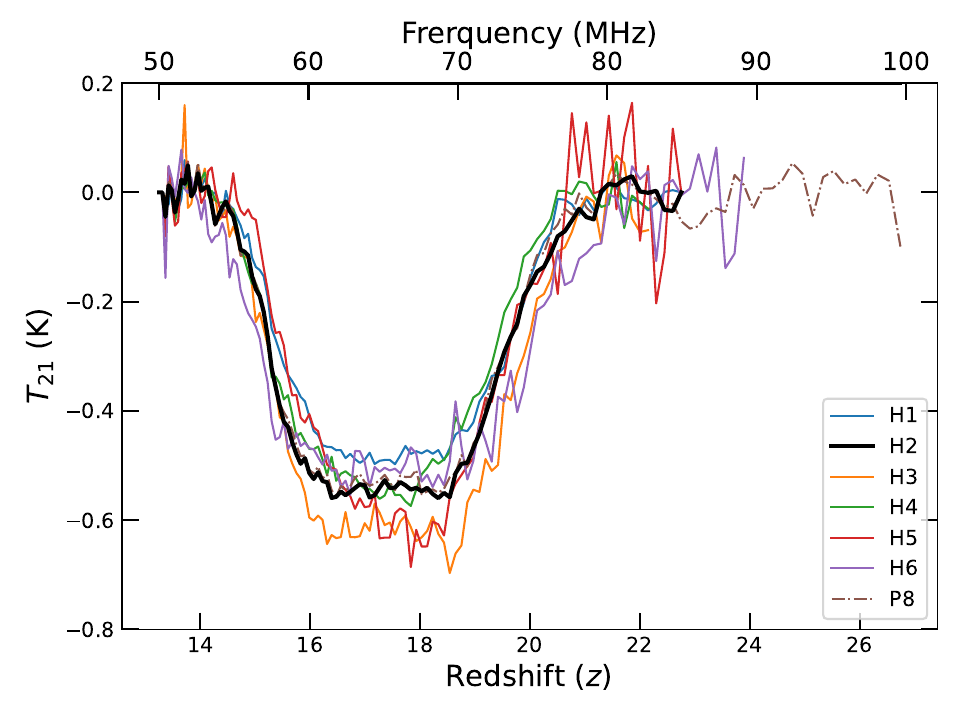}
    \caption[The EDGES observation of global HI 21-cm signal.]{The EDGES observation of global HI 21-cm signal. Each profile denotes the brightness temperature $T_{21}$ for different hardware configurations that were used for the observations denoted as H1, H2,.....P8. The black thick line is the model fit with the highest signal-to-noise ratio. The graph is plotted using the publicly available data at \href{http://loco.lab.asu.edu/edges/edges-data-release/}{http://loco.lab.asu.edu/edges/edges-data-release/}.}
    \label{fig:edges}
\end{figure}
The observations of global HI 21-cm absorption signal by experiments such as the Experiment to Detect the Epoch of Reionization Signature (EDGES) \citep{EDGES18} have opened up the possibilities to study the evolution of the early stars and galaxies and the thermal state of the intergalactic medium during cosmic dawn. The detected signal shown in fig.~\ref{fig:edges} by black thick line has an absorption depth of $0.5^{+0.5}_{-0.2}$~K centred at frequency $78\pm 1$~MHz or redshift $z \sim 17$. The `U' shaped 21-cm signal caries signatures of early Lyman-$\alpha$ (Ly-$\alpha$) coupling and heating of the IGM in the redshift range of $\sim 14-22$. However, there seem to be quite controversies exist with the detected EDGES signal. For example, a recent observation by Shaped Antenna measurement of the background RAdio Spectrum (SARAS~3) \citep{Saurabh_2021} has claimed that the EDGES profile may not be of astrophysical origin. 
This new measurement rejects the best-fitting profile found by \citet{EDGES18} with 95.3\% confidence. This observation suggests that \citet{EDGES18} observation is not an evidence for new astrophysics or non-standard cosmology.
There are also concerns that the unusual EDGES signal may arise due to unaccounted systematics \citep{Hills_2018, Bradley2019, Singh2019, Sims2020}. Nonetheless, several explanations have been proposed in order to explain the unusual absorption trough as the absorption depth of the detected signal is almost two times larger than the strongest theoretical prediction. 
The 21-cm differential brightness temperature depends on the background radio signal, the hydrogen spin temperature and the kinetic temperature of the IGM, explaining the detection of the absorption trough requires additional physics that can lead, for instance, to a high SFR density \citep{Mirocha_2019, Mebane_2020, 2022MNRAS.515.2901M} or high UV luminosity density \citep{2023arXiv230502703H}. The depth of the detected signal at $z\sim 17$ could possibly be explained either by a colder IGM achieved via cold dark matter and baryon interaction during cosmic dawn \citep[see e.g.,][]{Barkana18Nature, Barkana18PRD, Munoz18, Slatyer2018}, and the excess radio background over the cosmic microwave background radiation \citep[e.g.,][]{Fraser_2018, Pospelov_2018, Feng_2018, Ewall_2018, Fialkov_2019, Ewall2020}. 

However, the rise of the absorption signal in the redshift range $z\sim 16-14$ requires the IGM temperature to increase very rapidly to match the background radio temperature. {The X-ray heating by the first generation of galaxies/black holes is one such possibility that has been widely studied \citep{pritchard2007, baek09, mesinger13, ghara14, Pacucci2014, Fialkov2014, fialkov14b, Arpan2017, Ma2021}. Other possibilities such as heating due to the primordial magnetic field \citep{SS05, Bera_2020}, shocks \citep{xu21}, Ly-$\alpha$ photons \citep{Madau1997, Chuzhoy2006, ghara2020, Ciardi2010, Reis2021, mittal21}, CMB photons \citep{Venumadhav2018}  have also been explored.} However, all these mechanisms have their own parameter spaces that are poorly constrained in the high redshift universe. Thus, how and when did the IGM heating take place during cosmic dawn still remains an unsettled issue.

\section{Outline of the thesis}
Inspired by the upcoming and future observations, we devoted our work to understanding the physical mechanisms that impact the global HI 21-cm signal. Furthermore, we combine constraints derived from the observation of the global 21-cm signal along with the observations from the reionization and post-reionization epoch to shed light on the faint end of the UV luminosity function.

In \textbf{chapter~\ref{chap:PMF}}, our study on one of the heating sources during the dark ages and cosmic dawn namely, primordial magnetic field (PMF) is described. We investigate the decaying of magnetic fields through ambipolar diffusion (AD) and decaying magneto-hydrodynamic turbulence (DT) processes at different redshifts. We further present the impact of the decaying magnetic field on IGM temperature. Finally, we provide the upper limit on PMFs in the colder IGM background achieved via dark matter-baryon interaction using EDGES constraints on the global 21-cm signal.

We present our semi-analytical models for Population~III and Population~II star formations in \textbf{chapter~\ref{sec:SFRD_th}}. Our models include several feedbacks that impact the star formations at high redshifts, in particular, Lyman-Werner, AGN, and Supernovae feedback. We further describe the Ly-$\alpha$ background due to the generation of Ly-$\alpha$ photons from the first stars.

In \textbf{chapter~\ref{chap:CR}}, we present the detailed mechanisms of the heating processes due to cosmic ray protons and their impact on gas kinetic temperature, $T_g$, spin temperature, $T_s$, and the corresponding brightness temperature, $T_{21}$ during cosmic dawn and EoR. We calculate the evolution of CR particles and their energy deposition in IGM during cosmic dawn. We also describe the impact of CR protons in the presence of dark matter-baryon interaction in light of EDGES 21-cm signal detection. Moreover, we present a comparison of heating due to cosmic rays with X-ray heating.

In \textbf{chapter~\ref{chap:Bridging}}, we explore a source model derived from radiative transfer hydrodynamic simulation. We then calibrate our model with recent measurements of global HI 21-cm signal from CD and with several reionization observables to explore the joint parameter space to determine the range of models that naturally reproduce the combined constraints from reionization and cosmic dawn. 

In \textbf{chapter~\ref{chap:conclusion}}, we discuss and summarise our main results obtained. Finally, in \textbf{chapter~\ref{chap:future}}, we briefly outline the future outlook of this thesis work.
Throughout this article we assume a flat, $\Lambda$CDM cosmology  with the cosmological parameters obtained from recent Planck 2018 \citep{Planck18} observation, i.e. $\Omega_{\Lambda} = 0.69$, $\Omega_{{\textrm m}} = 0.31$, $\Omega_{{\textrm b}} = 0.049$, and the Hubble parameter $H_0 = 67.66$~km/s/Mpc, unless otherwise mentioned.

\fancyhead[RO]{\leftmark}
\fancyhead[LE]{\textsc{\chaptername~\thechapter}}

\chapter[IGM heating due to PMF]{IGM heating due to primordial magnetic fields during cosmic dawn\footnote{This chapter is adapted from the paper, "Primordial magnetic fields during the cosmic dawn in light of EDGES 21-cm signal" by \textcite{Bera_2020}.}}

\epigraph{\itshape  The primordial magnetic field is a tantalizing clue to the physics of the early universe, and a window into the processes that shaped the cosmos we observe today.}{-- J. Richard Bond}

\label{chap:PMF}

\startcontents[chapters]
\printmyminitoc{
The global redshifted HI 21-cm signal from the dark ages and cosmic dawn is a promising tool to study the primordial magnetic field \citep[see][for a review]{kandu2016}. The primordial magnetic field can heat up the Hydrogen and Helium gas in the intergalactic medium by processes such as the ambipolar diffusion (AD) and decaying turbulence (DT) \citep{1998PhRvD..57.3264J, kandu98, KK14, Chluba15}. This indirectly affects the spin temperature and the globally averaged redshifted HI 21-cm  signal \citep{sethi05}. Furthermore,  the growth of structures during the cosmic dawn gets accelerated in  presence of magnetic field in the IGM. As a consequence, the primordial magnetic field can have an important impact on the formations of the first luminous sources \citep{sethi08, Schleicher08}. A substantial amount of theoretical work has been carried out to understand, in detail, the role of the primordial field on the HI 21-cm signal \citep{tashiro2006, Schleicher09, Venumadhav2017, kunze2019}, early structure formation during the cosmic dawn and reionization \citep{kim1996, yamazaki2006, pandey2015}. 

The measurements of the global HI 21-cm absorption signal by the EDGES experiments in the redshift range $z \sim 14$ to $20$ \citep{EDGES18} have opened up a possibility to constrain the primordial magnetic field and understand its evolution during the cosmic dawn and dark ages. The details of this observation are given in the previous chapter. In a recent work, \citet{Minoda19} has exploited the EDGES data to put an upper limit on the primordial magnetic field. The analysis has been carried out on the backdrop of the standard cosmological model and baryonic interactions of the IGM. However, the measured EDGES absorption signal is  $\sim 2 -3$ times stronger compared to predictions by the standard model. If the measurements are confirmed, one promising way to explain the measured signal is to consider the IGM to be significantly  `colder' compared to the IGM kinetic temperature predicted by the standard scenario. As mentioned before, one needs to consider a non-standard cooling mechanism such as the DM-baryon interaction in order to make the IGM colder \citep{Tashiro14, Munoz15}. 

In this chapter, we present the constraints on the primordial magnetic field using the EDGES 21-cm absorption profile on the backdrop of the colder IGM scenario. We consider interactions between cold DM particles and baryons \citep{Tashiro14, Munoz15} which makes the IGM colder as compared to that in the standard predictions. If one considers DM-baryon interaction in order to make the IGM colder, the exact constraints on the primordial magnetic field should also depend on the mass of the DM particles and the interaction cross-section between the DM particles and baryons. Thus, it is important to highlight these aspects in order to understand the role of the primordial magnetic field on the 21-cm absorption signal and put limits on the primordial magnetic field. Recently, \citet{bhatt2020} has used the EDGES low band measurements to study constraints on the primordial magnetic field in the presence of DM-baryonic interaction. Various other observations such as the CMBR, the Sunyaev-Zel'dovich effect, the star formation, blazar light curve have been exploited to constrain the primordial magnetic field \citep{planck15b, saga20, Minoda17, marinacci2016, takahashi2013}. In addition, we explore the redshift evolution of the primordial magnetic field during the dark ages and cosmic dawn. The differential brightness temperature in the EDGES absorption profile starts increasing at redshift $z \sim 16$ which suggests that heating of the IGM started around that redshift. Here, we also investigate if primordial magnetic heating is able to explain this behavior. Our analysis also allows us to study the constraints on the mass of the DM particle and the interaction cross-section in presence of the primordial magnetic field. 

We start with a brief discussion and essential equations regarding the heating due to the primordial magnetic field and heating/cooling due to the interactions between dark-matter and baryons. We then present our results on redshift evolution of IGM temperature and ionization fraction in presence of the primordial magnetic field, DM-baryon interaction, and the combined impact in sections~\ref{subsec:effect of DM}, \ref{subsec:effect-of-PMF} and \ref{subsec:effect of DM-PMF} respectively. We discuss constraints on the dark-matter mass and interaction cross-section in section~\ref{subsec:DM-b-constraints} and present summary in section~\ref{sec:PMF_conclusion}.
}

\section{Energy deposition due to primordial magnetic field}
\label{sec:PMF}
Magnetic field exerts Lorentz force on the ionized component of the IGM. This causes rise in the IGM temperature, $T_g$. There are mainly two processes namely the ambipolar diffusion and decaying turbulence by which the magnetic field can heat up the IGM during the cosmic dawn and dark ages. We follow the prescription presented in \textcite{SS05} and \textcite{Chluba15} to calculate the rate of heating due to these two processes.  The heating rate (in unit of energy per unit time per unit volume)  due to the ambipolar diffusion is given by,
\begin{eqnarray}
    \Gamma_{\rm AD} = \frac{(1-x_e)}{\gamma x_e \rho_b^2} \frac{\Big \langle \vert (\nabla\times \boldsymbol{B})\times \boldsymbol{B} \vert^2 \Big \rangle}{16\pi^2},
    \label{eq:gamma_AD}
\end{eqnarray}
where $x_e = n_e/n_{\rm H}$ is the residual free electron fraction and $n_{\rm H} = n_{\rm HI} + n_{\rm HII}$. We assume $n_{\rm HII} = n_e$ as Helium is considered to be fully neutral in the redshift range of our interest. Further, $\rho_b$ is the baryon mass density at redshift $z$, and the coupling coefficient between the ionized and neutral components is $\gamma=\langle \sigma v \rangle _{HH^+}/{2m_H}= 1.94 \times 10^{14} \, (T_g/{\rm K})^{0.375} \, {\rm cm}^3 {\rm gm}^{-1} {\rm s}^{-1}$. The Lorentz force can be approximated as $ \Big \langle \vert (\nabla\times \boldsymbol{B})\times \boldsymbol{B} \vert^2 \Big \rangle \approx 16 \pi^2 \, \rho_B(z)^2 \, l_d(z)^{-2} f_L(n_B+3)$ \autocite{Chluba15}, where $\rho_B(z)=|{\bf B}|^2/8\pi$ is the magnetic field energy density at redshift $z$, $f_L(p)=0.8313[1-1.02 \times 10^{-2} p]p^{1.105}$, and  $l_d^{-1}= (1+z) \,k_D$. The damping scale is given by $k_D \approx 286.91 \, (B_0/{\rm nG})^{-1}\, {\rm Mpc}^{-1}$ \autocite{KK14}. We note that the above heating rate is inversely proportional to the coupling coefficient  $\gamma$ and the residual electron fraction $x_e$. Furthermore, both $\gamma$ and the ionization fraction, $x_e$ get suppressed when the IGM is colder compared to the standard scenario. As a result, the ambipolar heating rate becomes more efficient during the cosmic dawn and dark ages.

The heating rate due to the  decaying turbulence is described by,
\begin{eqnarray}
    \Gamma_{\rm DT} = \frac{3m}{2} \frac{\Big\lbrack \ln{\left(1+\frac{t_i}{t_d} \right)}\Big\rbrack ^m}{\Big \lbrack \ln{\left(1+\frac{t_i}{t_d}\right)} + \frac{3}{2} \ln{\left(\frac{1+z_i}{1+z}\right)}\Big \rbrack^{m+1}} H(z)\, \rho_B(z),
    \label{eq:gamma_DT}
\end{eqnarray}
where $m=2 \, (n_B+3)/(n_B+5)$, and $n_B$ is the spectral index corresponding to the primordial magnetic field.  The physical decay time scale ($t_d$) for turbulence and the time ($t_i$) at which decaying magnetic turbulence becomes dominant are related as \autocite{Chluba15},
\begin{equation}
    t_i/t_d \simeq 14.8(B_0/{\rm nG})^{-1}(k_D/{\rm Mpc}^{-1})^{-1} .
\end{equation}
The heating rate due to the decaying turbulence is more efficient at early times as it is proportional to the Hubble rate, $H(z)$, and the primordial magnetic energy density. The effect monotonically decreases at lower redshifts and becomes sub-dominant during the cosmic dawn and dark ages. 

It is often assumed that, like the CMBR energy density,  the primordial magnetic field and energy density scale with redshift $z$ as $B(z) = B_0 (1+z)^2$ and $\rho_B(z) \sim (1+z)^4$ respectively under magnetic flux freezing condition.  However, the magnetic field energy continuously gets transferred to the IGM through the ambipolar diffusion and decaying turbulence processes. For the magnetic field with $B_0 \gtrsim 1 \, {\rm nG}$, the transfer may be insignificant compared to the total magnetic field energy and the above scalings holds. However, this may not be a valid assumption for lower magnetic field, $B_0 \lesssim  0.1 $ nG.  Therefore, we self-consistently calculate the redshift evolution of the magnetic field energy using the following equation,
\begin{equation}
    \frac{d}{dz} \left( \frac{\vert \boldsymbol{B} \vert^2}{8 \pi} \right) = \frac{4}{1+z} \left( \frac{\vert \boldsymbol{B} \vert^2}{8 \pi} \right) + \frac{1}{H(z)\,(1+z)} (\Gamma_{\rm DT} + \Gamma_{\rm AD}). 
    \label{eq:mag_density}
\end{equation}
The first term in the right hand side of the above equation quantifies the effect due to the adiabatic expansion of universe, and the second term quantifies the loss of the magnetic energy due to the IGM heating due to the processes described above. 

\section{Dark matter-baryon interaction}
\label{subsec:dark-matter}
As mentioned before, it has been shown in a few recent studies that the interactions between the cold dark matter particles and baryons could help the IGM to cool faster than the standard adiabatic cooling \citep{Barkana18Nature,PhysRevLett.121.011102,Munoz18,Liu2019prd}.
In our analysis, we consider this kind of interaction between the cold DM particles and baryons to explain the unusually strong absorption signal found by the EDGES experiments \citep{Barkana18Nature}. We adopt the Rutherford-like velocity-dependent interaction model presented in \citet{Munoz15} where the interaction cross-section depends on the dark matter-baryon relative velocity as  $\sigma = \sigma_0 (v/c)^{-4}$. 
The dark matter-baryon interaction models are highly constrained by structure formation \citep{Boehm2005}, primordial nucleosynthesis and cosmic rays \citep{Cyburt2002}, CMB anisotropy \citep{Cora2014}, spectral distortions \citet{Yacine2015, James2017}, galaxy clusters \citet{Chuzhoy2006, Hu2008, Qin2001}, gravitational lensing \citet{Natarajan2002, Markevitch2004}, thermal history of the intergalactic medium \citet{Julian2017, Cirelli2009}, 21-cm observations \citep{Tashiro2014}, and so on.
We note that our model satisfies the current limits \citep{Xu2018, Slatyer2018} given on the elastic scattering cross-section between dark matter and baryons using the measurements of CMB temperature and polarization power spectra by the Planck satellite \citep{planck15}, and the Lyman-$\alpha$ forest flux power spectrum by the Sloan Digital Sky Survey (SDSS) \citep{SDSS2004}. 

The energy transfer rate from dark matters to baryons due to such interactions can be modeled as \citet{Munoz15}, 
\begin{equation}
    \frac{dQ_b}{dt} = \frac{2 m_b \rho_{\chi} \sigma_0 e^{-r^{2} / 2} (T_{\chi} - T_g) k_B c^4}{(m_b + m_{\chi})^2 \sqrt{2\pi} u^{3}_{th}} + \frac{\rho_{\chi}}{\rho_m} \frac{m_{\chi} m_b}{m_{\chi} + m_b} V_{\chi b} \frac{D(V_{\chi b})}{c^2}.
    \label{eq:dQ_b}
\end{equation}
Similarly, the heating rate of the DM, $\Dot{Q_\chi}$ can be obtained by just replacing $b \leftrightarrow \chi$ in the above expression due to symmetry.  Here, $m_\chi$, $m_b$ and $\rho_\chi$, $\rho_b$ are the masses and energy densities of dark matter and baryon respectively.  We can see from equation~(\ref{eq:dQ_b}) that the heating rate is proportional to the temperature difference between two fluids i.e. $(T_{\chi} - T_g)$. The second term in equation~(\ref{eq:dQ_b}) arises due to the friction between dark matter and baryon fluids as they flow at different velocities. Hence both the fluids get heated up depending on their relative velocity $V_{\chi b}$ and the drag term $D(V_{\chi b})$, given as,
\begin{equation}
\frac{dV_{\chi b}}{dz} = \frac{V_{\chi b}}{1+z} + \frac{D(V_{\chi b})}{H(z)(1+z)}
\label{eq:vxb}
\end{equation}
and
\begin{equation}
D(V_{\chi b})  = \frac{\rho_m \sigma_0 c^4}{m_b + m_\chi} \frac{1}{V^2_{\chi b}} F(r).
\end{equation}
The variance of the thermal relative motion of dark matter and baryon fluids $u_{th}^2 = k_B(T_b/m_b + T_\chi/m_\chi$)  and $r =V_{\chi b}/u_{th}$. The function $F(r)$ is given by
\begin{equation}
F(r) = erf \Big( \frac{r}{\sqrt{2}} \Big) - \sqrt{ \frac{2}{\pi}} r e^{-r^2/2}.
\end{equation}
We see that $F(r)$ grows with r, $F(0) = 0$ when $r = 0$ and $F(r) \rightarrow 1$ when $r \rightarrow \infty$. This ensures that the heating due to the friction is negligible when the relative velocity $V_{\chi b}$ is smaller compared to the thermal motion of dark matter and baryon fluid $u_{th}$. However, it can be significant if $V_{\chi b}$ is higher than $u_{th}$. 
The evolution of the DM temperature $T_{\chi}$ can be calculated using,
\begin{equation}
    \frac{dT_\chi}{dz} = \frac{2T_\chi}{1+z} - \frac{2}{3k_B} \frac{\Dot{Q_\chi}}{H(z)\, (1+z)}.
    \label{eq:Tchi}
\end{equation}
The first and second terms on the r.h.s quantify the adiabatic cooling and heating rate per dark matter particle due to its interactions with baryons respectively. 

We see from equations~(\ref{eq:Tg}) and (\ref{eq:dQ_b}), that the IGM temperature becomes velocity ($V_{\chi b}$) dependent as soon as  the dark matter-baryon interaction is taken into consideration which, in turn, modifies the brightness temperature $T_{21}$. Therefore, the observable global HI 21-cm brightness temperature is calculated by averaging over the velocity $V_{\chi b}$ as,
\begin{equation}
\langle T_{21} (z) \rangle = \int d^3 V_{\chi b} T_{21} (V_{\chi b}) P(V_{\chi b}),
\label{eq:average}
\end{equation}
where the initial velocity $V_{\chi b, 0}$ follows the probability distribution 
\begin{equation}
P(V_{\chi b, 0}) = \frac{e^{-3 V^2_{\chi b, 0}/( 2 V^2_{\rm rms})}}{(\frac{2 \pi}{3} V^2_{\rm rms})^{3/2}}.
\label{eq:probability}
\end{equation}
In order to calculate the velocity averaged IGM temperature $\langle T_g(z) \rangle$ and ionization fraction $\langle x_e \rangle$, the same procedure is followed.

We simultaneously solve equations~(\ref{eq:ion_frac}), (\ref{eq:Tg}), and (\ref{eq:mag_density} -- \ref{eq:Tchi}) to get the evaluations in $T_g$ and $x_e$ for a range possible values of the dark matter particle mass $m_{\chi}$, the interaction cross-section $\sigma_{45}=\frac{\sigma_0}{10^{-45} \, {\rm m^2}}$, and the initial magnetic field ($B_0$) for a given $V_{\chi b}$. We then use eq. \ref{eq:average} to calculate the averaged quantities such as, $\langle T_{21} (z) \rangle$, $\langle T_g(z) \rangle$ and $\langle x_e \rangle$. Note that all values/results quoted below are these average quantities even if we don't mention them explicitly. Below we discuss our results on the heating due to the primordial magnetic field, the impacts of the DM-baryonic interaction in presence/absence of the magnetic field. In addition, we present our studies on the role of the residual free electron fraction $x_e$ in the context of the colder IGM background, the evolution of the primordial magnetic field, and the upper limit on the primordial magnetic field using EDGES absorption profile.  We set the following initial conditions at redshift $z_i=1010$: $T_{\rm gi}  = 2.725(1+z_i) \, {\rm K};\, T_{\chi i}= 0,\, V_{\chi b,i} = V_{\rm rms i} = 29 \, {\rm km/s}$, $B_i=B_0(1+z_i)^2$ and $x_{ei}=0.055$ \autocite[obtained from RECFAST code\footnote{\url{http://www.astro.ubc.ca/people/scott/recfast.html}}][]{Seager1999, Seager2000}.

\section{Impact on heating due to the primordial magnetic field}
\label{subsec:effect-of-PMF}
\begin{figure}
	\includegraphics[width=\columnwidth]{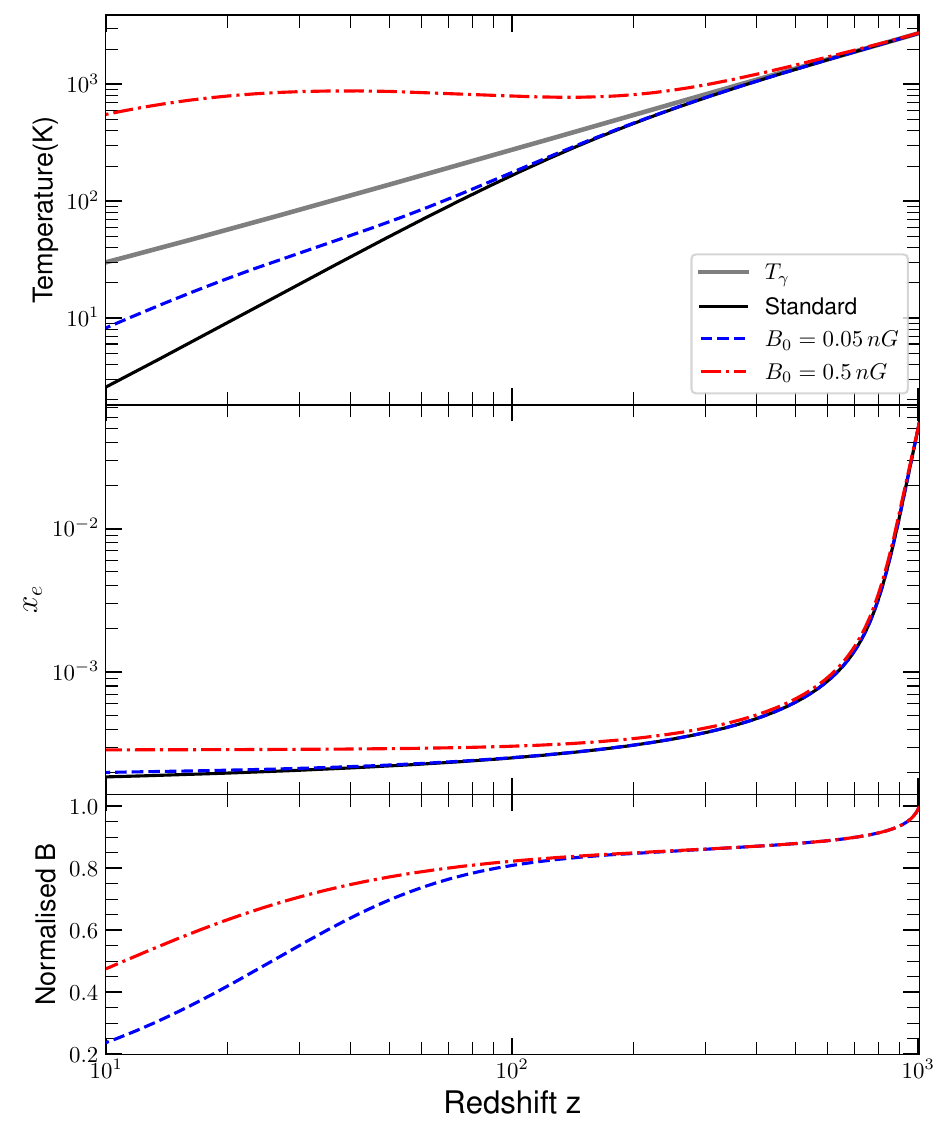}
    \caption[IGM kinetic temperature $T_g$, residual free electron fraction $x_e$, and normalized magnetic field as a function of redshift are shown in presence of the primordial magnetic field.]{The upper and middle panels show the IGM kinetic temperature $T_g$ and residual free electron fraction $x_e$ as a function of redshift  in presence of the primordial magnetic field.  The solid (black), dashed-dotted (red) and dashed (blue) lines correspond to the primordial magnetic field with $B_0 = 0$,  $0.05$ and $0.5\, {\rm nG}$ respectively. The grey thick solid line shows the CMBR temperature $T_{\gamma}$. We do not include the DM-baryon interaction here. The lower panel shows the normalised magnetic field i.e. $\frac{B(z)}{B_0 (1+z)^2}$ for the same $B_0$ values mentioned above.}
    \label{fig:PMF}
\end{figure}
The upper panel of Fig.~\ref{fig:PMF} shows the evolution of IGM temperature $T_g$ in presence of the primordial magnetic field with $B_0=0.05 \, {\rm nG}$ and $0.5 \, {\rm nG}$.  We fix $n_B = -2.9$ throughout our analysis. In order to understand the role of the primordial magnetic field alone we do not include the DM-baryon interaction in Fig.~\ref{fig:PMF}. Note that our results for $B_0=3 \, {\rm nG}$ and Hydrogen only scenario is very similar to that presented in \textcite{Chluba15} for the similar scenario. We find that the primordial magnetic field makes a noticeable change in the IGM temperature during the cosmic dawn and dark ages ($z \lesssim 100$) for $B_0 \gtrsim 0.03 \, {\rm nG}$. This is because the ambipolar diffusion becomes very active at lower redshifts as  it is inversely proportional to the square of the baryon density, $\rho_b$. It also scales with the IGM temperature as $T^{-0.375}_{g}$ (see eq. \ref{eq:gamma_AD}).  The Effects due to the decaying turbulence, which scales as $\Gamma_{DT} \propto H(z) \, \rho_B(z)$, get diluted at lower redshifts. We find that for $B_0 \sim 0.1 \, {\rm nG}$, the IGM temperature rises to the CMBR temperature and, consequently,  the global differential brightness temperature  $T_{21}$ becomes nearly zero. Further increase of the primordial magnetic field causes the IGM temperature to go above the CMBR temperature and $T_{21}$ becomes positive. This is completely ruled out as the EDGES measured the HI 21-cm signal in absorption i.e.,  $T_{21}$ is negative. This put an upper limit on the primordial magnetic field and we find  $B_0 \lesssim 0.1 \, {\rm nG}$, similar to the upper limit found by \textcite{Minoda19}.

The middle panel of Fig.~\ref{fig:PMF} shows the history of residual free electron fraction, $x_e$. We see that $x_e$ increases if we increase the magnetic field $B_0$. This is because of  suppression in the Hydrogen recombination rate $\alpha_e$ due to an increase in the IGM temperature $T_g$. The increase is more prominent during the cosmic dawn and dark ages. For example, $x_e$ increases by a factor of $\sim 1.5$ as compared to the standard prediction at redshift $z=17$ if $B_0=0.5 \, {\rm nG}$. Conversely, the residual free electron fraction $x_e$  directly influences the magnetic heating and its evolution through the ambipolar diffusion process (eq. \ref{eq:gamma_AD} and \ref{eq:mag_density}) which is dominant over the decaying turbulence during the cosmic dawn and dark ages. Therefore it is important to highlight the role of $x_e$ in constraining the primordial magnetic field using the global 21-cm signal. Moreover, $x_e$ also affects the standard Compton heating (see eq. \ref{eq:Tg}). 

The bottom panel of  Fig.~\ref{fig:PMF} shows the evolution of the primordial magnetic field. The normalised primordial magnetic field ($\frac{B(z)}{B_0(1+z)^2}$ ) has been plotted here to highlight any departure  from the simple $B_0(1+z)^2$ scaling. We find that the primordial normalised magnetic field maintains a constant value at higher redshifts $z \gtrsim 100$, and then decays at lower redshifts during the cosmic dawn and dark ages as a considerable fraction of the magnetic field energy is transferred to the IGM for its heating through the ambipolar diffusion process. The ambipolar diffusion becomes very active at lower redshifts for reasons explained in subsection \ref{sec:PMF}. We also notice that the amount of decay of the magnetic field depends on $B_0$. For example, the normalised primordial magnetic field goes down to $\sim 0.4$ and $\sim 0.6$ for $B_0=0.05 \, {\rm nG}$ and $0.5 \, {\rm nG}$ at redshift $z \sim 17.2$. This implies that the fractional decay of the magnetic energy is more when the primordial magnetic field is weaker. For a higher primordial magnetic field with  $B_0 \gtrsim 1 \, {\rm nG}$, the fractional decay is not significant and it can be safely assumed to scale as $(1+z)^2$.

\section{Effect of dark matter-baryon interaction}
\label{subsec:effect of DM} 
\begin{figure}
	\includegraphics[width=\columnwidth]{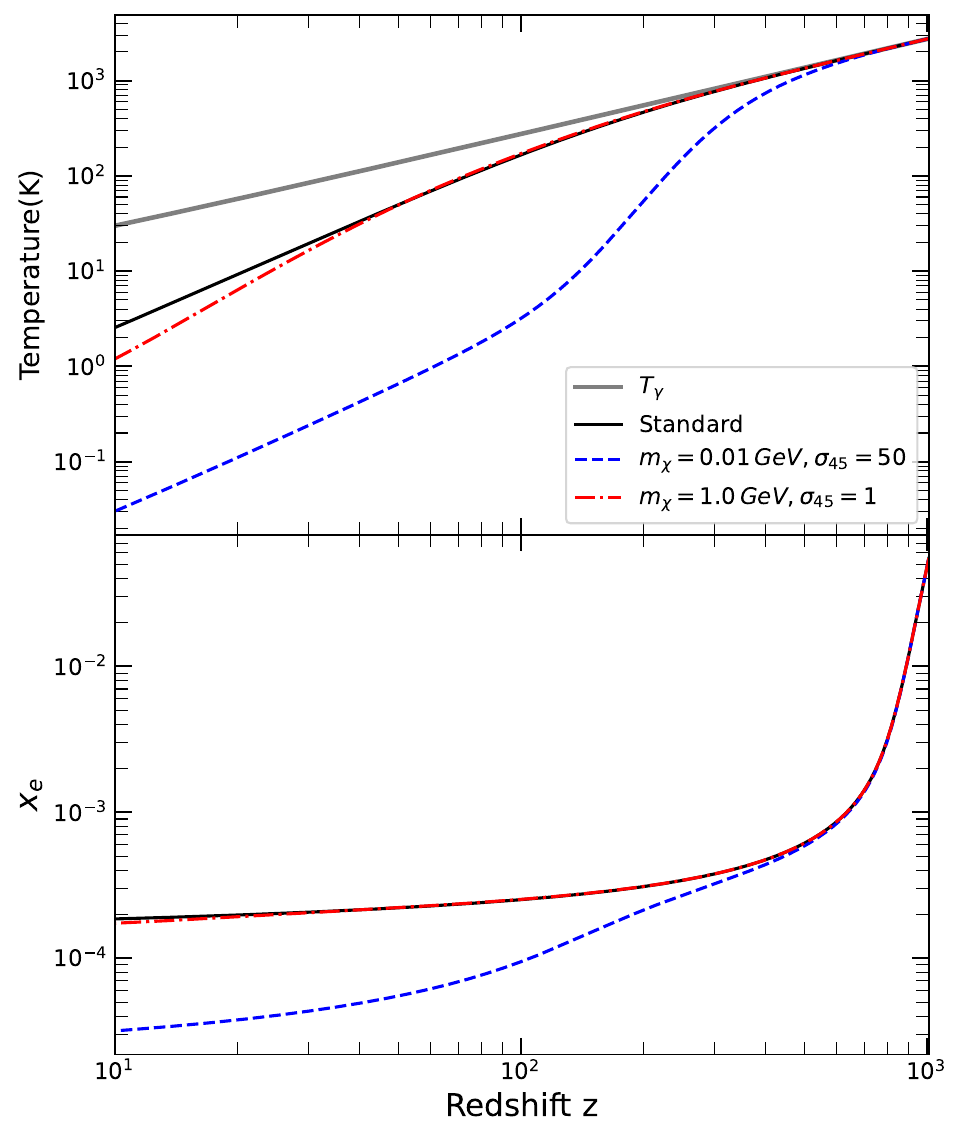}
    \caption[Redshift evolution of the IGM kinetic temperature, $T_g$ and residual free electron fraction, $x_e$ are shown when the DM-baryon interaction is considered.]{The upper and lower panels show the IGM kinetic temperature, $T_g$ and residual free electron fraction, $x_e$ as a function of redshift  when the DM-baryon interaction is considered.  The solid (black), dashed (blue), and dashed-dotted (red) lines correspond to  $(m_\chi/{\rm GeV}, \, \sigma_{45}) = (0,0),\, (0.01, 50)$ and $(1, 1)$ respectively. The effect due to the primordial magnetic field is absent here. The grey thick solid line represents the CMBR temperature $T_{\gamma}$.}
    \label{fig:DM}
\end{figure}

We consider the DM-baryon interaction model that was discussed in Sec.~\ref{subsec:dark-matter}. As mentioned there, the model has two free parameters i.e., the mass of the dark matter particle, $m_{\chi}$ and the interaction cross-section between the dark matter particles and baryons, $\sigma_{45}$. Below we briefly discuss the impact of the DM-baryon interaction on the IGM temperature, $T_g$, and residual free electron fraction, $x_e$. We refer readers to \textcite{Datta20} for a more elaborate discussion.

The upper panel of Fig.~\ref{fig:DM} shows the evolution of IGM temperature for two sets of dark matter mass $m_\chi$ and interaction cross-section $\sigma_{45}$ i.e.,  $(1\, {\rm GeV},\, 1)$ and $(0.01\, {\rm GeV},\, 50)$. We also plot the IGM temperature as predicted in the standard model. As expected,  the interaction helps the IGM to cool faster and the IGM temperature becomes lower than the standard scenario during the cosmic dawn. Lower the dark matter mass, $m_\chi$ and/or larger the cross-section  $\sigma_{45}$, more the rate of IGM cooling is and, consequently, lower the IGM temperature. We note that for higher cross-section $\sigma_{45}$, the IGM temperature gets decoupled from the CMBR temperature early and coupled to the dark matter temperature $T_{\chi}$. This helps the IGM and the dark matter to reach the  thermal equilibrium. After that, both the IGM and dark matter temperatures scale as $(1+z)^2$ which is seen at redshifts $z \lesssim 100$ for  $m_{\chi}=0.01\, {\rm GeV}$ and $\sigma_{45}=50$ (the blue-dashed curve in Fig. ~\ref{fig:DM}). Here we note that there are mainly two effects arising due to the interaction between the cold DM and baryon. First, it helps to cool down the IGM faster (first term of r.h.s. of eq.~\ref{eq:dQ_b}). Second, the friction due to the relative velocity between the DM and baryon can heat up both the DM and IGM (second term of r.h.s. of eq.~\ref{eq:dQ_b}). We find that the friction heating dominates over the cooling for the DM particle mass $m_{\chi} \gtrsim 1\, {\rm GeV}$, and instead of cooling, the IGM gets heated due to the DM-baryon interaction for higher DM particle mass. However, in our case, we need faster cooling off the IGM. Therefore, friction heating always remains subdominant in our case.  The bottom panel of Fig.~\ref{fig:DM} shows the  evolution of the residual free electron fraction, $x_e$.  As expected the residual free electron fraction is lower when the DM-baryon interaction comes into play. This is because the Hydrogen recombination rate $\alpha_e$ is increased when the IGM temperature is lower. The change in $x_e$ is not significant for $m_{\chi}=1$~GeV and $\sigma_{45}=1$ (red curve). However,   $x_e$  is reduced by factor of $\sim 5$  for $m_{\chi}=0.01\, {\rm GeV}$ and $\sigma_{45}=50$ (blue curve). The reduced $x_e$ enhances the rate of IGM heating through the ambipolar diffusion. At the same time, lower IGM temperature reduces the coupling co-efficient $\gamma(T_g)$ (eq. \ref{eq:gamma_AD}), which again enhances the heating rate. Moreover, heating due to the Compton process, which is proportional to $x_e (T_{\gamma}-T_g)$ (second term on the rhs of eq. \ref{eq:Tg}),  gets affected when the IGM is colder compared to the standard scenario.

\section{Combined impact of primordial magnetic field and dark matter-baryon interaction}
\label{subsec:effect of DM-PMF}
\begin{table}
    \renewcommand{\arraystretch}{2.0}
	\centering
	\caption{The table shows the globally averaged differential brightness temperature $T_{21}$ at redshift $z=17.2$ for various set of the model parameters $m_{\chi}$, $\sigma_{45}$ and $B_0$. The allowed range of $T_{21}$ at redshift $z=17.2$ as measured by the EDGES is  $-0.3 \, {\rm K}$ to $-1.0 \, {\rm K}$. \\}
	\label{tab:table1}
    \begin{adjustbox}{width=0.8\textwidth}
    \small
	\begin{tabular}{||c|c|c|c|c||} 
		\hline \hline
		$m_\chi$ & $\sigma_{45}$ & $B_0$ & $T_{21}$ & Allowed\\
		$({\rm GeV})$ &  & $({\rm nG})$ & $({\rm K})$ &\\
		\hline \hline
		$\times$ & $\times$ & $\times$ & -0.22 & $\times$\\
		$\times$ & $\times$ & 0.1 & 0.00 & $\times$\\
		0.1 & 5 & 0.1 & -0.87 & $\surd$\\
		1.0 & 1 & $\times$ & -0.62 & $\surd$\\
		1.0 & 1 & 0.05 & -0.15 & $\times$\\
		0.001 & 300 & 0.4 & -0.33 & $\surd$\\
		0.1 & 50 & 0.1 & -1.08 & $\times$\\
		\hline
	\end{tabular}
    \end{adjustbox}
\end{table}

\begin{figure}
	\includegraphics[width=\columnwidth]{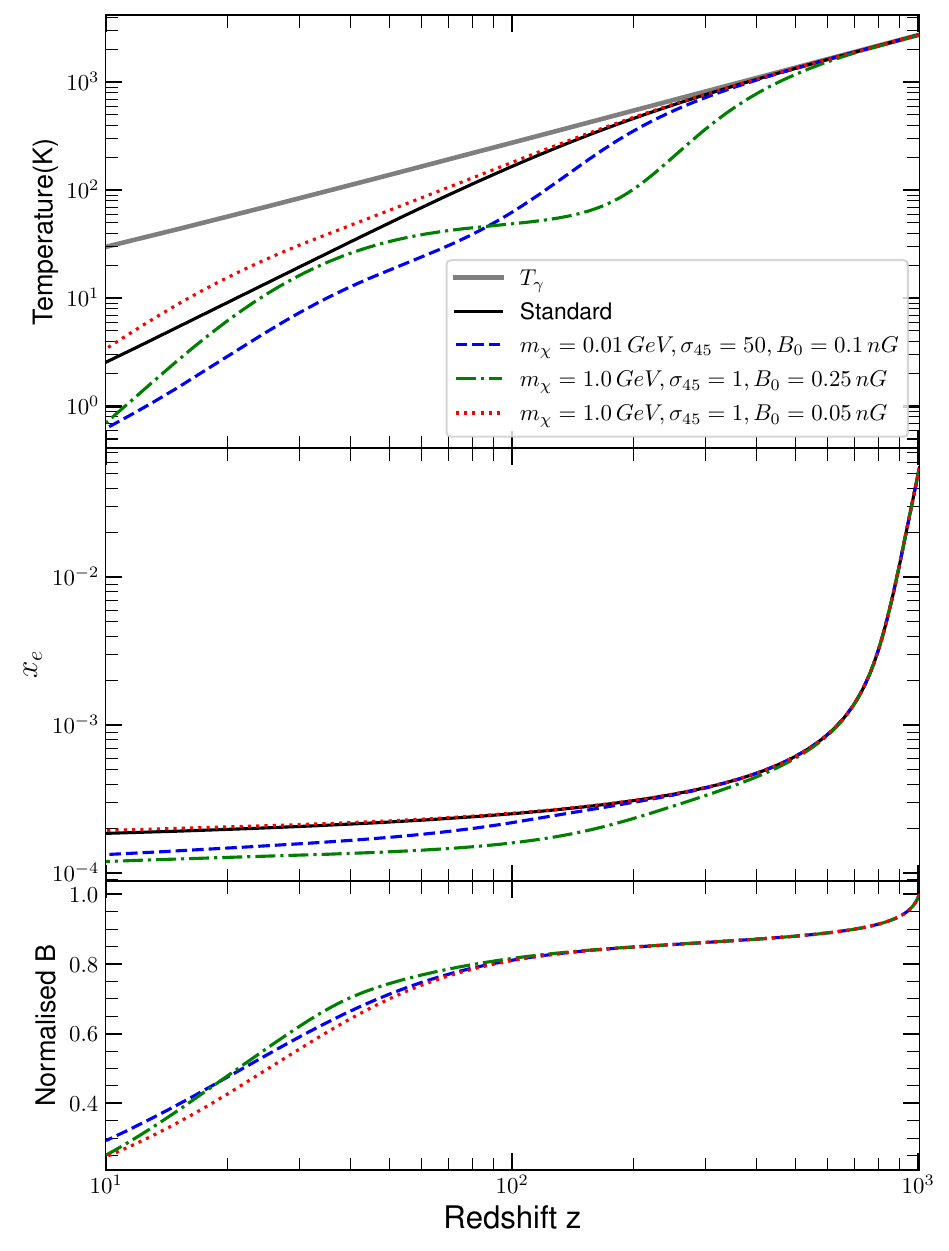}
    \caption[Redshift evolution of the IGM kinetic temperature $T_g$, residual free electron fraction $x_e$ and normalized magnetic field are shown when both the primordial magnetic field and DM-baryon interaction are considered.]{Same as Fig. \ref{fig:PMF}, however both the primordial magnetic field and DM-baryon interaction are considered here.}
    \label{fig:DM_PMF}
\end{figure}

Here we discuss results on the combined impact of the primordial magnetic field and DM-baryon interaction on the IGM temperature evolution.  Fig. \ref{fig:DM_PMF} shows the evolution of the IGM temperature when both the  primordial magnetic field and DM-baryon interactions are considered. In Table~\ref{tab:table1} we have mentioned $T_{21}$ at $z=17.2$ as predicted by our models with different model parameters and shown which parameter set is allowed or not allowed determined by the EDGES measurements. We see that the differential brightness temperature $T_{21}$ at $z=17.2$,  for the parameter set $m_{\chi}=0.001\, {\rm GeV}$,  $\sigma_{45}=30$, is within the allowed range  when the primordial magnetic field with $B_0$ as high as $B_0=0.4 \, {\rm nG}$ is active, although $T_{21}$ is much lower when the magnetic field is kept off. Similarly, the parameter set $m_{\chi}=0.1\, {\rm GeV}$,  $\sigma_{45}=5$ is ruled out as it predicts much lower $T_{21}$ than what  is allowed by the EDGES data. However,  if we include the primordial magnetic field, say, $B_0=0.1 \, {\rm nG}$, the above parameter set becomes allowed. Contrary to this, $T_{21}$ predicted by some combinations of $m_{\chi}$,  $\sigma_{45}$  could be well within the allowed range when there is no primordial magnetic field, but ruled out when the magnetic field is applied. For example $T_{21}=-0.62 \, {\rm K}$ for  $m_{\chi}=1 \, {\rm GeV}$,  $\sigma_{45}=1$ when $B_0=0$, but goes to $-0.15 \, {\rm K}$ which is above the allowed range for $B_0=0.05  \, {\rm nG}$. 

We discussed in section \ref{subsec:effect-of-PMF} that the primordial magnetic field with $B_0 \gtrsim 0.1  \, {\rm nG}$ is ruled out in the standard scenario, but it can be well within the allowed range when the interaction between DM and baryon with an appropriate parameter sets comes into play.  In general, we find that the exact upper limit on the primordial magnetic field depends on the mass of the DM particles $m_{\chi}$ and the DM-baryonic interaction cross-section $\sigma_{45}$. We see that the primordial magnetic field with $B_0 \sim 0.4 \, {\rm nG}$ is allowed for an appropriate set of $m_{\chi}$ and $\sigma_{45}$. Note that this primordial magnetic field is ruled out in the standard scenario.

The upper panel of Fig. \ref{fig:DM_PMF} shows that the primordial magnetic field and DM-baryonic interaction together introduces a `plateau-like feature' in the redshift evolution of the IGM temperature for a certain range of model parameters $m_{\chi}$, $\sigma_{45}$  and $B_0$. One such example can be seen for $m_{\chi}=0.01\, {\rm GeV}$,  $\sigma_{45}=50$  and $B_0=0.25 \, {\rm nG}$ where the plateau like feature is seen in redshift range $\sim 50-150$.  The cooling rate due to the DM-baryonic interaction and heating rate due to the primordial magnetic field compensates each other for a certain redshift range which gives the plateau-like feature. At lower redshifts, the heating due to the primordial magnetic field, which scales as $ B^4(z)$, becomes ineffective as the primordial magnetic field decays very fast. This is both due to the adiabatic expansion of universe and loss of the magnetic energy due to heating. We notice that this plateau-like feature is not so prominent for lower primordial magnetic field. The `plateau-like feature' is a unique signature of the DM-baryonic interaction in presence of the primordial magnetic field. However, it can only be probed by space-based experiments as it appears at redshift range $\sim 50-150$.  

The middle and lower panels of Fig. \ref{fig:DM_PMF}  show the residual electron fraction, $x_e$, and primordial magnetic field, $B(z)$ as a function of redshift respectively. Like in previous cases, the residual electron fraction $x_e$ is suppressed  when both the DM-baryonic interactions and primordial magnetic field are active. The suppressed residual electron fraction enhances the heating rate occurring due to the ambipolar diffusion. The primordial magnetic  field loses its energy (other than the adiabatic loss because of the expansion of universe) due to the transfer of energy to IGM heating through the ambipolar diffusion process. This loss starts becoming important at lower redshifts $z \lesssim 100$.  As the primordial magnetic field decays very fast, the magnetic heating becomes ineffective at lower redshifts.  The EDGES absorption spectra show that the IGM temperature is rising  at  redshifts $z \lesssim 17$. There are several possible mechanisms by which the IGM can be heated up such as heating due to soft X-ray, Ly-$\alpha$, DM decay/annihilation \autocite{pritchard2007, ghara14, ghara2020, sethi05, furlanetto2006, liu2018}. However, we find that the primordial magnetic field is not able to considerably heat up the IGM at the later phase of the cosmic dawn and, therefore, can not explain the heating part of the EDGES absorption profile. 

\section{Constraints on dark matter-baryon interaction in presence of the primordial magnetic field}
\label{subsec:DM-b-constraints}
\begin{figure*}
	\includegraphics[width=\textwidth]{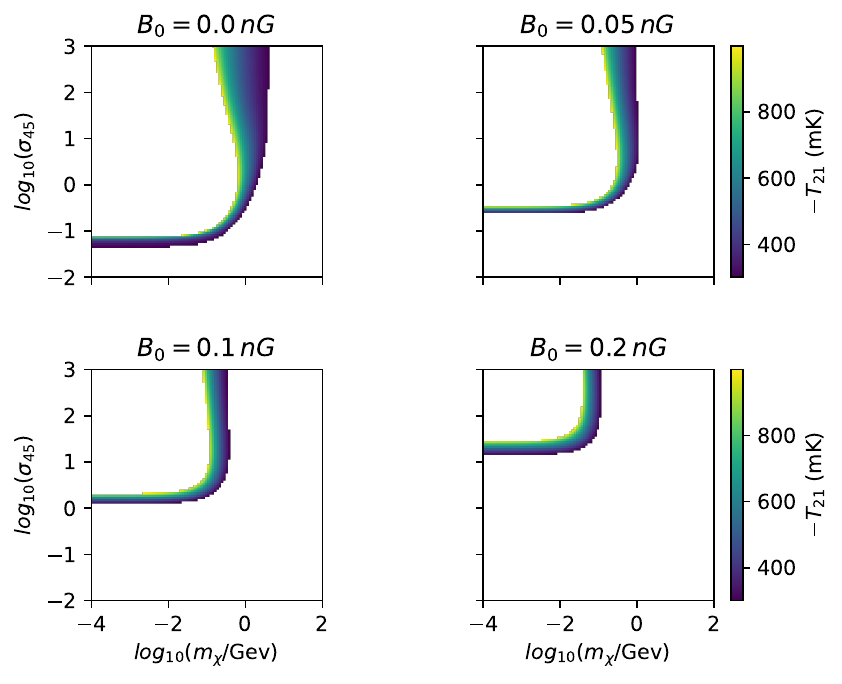}
    \caption[Bounds on dark matter mass, $m_{\chi}$ and interaction cross-section, $\sigma_{45}$ in presence of the primordial magnetic field considering the constraints on $T_{21}$ given by EDGES.]{Bounds on dark matter mass, $m_{\chi}$ and interaction cross-section, $\sigma_{45}$ in presence of the primordial magnetic field considering the constraints on $T_{21}$ given by EDGES. As the amplitude of magnetic field, $B_0$ increases, the allowed dark-matter mass decreases and cross-section increases. The colour bar denotes the allowed $T_{21}$ constrained by EDGES experiment.}
    \label{fig:contour}
\end{figure*}
We now provide the constraints on the allowed ranges of the dark-matter mass and interaction cross-section using EDGES observation.
Fig. \ref{fig:contour} demonstrates the  constraints on the DM-baryon interaction  in presence of the primordial magnetic field. The top left panel presents constraints on the model parameters $m_{\chi}$ and $\sigma_{45}$  when there is no magnetic field i.e., $B_0=0$. This is quite similar to constraints obtained by \textcite{Barkana18Nature}.  Note that the constraints are obtained by restricting the differential brightness temperature $T_{21}$ within $-0.3 \, {\rm mK}$ to $-1.0 \, {\rm K}$ as suggested by the EDGES measurements.  The DM particle with mass higher than a few ${\rm GeV}$ is ruled out because the cooling due to the DM-baryonic interaction becomes inefficient and the drag heating due to the friction between the DM and baryon starts to dominate for higher DM particle mass. Therefore, the drag heating is found to have a very negligible role in the case considered here. The top right, bottom left and the bottom right panels show constraints on the model parameters $m_{\chi}$ and $\sigma_{45}$  in presence of the primordial magnetic field with $B_0=0.05, \, 0.1 $ and $0.2 \, {\rm nG}$ respectively. We see that the allowed range of the DM-baryon cross-section $\sigma_{45}$  gradually increases as $B_0$ is increased. For example, the lowest allowed $\sigma_{45}$  moves up, from $\sim 4 \times 10^{-47} \, {\rm m^2}$,  to $\sim 2.5 \times 10^{-46} \, {\rm m^2}$, $\sim 1.5 \times 10^{-45} \, {\rm m^2}$ and $\sim 1.5 \times 10^{-44} \, {\rm m^2}$ for $B_0=0.05, \, 0.1 $ and $0.2 \, {\rm nG}$ respectively. On the other hand, the maximum allowed mass of the DM particle $m_{\chi}$ gradually decreases for higher magnetic field. In Fig. \ref{fig:contour} we find that the highest allowed $m_{\chi}$ goes down, from $\sim 5  \, {\rm GeV}$, to $ \sim 1\, {\rm GeV}$,  $ \sim 0.3 \, {\rm GeV}$ and $\sim 0.1 \, {\rm GeV}$ for $B_0=0.05, \, 0.1 $ and $0.2 \, {\rm nG}$ respectively. The primordial magnetic field heats up the IGM and the heating is more for higher values of $B_0$. The DM-baryonic interaction needs to be more efficient  to compensate for this extra heating which can be achieved either by increasing the cross-section $\sigma_{45}$ or/and lowering the mass of the dark-matter particle $m_{\chi}$.

The above discussion also tells that the exact upper limit on the primordial magnetic field parameter $B_0$ depends on the mass $m_{\chi}$ and the cross-section $\sigma_{45}$. Higher primordial magnetic field is allowed if $\sigma_{45}$  is increased and/or  $m_{\chi}$ is decreased. We see that the primordial magnetic field with $B_0 \sim 0.4 \, {\rm nG}$ (Table \ref{tab:table1}) is allowed for an appropriate set of $m_{\chi}$ and $\sigma_{45}$. Note that $B_0 \gtrsim 0.1 \, {\rm nG}$ is ruled out in the standard scenario. However, we find that the primordial magnetic field with $B_0 \gtrsim 1 \, {\rm nG}$ may not be allowed as this requires very efficient cooling of the IGM which is unlikely even for very high cross-section and lower DM particle mass. Although, we note that a recent study by \textcite{bhatt2020}, which has used the EDGES measurements, finds an upper limit of $\sim 10^{-6} \, {\rm G}$ on the primordial magnetic field  for $m_{\chi} \lesssim 10^{-2} \, {\rm GeV}$ in presence of the DM-baryonic interaction.
 
\section{Summary}
\label{sec:PMF_conclusion}
Our analysis demonstrates the prospects of constraining the primordial magnetic field in light of the EDGES low band 21-cm absorption spectra during the cosmic dawn. Our study is carried out on the background of `colder IGM'  which is a promising avenue to explain the strong absorption signal found by the EDGES. We consider an interaction between baryons and cold DM particles which makes the IGM colder than in the standard scenario. The primordial magnetic field heats up the IGM through the ambipolar diffusion and decaying turbulence which, in turn, influences the 21-cm differential brightness temperature. We highlight the role of the residual electron fraction. We also study constraints on the DM-baryon interaction in presence of the primordial magnetic field, features in the redshift evolution of IGM temperature. In addition, we consider the redshift evolution of the primordial magnetic field during dark ages and cosmic dawn. In particular, we focus on the departure from the simple adiabatic scaling of the primordial magnetic field ( i.e. $B(z) \propto (1+z)^2$ ) due to the transfer of magnetic energy to the IGM.

Studying the role of the primordial magnetic field on the background of colder IGM is important for several reasons. First, it suppresses the abundance of the residual free electron fraction $x_e$  \autocite{Datta20} which, in turn, enhances the rate of IGM heating through the ambipolar diffusion. Second, the coupling coefficient between the ionized  and neutral components $\gamma(T_g)$ decreases with the IGM temperature, which again results in the increased heating rate  (see eq. \ref{eq:gamma_AD}). Third, the heating rate due to the Compton process, which is proportional to  $(T_{\gamma} -T_g)$ and $x_e$, too gets affected when the background IGM temperature $T_g$ is lower (eq. \ref{eq:Tg}).  We find that collectively all these effects make the heating rate due the magnetic field faster in the colder background in comparison to the heating rate in the standard scenario. Consequently, the primordial magnetic field decays, with redshift, at a much faster rate compared to the simple $(1+z)^2$ scaling during the dark ages and cosmic dawn. The decay is particularly significant for $B_0 \lesssim 0.5 \, {\rm nG}$ when the fractional change in the magnetic field due to the heating loss could be $\sim 50 \%$ or higher. This is unique in the colder IGM scenario.  

Next, we find that the upper limit on the primordial magnetic field using the EDGES measurements is determined by the underlying non-standard cooling process, i.e., the DM-baryon interaction here.  Higher primordial magnetic field may be allowed when the underlying DM-baryon interaction cross section is higher and/or the DM particle mass is lower, i.e.,  the exact upper limit on $B_0$ depends on the DM mass and the interaction cross-section. For example, the primordial magnetic filed with $B_0 \sim 0.4 \, {\rm nG}$  which is ruled out in the standard model \autocite{Minoda19}, may be allowed if the DM-baryon interaction with   $m_{\chi}=0.01 \, {\rm GeV}$ and $\sigma_{45}=100$ is included. However, we find that the primordial magnetic field with $B_0 \gtrsim 1 \, {\rm nG}$ may not be allowed as this requires very efficient cooling of the IGM which is unlikely to occur  even for very strong possible DM-baryon interaction.

Furthermore, we observe that the primordial magnetic field and DM-baryonic interaction together introduces `a plateau-like feature' in the redshift evolution of the IGM temperature for a certain range of model parameters $m_{\chi}$, $\sigma_{45}$  and $B_0$. The cooling rate due to the DM-baryonic interaction and heating rate due to the primordial magnetic field compensates each other for a certain redshift range which produces the plateau-like feature. However, this kind of plateau is not prominent for lower primordial magnetic field with $B_0 \lesssim 0.1 \, {\rm nG}$.

The EDGES absorption spectra suggest that the IGM temperature has possibly  gone up from $\sim 3 \, {\rm K}$ at redshift $z \approx 16$ to $\sim 40 \, {\rm K}$ at redshift $z \approx 14.5$. There are several possible candidates such as soft X-ray photons from the first generation of X-ray binaries, mini-quasars, high energy photons from DM-decay/annihilations, primordial magnetic field, etc. which could heat up the IGM during the cosmic dawn. However, our study shows that the heating due the primordial magnetic field becomes very weak during the above redshift range. Because the magnetic energy density decreases very fast prior to the cosmic dawn both due to the adiabatic expansion of universe and the loss due to IGM heating. Therefore, it is unlikely that the primordial magnetic field contributes to the heating of the IGM during the late phase of the cosmic dawn as indicated by the EDGES measurements. 

Finally, we see that the allowed DM-baryon cross-section $\sigma_{45}$ gradually shifts towards higher values as $B_0$ is increases. On the other hand, the allowed mass of the DM particle $m_{\chi}$ gradually decreases for higher values of the primordial magnetic field. Because, the DM-baryon interaction needs to be more efficient to compensate for the excess heating caused due to higher magnetic field, which can be achieved either by increasing the cross-section or lowering the mass of the Dark matter particle.

There could be various other models of the DM-baryon interactions, for which the exact upper limit on the primordial magnetic field, and all other results  discussed above might change to some extent. However, the general conclusions regarding the role of the primordial magnetic field on a colder IGM background are likely to remain valid for any mechanism providing faster cooling off the IGM.  
We now focus on different radiation backgrounds that are generated due to the formation of first stars or early galaxies during cosmic dawn and epoch of reionization.

\chapter[First stars and early galaxies]{First stars and galaxies in the early universe}
\epigraph{\itshape When we look out into space, we're looking back in time; the light from a galaxy a billion light-years away, for instance, will take a billion years to reach us. The history is there for us to see. It's not mushed up like the geologic record of Earth. You can just see it exactly as it was.}{-- Margaret Geller}
\label{sec:SFRD_th}

\startcontents[chapters]
\printmyminitoc{
After the epoch of recombination, for the first few hundred million years, universe was mostly filled with neutral hydrogen and helium atoms.
In the hierarchical structure formation, the dark matter halos were formed by gravitational collapse. The baryons (primordial hydrogen and helium) were pulled into the potential wells created by the dark matter halos. 
The halo mass function can provide the number density of such halos at different redshifts.
The differential halo mass function or, in short, halo mass function is the number of halos in the mass range  of $M$ and $M$+$dM$ per unit comoving volume. This can be written as \citep[see][for a review]{Reed_2007}, 
\begin{equation}
    \frac{dn(M, z)}{dM} dM= \frac{\bar \rho_0}{M} f(\sigma) \left| \frac{d \ln \sigma}{dM} \right|dM.
    \label{eq:dndM}
\end{equation}
Here $\bar \rho_0$ is the total matter density of the universe at present
and $\sigma(M,z) $ is the RMS fluctuations in the {dark matter} density field at scales corresponding to mass $M$ and redshift $z$.
Note that, the function $f(\sigma)$ depends on the particular form of the halo mass function that one adopts. There are several prescriptions of the halo mass function and among them Press–Schechter formalism \citep{press1974} and Sheth-Tormen formalism \citep{S-T_1999} are widely used.
In this work, we adopt the Sheth-Tormen mass function as it fits well with the mass functions obtained from numerical simulations for a broader range of masses and redshifts. Once the halo collapsed,
the baryonic gas inside the dark matter halos would form stars if the gas cools. 

The comoving global star formation rate densities can be calculated as \citep{Samui_2007},
\begin{equation}
    {\rm SFRD}(z) = \int_{z}^{\infty} dz_c \int_{M_{\rm min}}^{\infty} dM^{\prime} \dot M_{*}(M^{\prime},z,z_c) {\frac{d^2n(M^{\prime},z_c)} {dz_c dM^{\prime}}}, 
    \label{eq:SFR_th}
\end{equation}
where $z_c$ is the collapse redshift of the dark matter halos which is forming stars at a rate $\dot M_{*}$ at redshift $z<z_c$. 
Here, we assume that the redshift derivative of $\frac{dn(M, z)}{dM}$ (see Appendix for the calculation) in equation~\ref{eq:SFR_th} provides the formation rate of dark matter halos \citep{Samui_2009}. Further, $M_{\rm min}$ is the lower mass cut-off of halos that can form stars. Note that, $\dot M_{*}$ depends on the halo mass $M$ and the cooling mechanisms.
As the gas cools and gas mass exceeds the Jeans mass, first stars started to form. Existing theoretical studies \citep{bromm04,abel02} suggest that the first stars thus formed were massive, luminous, and metal-free, known as Population~III (hereafter Pop~III) stars. They are likely to produce copious amounts of UV photons and strongly affect the high redshift IGM, in turn, the cosmic 21-cm signal during cosmic dawn \citep{Fialkov2014, 2015MNRAS.448..654Y, 2018MNRAS.478.5591M, Mebane_2018, 2018MNRAS.480.1925T, 2019ApJ...877L...5S, atri20, Bera2022,  2022MNRAS.511.3657M, 2022arXiv220102638H}. 
Our models of Population~III (Pop~III) and Population~II (Pop~II) stars are described below in detail.
}


\section{Pop~III star formation}
\label{subsec:PopIII}
In this section, we describe the Pop~III star formation model in  small mass halos.
We mostly follow \citet{Mebane_2018} for this purpose. 
Pop~III stars are the first generation metal-free stars that form inside the minihalos where the gas can cool by the H$_2$ cooling only.
The star formation mechanism in these halos is not fully understood and there is no consensus about the initial mass function of Pop~III stars yet \citep{abe2021,parsons2021,lazar2022}. Pop~III stars could be very massive ($>100 M_{\odot}$) if the Jeans mass clump does not experience several fragmentation \citep{bromm99,abel02}. Due to the small size of their host halos, these stars are likely to form in small numbers, possibly in isolation.
{Most of them are expected to be short-lived and either explode as a supernova or directly collapse into a black hole depending on their masses \citep{Mebane_2018}. Stars, having a mass in the range $40$~M$_{\odot}$ to $140$~M$_{\odot}$ and also above $260$~$\rm M_{\odot}$, are likely to collapse into a black hole. On the other hand, stars below $40$~M$_{\odot}$ preferably end their lives in a core-collapse supernova of energy $\sim 10^{51}$ erg. The intermediate mass range ($140$~$\rm M_{\odot}$ - $260$~$\rm M_{\odot}$) stars are likely to explode as a pair-instability supernova that releases a kinetic energy of $\sim 10^{52}$ erg \citep{Wise2008, Greif2010}.}
{Here for the demonstration purpose, we assume that a Pop~III star of mass $145$~$\rm M_{\odot}$ is being formed per galaxy which is likely to explode as a SNe of energy $\sim 10^{52}$ erg \citep[mid mass range of ][]{Mebane_2018}. The variation of the IMF will be taken care of by the parameter $\epsilon_{\rm III}$ described later. } 

Note that, the H$_2$ cooling depends on the amount of molecular hydrogen present in the halo.
{The molecular hydrogen forms through $\rm H^-$ and $e+$ catalysis. It can be destroyed by CMBR photons at high redshift but it gets destroyed more by the presence of Lyman-Werner (LW) photons during the cosmic dawn that we model later in this section.}
Hence, the formation and destruction rates of $\rm H^-$ need to be balanced to get the desired H$_2$ in a halo.
\citet{Tegmark_1997} found that this fraction of molecular hydrogen varies with the halo's virial mass/temperature, and can be approximated as,
\begin{equation}
    f_{\rm H_2} \approx 3.5 \times 10^{-4} T_3^{1.52},
    \label{eq:fH2}
\end{equation}
where $T_3 = T_{\rm vir}/10^{3}$ K. Further note that there is a critical threshold fraction of molecular hydrogen required for the cooling to be efficient to form Pop~III stars. This critical threshold fraction is given by \citep{Tegmark_1997}, 
\begin{equation}
    f_{\rm crit,H_2} \approx 1.6 \times 10^{-4} \left( \frac{1+z}{20} \right)^{-3/2} \left( 1+ \frac{10 T_3^{7/2}}{60+T_3^4} \right)^{-1} exp\left( \frac{0.512 K}{T_3} \right).
    \label{eq:f_cH2}
\end{equation}
The molecular cooling becomes efficient once $f_{\rm H_2} > f_{\rm crit,H_2}$ and this sets the minimum mass of halo that can host Pop~III stars (i.e. $M_{\rm min}$ in Eq.~\ref{eq:SFR_th}) in absence of any feedback.

{
The Pop~III stars are the first generation of metal-free stars likely to form in the small mass haloes at high redshifts. The transition from Pop~III to Pop~II star formation is governed by the amount of metal present in the star forming haloes. Several authors have modeled such transition with different model prescriptions. For example, \citet{sun_2021} have used a critical halo mass governed by the metal mixing time scale for the maximum halo mass that can host Pop~III stars and have shown such simple model can reproduce the Pop~III star formation reasonably well with detailed metal mixing model such as \citet{Mebane_2018}. They have also shown that the Pop~III star formation in two extreme models vary less than an order of magnitude. Motivated by that we assume here that the critical mass for transition from Pop~III to Pop~II stars occurs at a fixed mass which is the atomic cooling mass (corresponding to a virial temperature $T_{\rm vir}=10^4~$K). A detailed modeling may be more appropriate. However, as we will show that even with this simple assumption the resulting Pop~III star formation matches reasonable well with the detailed metal mixing models \citep[i.e.][]{Mebane_2018, sun_2021}.
}

As soon as the first generation of stars form they produce a background of Lyman-Werner radiation. The LW band consists of photons having energy in the range of {11.2}-13.6 eV which can photo-dissociate the molecular H$_2$. Thus, if a halo is present in a LW background, {it  reduces the concentration of H$_2$ to a level that H$_2$ cooling becomes inefficient and the halo cannot host any stars further.}
As a consequence, the minimum mass of a halo that can sustain the Pop~III star formation increases. Therefore, a self-consistent calculation is required to determine the minimum halo mass that can harbour Pop~III stars. 
Given the star formation, the background Lyman-Werner flux ($\rm J_{LW}$) can be calculated as \citep{visbal14},
\begin{equation}
    J_{\rm LW}(z) = \frac{c}{4 \pi} \int_z^{z_m} \frac{dt}{dz^{'}} (1+z)^3 \epsilon(z^{'}) dz^{'}.
    \label{eq:J_LW}
\end{equation}
Here $z_m$ is the maximum redshift that a LW photon can travel through IGM undisturbed and redshift into a Lyman series line, and can be written as, $\frac{1+z_m}{1+z} = 1.04$ \citep{visbal14}. $c$ is the speed of light and $\epsilon(z)$ is the specific LW comoving luminosity density which is given by,
\begin{equation}
    \epsilon(z) = \int_{z}^{\infty} dz_c \int_{M_{\rm min}}^\infty {\frac{d^2n(M,z_c)}{dz_c dM}} \frac{\dot M_{*}}{m_p} \left( \frac{N_{\rm LW} E_{\rm LW}}{\Delta \nu_{\rm LW}} \right) dM,
    \label{eq:eps_z}
\end{equation}
where $m_p$ is the proton mass, $E_{\rm LW} = 11.9\,{\rm eV}$ is the average energy of a LW photon and $\Delta \nu_{\rm LW} = 5.8 \times 10^{14}$ Hz \citep{Mebane_2018} is the LW frequency band.
Further, $N_{\rm LW}$ is the number of LW photons produced per baryon of stars. 
Note that, both Pop~III and Pop~II stars (that we discuss later) produce LW photons. For Pop~III stars we take $N_{\rm LW} = 4800$  and $N_{\rm LW} = 9690$ for Pop~II stars \citep{Barkana_2005, Pritchard_2006}. 

In the presence of the LW background
given above \citet{machacek01} showed that the mass of a
halo that can host Pop~III stars
must have virial temperature above a critical temperature given by,
\begin{equation}
    \frac{T_{\rm crit}}{1000 K} \sim 0.36 \left[ (\Omega_{\rm b} h^2)^{-1} (4 \pi J_{\rm LW}) \left( \frac{1+z}{20} \right)^{3/2} \right]^{0.22},
    \label{eq:LW_mass}
\end{equation}
where $J_{\rm LW}$ is in unit of $10^{-21} \rm{erg \, s^{-1}cm^{-2}Hz^{-1}sr^{-1}}$. We use this $T_{\rm crit}$ to calculate $M_{\rm min}$ in presence of LW background if that $M_{\rm min}$ is greater than the $M_{\rm min}$ obtained using the condition given in Eq.~\ref{eq:f_cH2}.

Once the first generation of stars enrich the IGM with metals, the metal-enriched Population~II stars began to form in the atomic cooling halos. The properties of these stars are quite similar to the stars that we see today, and their properties are much more constrained compared to Pop~III stars. There already exist several models of Pop~II stars that are well-constrained by different observational evidence \citep{Galform, 2005MNRAS.361..577C, 2006MNRAS.371L..55C, Galacticus,  Samui_2014,  dayal2018}.  
The number density of the Pop~III stars is expected to decrease rapidly at $z\lesssim 15$ once the LW feedback becomes significant. Thus, the major contribution to the ionization of the IGM neutral hydrogen is expected to come from Pop~II stars inside the first galaxies. Given a dark matter halo of mass $M_{\rm halo}$, the amount of stellar mass contained, intrinsic spectral energy distribution, IMF, escape fraction of these ionizing photons, etc. are still uncertain for these early galaxies. Thus, numerical simulations of EoR generally work under simplified pictures of Pop~II star formation. Most of these simulations assume simple scaling relations between the total stellar mass  and the hosting dark matter halo mass \citep{ghara15b, 2019MNRAS.487.1101R, 2021MNRAS.501....1G}. 

The impact of the fraction of baryons residing within the stars in a galaxy, the average number of ionizing photons per baryon produced in the stars,  and the escape fraction of the UV photons are degenerate on the ionization/thermal states of the IGM. Thus, many studies \citep[e.g.,][]{iliev12, ghara2020, 2022MNRAS.511.2239M} treat the product of all these quantities as a single parameter termed as ionization efficiency parameter. While semi-numerical simulations \citep[e.g.,][]{majumdar11, 2021MNRAS.501....1G, 2022MNRAS.511.2239M} do not have provision to adopt the spectral energy distributions (SEDs) of the galaxies, the radiative transfer simulations such as \citet{mellema06, 2011MNRAS.414..428P, ghara15b} use SEDs of such early galaxies assuming simple forms such as a blackbody spectrum or using a SED generated from population synthesis codes such as PEGASE \citep{Fioc97}. The ionizing photons emitted from the galaxies not only ionize the neutral hydrogen in the IGM but also suppress star formation in low-mass galaxies due to thermal feedback. Thus, the star formation inside dark matter halos with mass $M_{\textrm halo} \lesssim 10^9 ~M_{\odot}$ becomes inefficient if those halos remained inside ionized regions \citep{ghara18, 2021MNRAS.506.3717R}. The semi-analytical model for Pop~II star formation that we use in our work is given in the following section. 


\section{Pop~II star formation}
\label{subsec:PopII}
Here, we outline the star formation model in atomic cooling halos with the virial temperature greater than $10^4~$K (Pop~II stars) that we adopt from \citet{Samui_2014}. We choose this SNe feedback regulated star formation model as it can successfully describe
the galaxy luminosity functions upto $z=10$, and can explain the observed stellar mass in galaxies of mass ranges from $10^{7} M_{\odot} \leq M \leq 10^{13} M_{\odot}$. {Note that, the model is well calibrated till $z=10$ and it is extrapolated to high redshift which may introduce some biases.} We briefly describe the model here.

In the presence of SNe feedback, the star formation rate of a galaxy of total baryonic mass $M_b$, at a time $t$ after the formation of dark matter halo, can be written as,
\begin{equation}
    \dot M_* = \frac{M_b f_* f_t}{\tau [f_t (1+\eta_\omega)-1]} \left[e^{-\frac{t}{\tau}} - e^{-f_t (1+\eta_\omega)\frac{t}{\tau}}\right].
\end{equation}
Here, the amount of SNe feedback in the form of galactic outflows has been characterised by the parameter $\eta_{\omega}$ which is defined as $\dot M_{\omega}=\eta_{\omega} \dot M_*$, $\dot M_{\omega}$ being the mass outflow rate.
It depends on the circular velocity ($v_c$) of the galaxy as well as
the driving mechanism of the outflow. We consider $\eta_{\omega}=(v_c/100~{\rm km/s})^{-2}$ that describes the outflows driven by the cosmic rays along with the hot gas produced by the SNe. The normalization constant is chosen to fit the UV luminosity functions of high redshift galaxies \citep[see][for details]{Samui_2014}. 
The total baryonic mass is related to the halo mass as $M_b=(\Omega_{\rm b}/\Omega_{\rm m}) M$. Further, the dimensionless parameter $f_t$ fixes the duration of the star formation activity in terms of the dynamical time, $\tau$, of the halo.
It is assumed that once the dark matter halo virialises and accretes the baryonic matter, a fraction, $f_*$ of the gas gets cooled and becomes available for star formation.
Hence, $f_*$ characterises the star formation efficiency and has been optimized to fit various available observation of high redshift galaxies.
Further, note that, along with the SNe feedback, our Pop~II star formation model includes other feedbacks such as the radiative feedback from the ionizing photons, and AGN feedback as well. We are not describing all these feedback processes here in details, however the same can be found in \citet{Samui_2007}. As the ionization fraction is likely to be low at $z > 10$, the contribution from radiative feedback as well as AGN feedback are comparatively small at the redshift of our interest in present work.

\begin{figure*}
    \centering
	\includegraphics[width=0.7\columnwidth]{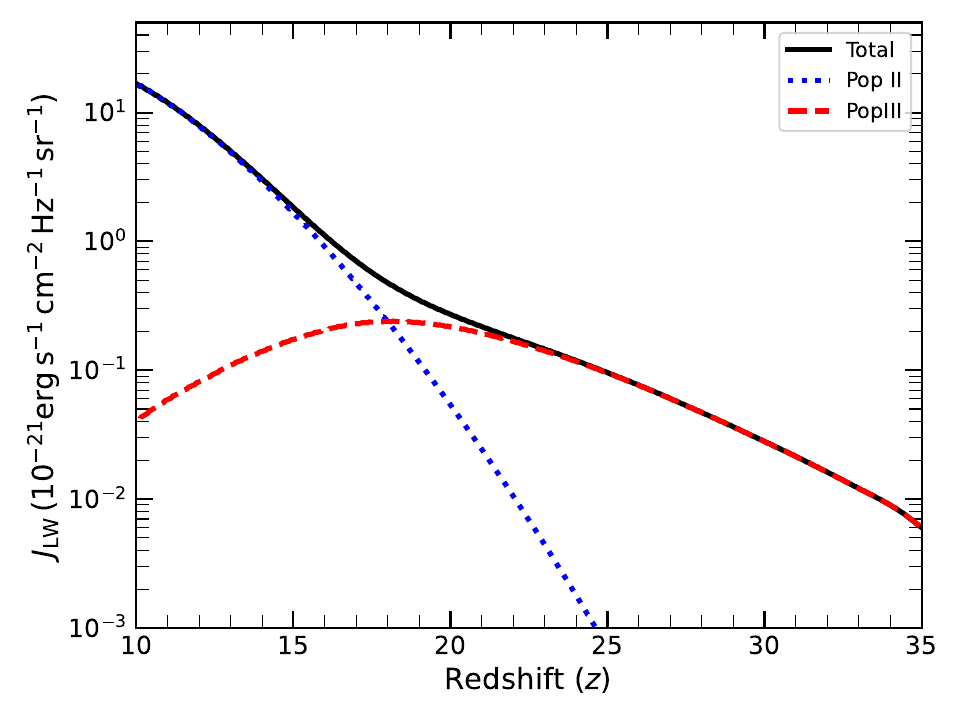}
	\includegraphics[width=0.7\columnwidth]{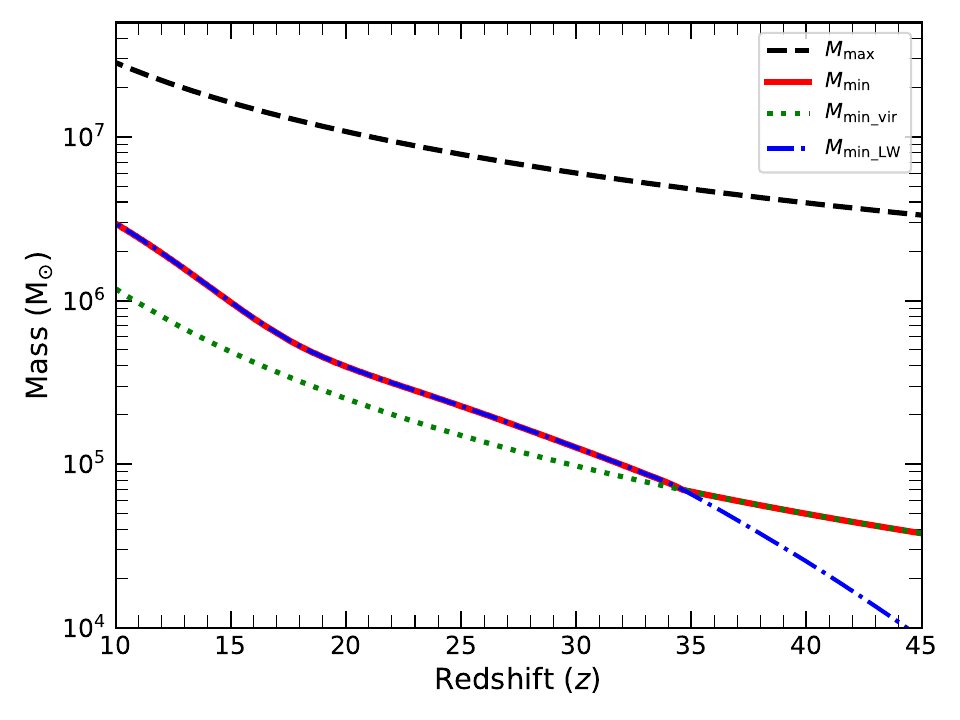}
    \caption[(a) The Lyman-Werner flux from Pop~III and Pop~II stars. (b) The maximum and minimum halo masses required for Pop~III and Pop~II stars to form.]{(a) The total Lyman-Werner flux, $J_{\rm LW}$, in units of $10^{-21} \, \rm erg \, s^{-1} cm^{-2} Hz^{-1} sr^{-1}$ at different redshift $z$ is plotted by black solid curve. The contribution from Pop~III and Pop~II stars are also plotted separately by red dashed and blue dotted curves respectively. (b) The black dashed curve represents the maximum halo mass corresponding to the molecular cooling cut-off for Pop~III star formation. The green curve represents the minimum halo mass required for Pop~III stars to form, whereas the blue dash-dotted one is the same considering Lyman-Werner feedback. At each redshift, the larger among the above two acts as the minimum halo mass at that redshift, represented here by the red solid curve.}
    \label{fig:JLW_mass}
\end{figure*}

\begin{figure}
    \centering
	\includegraphics[width=\columnwidth]{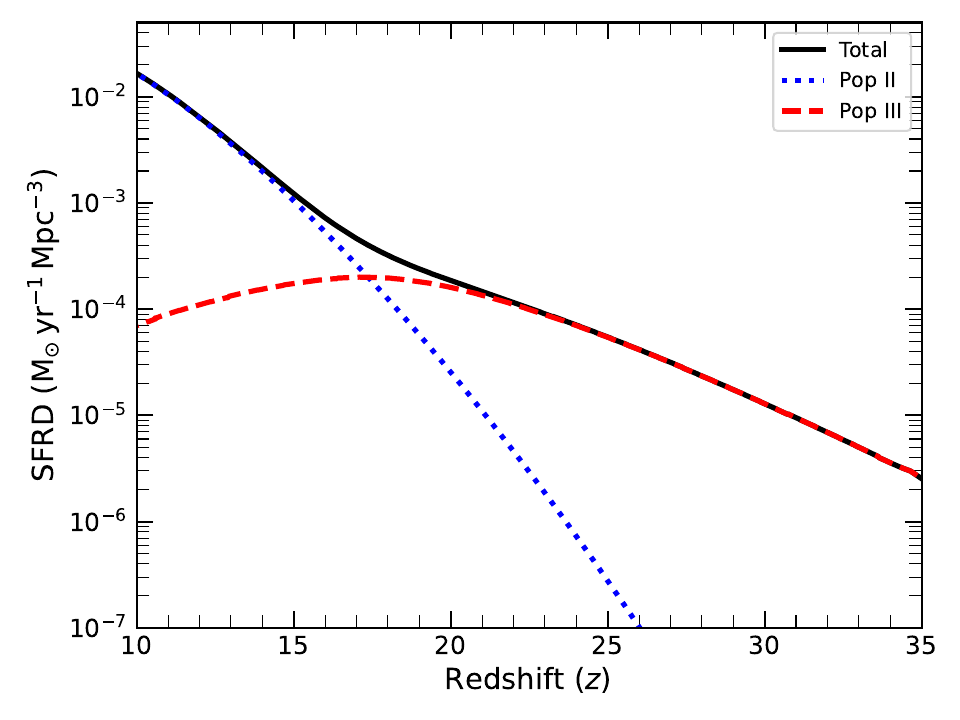}
    \caption[The total star formation rate densities considering both Pop~III and Pop~II stars.]{The total star formation rate density is represented by the black solid curve whereas the contribution of Pop~III and Pop~II stars are denoted by the red dashed and blue dotted curves respectively. 
    }
    \label{fig:SFR}
\end{figure}
\section{Lyman-Werner background and total star formation rate densities}
Here we show our results on star formation rates considering both metal-free Pop~III stars and metal-enriched Pop~II stars. We begin our numerical calculation from redshift $z\sim 50$. Pop~III star formation gets suppressed due to the presence of Lyman-Werner (LW) photons which, in turn, reduces the emission of LW photons from Pop~III stars itself. Therefore, the LW flux and Pop~III star formation rate density are calculated simultaneously and self-consistently.
The resulting LW specific intensity, $J_{\rm LW}$ as a function of redshift $z$ is shown in the upper panel of Fig.~\ref{fig:JLW_mass}. The total LW flux is represented by the black solid curve, whereas the contribution from Pop~III and Pop~II stars are shown separately by red dashed and blue dotted curves respectively. It is clear from the Fig.~\ref{fig:JLW_mass} that the LW contribution is dominated by Pop~III stars in early epochs. However, the contribution is taken over by the LW photons coming from the Pop~II stars at redshift $z \sim 18$. This rapid increase in LW background dissociates more and more H$_2$ molecules and, as a result, the minimum mass required to form Pop~III stars increases as can be seen in the lower panel of Fig.~\ref{fig:JLW_mass}. Here the minimum (red solid) and maximum masses (black dashed curve) of dark matter halos that could form Pop~III stars are plotted at different redshifts. The maximum halo mass for Pop~III star formation corresponds to the atomic cooling cut-off temperature, i.e. $10^4$~K. The minimum mass required for Pop~III star formation depends upon two factors, i) halos which satisfies the criteria $f_{\rm H_2} > f_{\rm crit,H_2}$ (discussed in Section~\ref{subsec:PopIII}) at each redshift (green dotted curve in the lower panel of Fig.~\ref{fig:JLW_mass}), ii) the minimum mass corresponding to Eq.~\ref{eq:LW_mass} determined by the LW flux (Eq.~\ref{eq:J_LW}) and plotted as blue dash-dotted curve in the lower panel of Fig~\ref{fig:JLW_mass}. At any redshift, the larger halo mass among the above two is considered to be the minimum halo mass (shown by red solid curve) for the calculation of Pop~III SFR density. As can be seen from the figure the minimum halo mass can be determined from the Eq.~\ref{eq:f_cH2} at redshifts i.e. $z \gtrsim 35$. However, in later epochs, the minimum mass is set by the LW background flux.

Note that, around $z\sim 17$ the total LW background increases very rapidly. This suppresses the Pop~III star formation rate and, as a consequence, the production of LW photons from Pop~III stars decreases at redshifts $z \lesssim 17$. It can be seen in Fig.~\ref{fig:SFR} where the star formation rate density as obtained in our model has been plotted.
Fig.~\ref{fig:SFR} shows the Pop~III (red dashed curve), Pop~II (blue dotted curve), and total (black solid curve) star formation rate densities.
We see in Fig.~\ref{fig:SFR} that the Pop~III star formation rate which initially rises with decreasing redshift dominates over the Pop~II star formation rate at redshifts $z \gtrsim 17$. As the LW background due to Pop~II stars increases, it narrows down the halo mass range that can form Pop~III stars. As a consequence, the Pop~III star formation decreases at redshift below $z \sim 17$ and the Pop~II star formation rate takes over the Pop~III at redshifts $z \lesssim 17$. 
Note that, the exact epoch of domination of Pop~II over Pop~III stars, in principle, depends, on several parameters such as the mass of the stars formed in Pop~III halos, the form of halo mass function one assumes, star formation efficiencies of Pop~III and Pop~II stars, etc. However, we do not expect the qualitative behavior of the above results to change significantly.

\section{Lyman-{\pdfmath{\alpha}} background during cosmic dawn}
\label{sec:ly_alpha}

As soon as the first generation of stars form, they emit photons at a range of frequencies, including at frequencies higher than Ly-$\alpha$ photons. Photons with  frequency between Ly-$\alpha$ and Ly-$\beta$ get redshifted directly into the Ly-$\alpha$ resonance. The photons having frequency between Ly-$\gamma$ and Ly-limit get redshifted into nearest Lyman series resonance and then  excite  ground state hydrogen atoms. 
These excited HI atoms decay back to $1s$ through radiative cascading process and eventually terminate either in  Ly-$\alpha$ photons or in two $2s \xrightarrow{} 1s$ photons.
Subsequently, the singlet and triplet HI hyperfine level population is altered through Ly-$\alpha$ absorption and emission which is known as the Wouthuysen-Field effect. As the HI spin temperature is determined by the relative population of these hyperfine states, $T_s$ gets affected due to the presence of Ly-$\alpha$ photons. This Ly-$\alpha$ mechanism determines the coupling and decoupling of spin temperature $T_s$ to the gas temperature $T_{g}$ which, in turn, determines the onset of 21-cm absorption signal. Thus, we require the Ly-$\alpha$ flux to calculate the 21-cm signal, and we mostly follow \citet{Barkana_2005, Hirata2006} and \citet{Pritchard_2006} in our work. Moreover, we incorporate the normalisation factor, $f_{{\rm esc}, \alpha}$ which takes care of the uncertainty in the IMF of the first stars and escape of Ly-$\alpha$ photons. Throughout our work, we consider $f_{{\rm esc}, \alpha}^{\rm III} = 0.1$ and $f_{{\rm esc}, \alpha}^{\rm II} = 1.0$ in the calculation of Ly-$\alpha$ flux for Pop~III and Pop~II stars respectively.
The Ly-$\alpha$ photon intensity or the spherically averaged number of photons striking a gas element per unit area, per unit frequency, per unit time, and per steradian is estimated as,
\begin{equation}
    J_{\alpha} = \frac{(1+z)^2}{4 \pi} \sum_{n=2}^{n_{\rm max}} f_{\rm recycle}(n) \int_{z}^{z_{\rm max}(n)} \frac{c dz^{'}}{H(z^{'})} \epsilon(\nu_n^{'},z^{'}) ,
    \label{eq:J_alpha}
\end{equation}
where, the sum is over all available energy states of hydrogen atom.
We consider the excited atoms upto $n=23$ and hence, the sum is truncated at $n_{\rm max} \simeq 23$ to exclude the levels for which the horizon resides within HII region of an isolated galaxy as pointed out in \citet{Pritchard_2006}. As the 21-cm absorption is governed by only the neutral hydrogen, we can safely use the above approximation. Further, $f_{\rm recycle}(n)$ is the probability that a Ly-$n$ photon will be able to generate a Ly-$\alpha$ photon and the values for different levels are tabulated in Table 1 of \citet{Pritchard_2006}. 
Moreover, a photon with frequency $\nu^{'}_n$ emitted at redshift $z'$  gets redshifted to a Ly series photon corresponding to a level $n$ and absorbed by a ground state HI at redshift $z$. Therefore, we can write \citep{Barkana_2005},
\begin{equation}
    \nu_n^{'} = \nu_{\rm LL} (1-n^{-2}) \frac{1+z^{'}}{1+z} ,
\end{equation}
where, $\nu_{\rm LL}$ is the Lyman limit frequency.
If the photon needs to be seen at Ly-$\alpha$ resonance at a redshift $z$, it should have been emitted at a redshift lower than $z_{\rm max}$ where, 
\begin{equation}
    z_{\rm max}(n) = (1 + z) \frac{[1 - (1+n)^{-2}]}{(1-n^{-2})} - 1 .
\end{equation}
In this way, the fluxes of photons which are emitted between consecutive atomic levels get summed up and contribute to the Ly-$\alpha$ flux.
Further, $\epsilon(\nu, z)$ in Eq.~\ref{eq:J_alpha}, is the comoving photon emissivity defined as the number of photons emitted per unit comoving volume, per proper time and frequency, at rest frame frequency $\nu$ and redshift $z$, and can be obtained from the star formation rate density as, 
\begin{equation}
    \epsilon(\nu, z) = \frac{{\rm SFRD}(z)}{m_p} \times \epsilon_b(\nu) ,
\end{equation}
where, $m_p$ is the mass of proton.
The spectral energy distribution function of the sources, $\epsilon_b(\nu)$, is modelled as a power law $\epsilon_b(\nu) \propto \nu^{\alpha_s - 1}$, where the spectral index $\alpha_s = 1.29$ and $0.14$ for Pop~III and Pop~II stars respectively. It is normalised to emit $4800$ photons per baryons between Ly-$\alpha$ and Lyman limit, out of which $2670$ photons are between Ly-$\alpha$ and Ly-$\beta$ for Pop~III stars. The corresponding numbers for Pop~II stars are $9690$ and $6520$ \citep{Barkana_2005, Pritchard_2006}.

\begin{figure}
    \centering
	\includegraphics[width=\columnwidth]{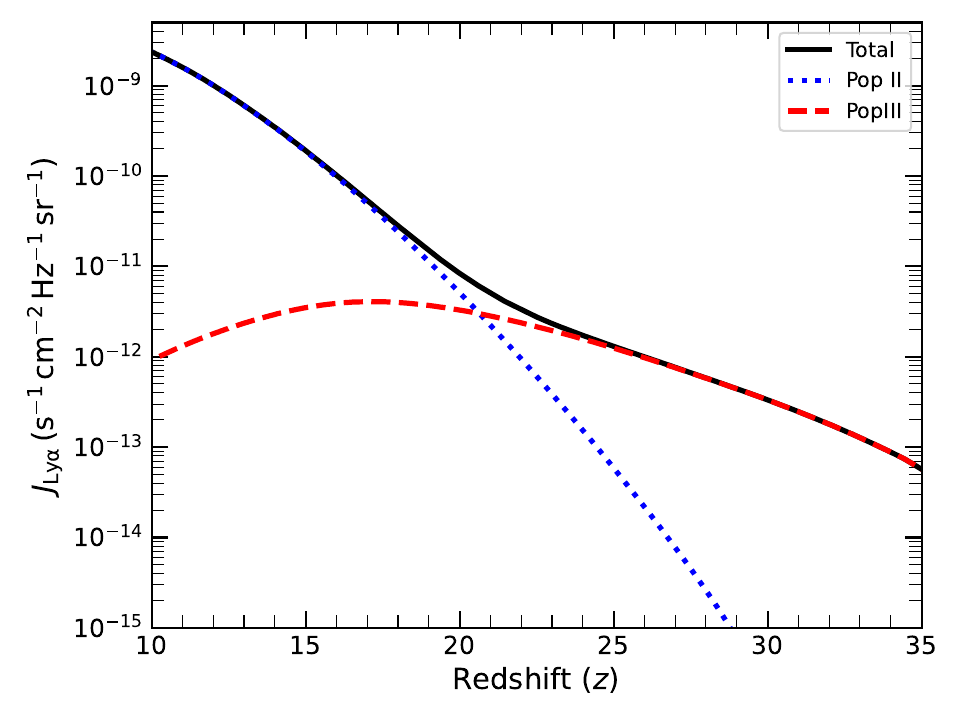}
    \caption[The redshift evolution of Lyman-$\alpha$ flux that arises from both Pop~III and Pop~II stars.]{Lyman-$\alpha$ flux as a function of redshift $z$ that arises from both Pop~III and Pop~II stars. Black solid curve denotes the total Ly-$\alpha$ flux and the representation of separate contributions of Pop~III and Pop~II stars is same as Fig.~\ref{fig:SFR}.}
    \label{fig:J_alpha}
\end{figure}
The Lyman-$\alpha$ flux resulting from the Pop~III and Pop~II stars is plotted in Fig.~\ref{fig:J_alpha}. It shows the Ly-$\alpha$ specific intensity as a function of redshift $z$. The total Ly-$\alpha$ photon flux is shown by the black solid curve, whereas the contributions from the Pop~III and Pop~II stars are also plotted by the red dashed and blue dotted curves respectively. The Ly-$\alpha$ flux is plotted in units of $\rm s^{-1} cm^{-2} Hz^{-1} sr^{-1}$. Our model accounts for the recycling of Ly-$n$ photons that ultimately cascade down to Ly-${\alpha}$ photons. We see that the contribution from the Pop~III stars dominates the total Ly-$\alpha$ intensity at redshift $z \gtrsim 20$. Due to strong Lyman-Werner feedback at lower redshifts, star formation enabled through molecular cooling gets suppressed and, as a consequence, Ly-$\alpha$ emission from these populations decreases gradually at lower redshifts. 

In the next chapter, we will focus on the impact of Ly-$\alpha$ background on the spin temperature, or in turn, on the cosmic 21-cm signal. Moreover, the cosmic rays are also expected to be generated from the first generation of galaxies and we will see how that impacts the IGM heating and thus the HI 21-cm signal.

\chapter[Impact of cosmic ray protons during CD \& EoR]{Impact of cosmic ray protons during cosmic dawn and epoch of reionization\footnote{This chapter is adapted from the paper, "Impact of cosmic rays on the global 21-cm signal during cosmic dawn" by \textcite{Bera2022}.}}

\epigraph{\itshape  On what can we now place our hopes of solving the many riddles which still exist as to the origin and composition of cosmic rays?}{-- Victor Francis Hess}
\label{chap:CR}

\startcontents[chapters]
\printmyminitoc{
As we discussed in the chapter~\ref{chap:PMF}, the decay of the primordial magnetic field prevents it to be a significant source of heating during cosmic dawn, 
in this chapter, we explore another possible source of IGM heating, namely, the cosmic rays generated from the first generation of galaxies.
The possibilities of heating and ionization of the IGM by cosmic rays from young galaxies are already discussed by \citet{Ginzburg_1966, Nath_Biermann_1993, Samui_2005, Samui_2018} in the context of reionization  and post-reionization era. Further, the impact of cosmic rays generated from microquasars on the reionization was discussed in \citet{Tueros_2014}. However, there has been only a handful of earlier works that discuss the effect of cosmic ray heating during the cosmic dawn and its consequences on the global 21-cm signal \citep{Sazonov_2015, Leite_2017, Jana_2019}. Among them, \citet{Sazonov_2015} has adopted a simplified approach by assuming that a small fraction of supernovae kinetic energy goes into the IGM heating without any detailed modeling of the energy transfer. They did not consider contributions from Pop~III and Pop~II stars separately as well. Moreover, the effects of cosmic rays coming from early epochs are also not considered by \citet{Sazonov_2015} as well as in \citet{Jana_2019}. On the other hand, \citet{Leite_2017} modeled the propagation of cosmic rays in the IGM but neither considered the detailed model of the star formation in Pop~III and Pop~II galaxies nor the 21-cm signal.
 
Here we investigate, in more detail, the heating of the IGM by cosmic rays from Pop~II and Pop~III stars and their impact on the global HI 21-cm signal during cosmic dawn. We model the contributions of cosmic rays from Pop~III stars and Pop~II stars separately as they are expected to produce different supernovae. Further, we consider more detailed modeling of star formation by taking into account various feedbacks like Lyman-Warner feedback, supernova, radiative feedback, etc., discussed in the previous chapter. Moreover, our model accounts for the evolution/propagation of cosmic ray particles from previous redshifts, and the energy deposition by these particles are computed in detail instead of adopting a simplified approach  as followed in previous studies. Our study also focuses on the standard IGM scenario along with a scenario where dark matter-baryon interaction is considered  in light of recent observations of the global 21-cm signal. Dark matter-baryon interaction makes the IGM colder as compared to the IGM in the standard scenario as discussed in sec.~\ref{subsec:dark-matter} and we refer to this as the `cold IGM' scenario. Finally, we show that our cosmic ray heating  model can explain the EDGES observations with a reasonable choice of model parameter and thus establishing the importance of detailed modeling of cosmic ray heating. The main purpose of this work is to investigate  cosmic rays from Pop III and Pop II stars as a potential source of IGM heating during cosmic dawn.  In addition, we present a comparison of the efficiency of cosmic rays heating with the more conventional IGM heating mechanism  by X-rays during cosmic dawn.

The structure of this chapter is as follows. In Section~\ref{sec:CR_th}, we outline the evolution of cosmic rays and the energy deposition by these particles into the IGM.  
We present our results of the spin temperature, $T_s$, the kinetic temperature, $T_g$, and finally, the differential brightness temperature, $T_{21}$ highlighting the effect of cosmic ray heating in Section~\ref{main_sec:results}. This section also highlights the changes in differential temperature due to the variation in assumed model parameters.
Finally in Section~\ref{main_sec:summary}, we present our conclusions including the summary of our results.
}

\section{Evolution of cosmic rays}
\label{sec:CR_th}

Cosmic rays are generated in the termination shock of the supernova explosions originating from both Pop~III and Pop~II stars. A significant fraction, $\epsilon \sim 0.15$ of the SNe kinetic energy ($E_{\rm SN}$) gets injected into the cosmic rays \autocite{Hillas_2005, Caprioli_2014}. Thus the average rate of energy injection per unit physical volume into the cosmic rays (in units of ${\rm erg\,s^{-1} cm^{-3}}$) can be calculated as \autocite{Samui_2005},
\begin{equation}
    \dot E_{\rm CR}(z) = 10^{-30} \epsilon \left( \frac{E_{\rm SN}}{10^{51} \, \rm erg} \right) f_{\rm SN} \left( \frac{{\rm SFRD}(z)}{\rm M_{\odot}\, yr^{-1} \, Mpc^{-3}} \right) (1+z)^3 .
    \label{eq:E_CR}
\end{equation}
Here the SFRD($z$) is obtained from Eq.~\ref{eq:SFR_th} for Pop~III and Pop~II stars. Further, $f_{\rm SN}$ is the number of SNe explosions per unit solar mass of star formation. {As already mentioned in the chapter~\ref{sec:SFRD_th}, we assume a single star of 145 $M_{\odot}$ is formed in minihalos and explodes as a supernova having energy $10^{52}$ erg \autocite[mid mass range of][]{Mebane_2018}.} Thus in our model, $f^{\rm SN} \approx 1/145$ and $E_{\rm SN} \sim 10^{52}$ erg for Pop~III stars.
{In case of Pop~II stars, we assume that a supernova of energy $E_{\rm SN} \sim 10^{51}$ erg is formed per 50 $M_{\odot}$ of a star having $1-100$ $M_{\odot}$ of Salpeter IMF and hence, $f_{\rm SN} = 0.02$ \autocite{Samui_2005}.}

We note that the cosmic ray protons mostly contribute to the heating of the IGM. In the case of a supernova exploding in minihalos, low energy protons ($\leq$30 MeV) can escape the halo and heat the intergalactic medium via collision with free $e^-$, and ionization of neutral hydrogen. These SNe are more energetic compared to the core-collapse SNe, and the shock front reaches the virial radius within the Sedov-Taylor (ST) phase itself. Thus the cosmic rays are generated outside the virial radius and get injected into the IGM easily \autocite{Sazonov_2015}. This is contrary to  massive atomic cooling halos hosting Pop~II stars where the low energy protons get confined within the halo and only high energy protons can escape into the IGM and contribute to the heating \autocite{Samui_2005}.
 
In general, the cosmic ray proton spectra can be modeled as a power law in the momentum space \autocite{schlickeiser2002cosmic} which is given by,
\begin{equation}
    {\frac{dn_{\rm CR}(p,z)}{dt}} dp = \dot N_0(z) \left( \frac{p}{p_0} \right)^{-q} dp,
    \label{eq:p_spectra}
\end{equation}
where $\dot N_0(z)$ (along with $p_0$) is the normalization factor which is determined from the total available energies of cosmic rays. It is calculated by integrating $E(p) \dot N_0 (p/p_0)^{-q} dp$  with a low energy cut-off of $10$ keV and equating to $\dot E_{\rm CR}(z)$ (Eq.~\ref{eq:E_CR}). The slope of the spectrum, $q$ has a typical value of $2.2$ \cite{schlickeiser2002cosmic} which matches quite well with observations.

Unlike UV photons, the number density of cosmic ray protons in the IGM at a redshift, $z$ is contributed by the cosmic rays generated at $z$, and cosmic rays that are injected and evolved from higher redshift, $z_i>z$. 
Therefore, the physical number density of cosmic rays at a redshift $z$, having velocity between $\beta$ and $\beta + d\beta$ is given by,
\begin{equation}
    N_{\rm CR}(\beta, z, z_0) = \int_{z_0}^{z} dz_i \frac{dn(z_i, p_i)}{dz_i} \frac{dp_i}{d\beta_i} \frac{d\beta_i}{d\beta} \left( \frac{1+z}{1+z_i} \right)^3 .
    \label{eq:N_cr}
\end{equation}
Here, $z_0$ is the initial redshift of cosmic ray injection by the first generation of stars and we have taken it to be $z= 50$. Cosmic ray protons are expected to loose their energy while propagating from redshift $z_i$ to $z$. 
The redshift evolution of velocity $v=\beta c$ of cosmic ray particles is governed by three processes, (i) the collision with free $e^-$, (ii) ionization of neutral hydrogen, and (iii) the adiabatic expansion of the universe, and given by \autocite{schlickeiser2002cosmic},
\begin{eqnarray}
    \frac{d\beta}{dz}& = & \frac{1}{H(z)\,(1+z)} \biggl[ 3.27 \times 10^{-16} n_e(z) \frac{(1-\beta^2)^{3/2}}{\beta} \frac{\beta^2}{x_m^3 + \beta^3} \nonumber \\ && ~~~~ + 1.94 \times 10^{-16} n_{\rm HI}(z) \frac{(1-\beta^2)^{3/2}}{\beta} \frac{2\beta^2}{(0.01)^{3} + 2\beta^3} \nonumber \\ &&~~~~ \times \{1 + 0.0185 \log \beta \, \Theta(\beta - 0.01)\} \biggr] + \frac{\beta (1-\beta^2)}{(1+z)}
    \label{eq:db_dz}
\end{eqnarray}
where, $n_e(z)$ and $n_{\rm HI}(z)$ are the electron and neutral hydrogen densities respectively in units of $\rm cm^{-3}$, $x_m = 0.0286 (T_g/(2\times10^6\, \rm K))^{1/2}$ and $\Theta$ is the Heaviside step function. 

We note that in Eq.~\ref{eq:db_dz} the two terms inside the square brackets are due to the collision with free electron and ionization of neutral hydrogen atoms respectively, both of which deposit energy to the IGM and contribute to heating. 
In case of a collision with free electrons, the complete energy loss by a cosmic ray proton becomes the thermal energy of the IGM. 
 
However, when the cosmic ray protons interact with the neutral intergalactic medium, it may result in primary ionization or excitation to a discrete level, and the entire energy of the cosmic ray proton does not get transferred to the free electron.  It is shown that the number of primary ion formations is almost proportional to the average loss by collision. In fact, one such ion pair gets formed if the primary loss is about $32$~eV \autocite[discussed in sec. 5.3.11 in][]{schlickeiser2002cosmic}. The total energy loss rate is then obtained by multiplying the number density of cosmic rays ($N_{\rm CR}$) with average energy loss and dividing by $32$ eV. It is shown that for each ionization process, $\Delta Q \simeq 20$~eV gets deposited as heat \autocite{spitzer_1969, Goldsmith_1978}. This leads to a factor of $5/8$ in the calculation of energy deposition by cosmic rays.
Hence, the heating rate of the IGM by cosmic ray particles due to ionization can be written as,
\begin{equation}
    \Gamma_{\rm CR}(z, z_0) = \frac{5}{8} \eta_1 \eta_2 \eta_3 \int \frac{dE(\beta)}{dz} N_{\rm CR}(\beta, z, z_0) \, d\beta ,
    \label{eq:gamma_CR}
\end{equation}
where, $\eta_1 = 5/3$ occurs due to the secondary ionization by $e^-$ which produces during ionization by primary cosmic ray particles, $\eta_2 \simeq 1.17$ accounts for the $10\%$ He abundance in the IGM, and $\eta_3 \simeq 1.43$ takes into account the contribution of heavy cosmic ray nuclei and cosmic ray electrons ($e^-$), positrons ($e^+$) \autocite{schlickeiser2002cosmic}.
These energy depositions due to the cosmic ray particles are incorporated in the temperature evolution of IGM i.e. in Eq.~\ref{eq:Tg} and as we will see they can contribute as a major heating source of IGM during the cosmic dawn.

As already mentioned, the low-energy protons that take part in collision and ionization do not escape from the massive atomic cooling Pop~II galaxies.
Only the high-energy protons can escape and may interact with high redshift IGM. It should be noted that a major part of the cosmic ray energy is carried by the high-energy protons ($\sim 1$~GeV).
If a sufficient magnetic field ($B$) is present in the IGM, these high-energy cosmic ray particles can gyrate along the magnetic field lines and generate Alfv\'en waves. When these waves get damped, the energy is transferred to the thermal gas \autocite{Kulsrud_1969, Skilling1975, Bell1978, Kulsrud_2004}. The energy deposition rate via this Alfv\'en wave generation is $|v_A.\nabla P_c|$, where $v_A= B/\sqrt{4\pi \rho}$ is the Alfv\'en velocity ($\rho $ is the plasma density) and $\nabla P_c$ is the cosmic ray pressure gradient.
Thus the time scale ($t_{\rm CR}$) for this process to influence the IGM of temperature $T_g$ can be calculated by
\begin{equation}
    t_{\rm CR}=\dfrac{3n k_B T_g}{2|v_A.\nabla P_c|}.
\end{equation}
This time scale should be compared to the Hubble time ($t_H$) if it can heat the IGM. Putting some reasonable numbers we find \autocite[also see][]{Samui_2018},
\begin{multline}
    \frac{t_{\rm CR}}{t_H} \approx 0.16 \left(\frac{h}{0.7}\right)^4 \left(\frac{\Omega_{\rm m}}{0.3}\right)^{1/2} \left(\frac{1+z}{16} \right)^4 \left(\frac{T_g}{10 \,\rm K} \right) \left(\frac{0.1\,{\rm nG}}{B_0} \right) \\ 
    \times \left( \frac{5 \times 10^{-5}\, {\rm eV/cm^3}}{E_{\rm CR}}\right)
     \left( \frac{L}{0.01\, {\rm Mpc}}\right) ,
\end{multline}
where, $L$ is the physical distance between galaxies and we have taken $L = 0.01$ Mpc as the average separation of typical galaxies at $z=15$.

{Thus we can see if a primordial magnetic field of present-day value, $B_0 = 0.1$~nG \autocite{Minoda19} is present,} and cosmic rays have energy density of $E_{\rm CR} = 5 \times 10^{-5}\, {\rm eV/cm^3}$ (this can easily be seen from Eq.~\ref{eq:E_CR} using SFRD at $z=15$), the cosmic rays can dissipate their energy to the IGM within Hubble time to influence the IGM temperature of $10$~K at $z\sim15$. Hence, this magnetosonic transfer of energy due to cosmic rays is an important source of IGM heating. To see its influence we assume a fraction {$Q_{\rm CR,II}$} of total cosmic ray energy density is transferred as the thermal energy of the IGM via the Alfv\'en waves from the high energy protons that escape from Pop~II galaxies. 
{It has been shown in \textcite{Samui_2018} that if only 10-20\% of cosmic rays energy can be transferred to the IGM it can significantly alter the thermal history of the IGM in the redshift range 2-4. Thus it is natural to consider the same for the thermal history of the IGM during the cosmic dawn. }

In late time, along with the ionizing photons, the cosmic rays from the first generation of galaxies are likely to alter the ionization fraction $x_e$ and it is determined by the second last term in eq.~\ref{eq:ion_frac} where,
\begin{equation}
    I_{\rm CR} = \int N_{\rm CR}(\beta, z, z_0) \sigma_{\rm HI} \beta d\beta .
\end{equation}
Here, $\sigma_{\rm HI} = \frac{1.23 \times 10^{-20}}{\beta^2} \left( 6.20 + \log \frac{\beta^2}{1-\beta^2} - 0.43 \beta^2 \right)$ \citep{1968ApJ...152..971S}.
However, we have checked that the cosmic rays can change the ionization fraction at most $10^{-3}$ for the supernova kinetic energy we considered here which is similar to \citet{Sazonov_2015}. Similar results were also obtained by \citet{Samui_2005}. Since we are interested in 21-cm signal in the redshift range 10 to 20, the 21-cm signal is mostly governed by the temperature differences between the IGM and the background radio signal rather than the ionization fraction of the hydrogen which is at most 0.1 by redshift $z=10$ thus only altering the 21-cm signal at most 10\% level \citep{furlanetto2006}. One should consider the contribution of UV photons from the first stars in order to model the ionization fraction more accurately that we are not taking into account in this particular work as we are interested in the cosmic dawn.

Here, we would like to mention that cosmic ray electrons from first sources can, in principle, generate radio background through their interactions with the intergalactic  magnetic field and alter the global 21-cm signal \autocite{Jana_2019}. However, we expect the resulting synchrotron radiation is expected to be insignificant during cosmic dawn given the current upper limit on the intergalactic magnetic field ($B_0 \lesssim 0.1$ nG) \autocite{Minoda19, Bera_2020}.

\section{Results and Discussion}
\label{main_sec:results}
In this section, we present our results of the global 21-cm signal highlighting the heating due to cosmic ray protons generated through supernova explosions from very early Pop~III and Pop~II stars. For better comprehension, we list below the default parameters that we used in our work.
The reason for considering such parameter values has already been discussed in the previous section.

Pop~III model :
\begin{itemize}
    \item {Each minihalo generates a single Pop~III star of $145$~$M_{\odot}$ which explodes as a core-collapse supernovae, so $f_{\rm SN,III} = 1/145$.}
    \item {The explosion energy of a Pop~III supernova is taken as $E_{\rm SN,III} = 10^{52}$~erg.}
    \item {{A fraction of SNe kinetic energy, $\epsilon_{\rm III} = 0.06$ gets utilised to accelerate the cosmic rays. This value is chosen in order to match the EDGES deep absorption profile, though it is a free parameter in our model and we show the variation of this parameter in the later section.}}
    \item {The slope, $q$ of the cosmic ray spectra (Eq.~\ref{eq:p_spectra}) generated by the supernova, is taken as $2.2$.}
\end{itemize}

Pop~II model : 
\begin{itemize}
    \item {We consider one supernova explosion per $50$~$M_{\odot}$ of star formation, i.e. $f_{\rm SN,II} = 0.02$}
    \item {The typical energy of Pop~II supernova is $E_{\rm SN,II} = 10^{51}$~erg.}
    \item {The fraction of SNe kinetic energy carried by the cosmic rays is $\epsilon_{\rm II} = 0.15$.}
    \item {{The amount of cosmic ray energy that gets transferred to the IGM and heats the surrounding gas is $Q_{\rm CR,II} = 0.15$. This value is chosen to explain the sharp rise of the EDGES profile. The variation of this parameter is shown in the Fig.~\ref{fig:T21_noDM}}}.
\end{itemize}

These are the default parameter values adopted from \textcite{Mebane_2018} and \textcite{Samui_2014} for Pop~III and Pop~II star formation models respectively, and from \textcite{schlickeiser2002cosmic} $\&$ \textcite{Samui_2005} for cosmic rays energy deposition. These parameters are chosen keeping in mind the EDGES observation. We also show results by varying parameters such as $q, \epsilon_{\rm III}, \epsilon_{\rm II}$ {and $Q_{\rm CR,II}$. As the impact of $\epsilon_{\rm II}$ and $Q_{\rm CR,II}$ are similar, we don't vary them separately rather we vary only $\epsilon_{\rm II}$, and represent it as $\epsilon_{\rm II}$ and/or $Q_{\rm CR,II}$.}

\subsection{Ly-{\ensuremath{\alpha}} coupling and resulting spin temperature}
In this section, we discuss the Lyman-$\alpha$ coupling of hydrogen spin temperature and IGM temperature due to the Lyman-$\alpha$ flux resulting from the Pop~III and Pop~II star formation as discussed in section~\ref{sec:ly_alpha}. 

\begin{figure*}
    \centering
	\includegraphics[width=0.7\columnwidth]{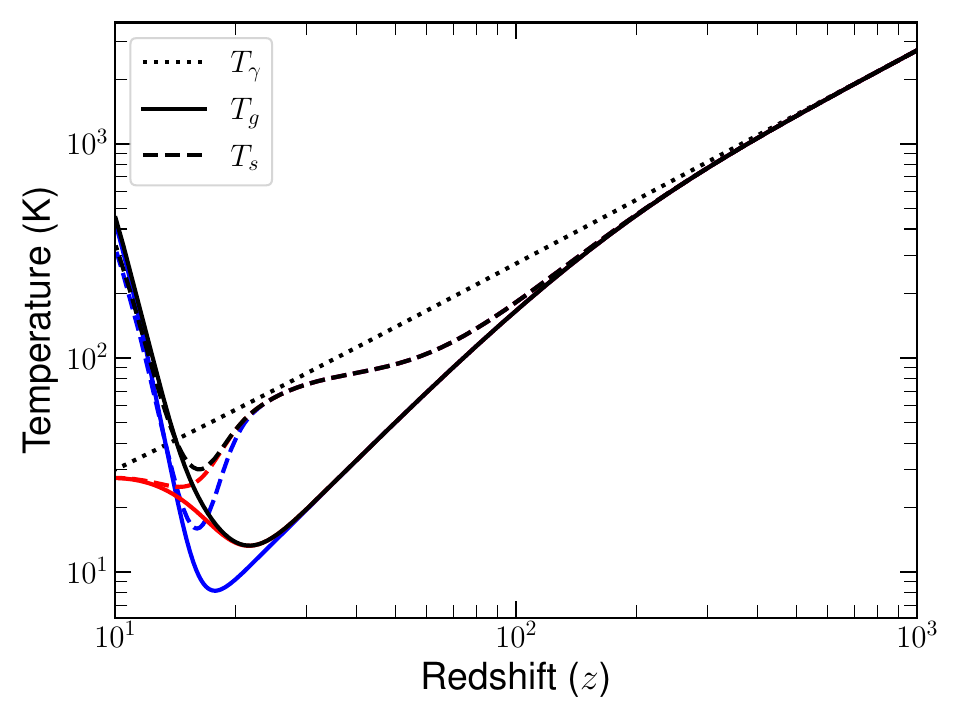}
	\includegraphics[width=0.7\columnwidth]{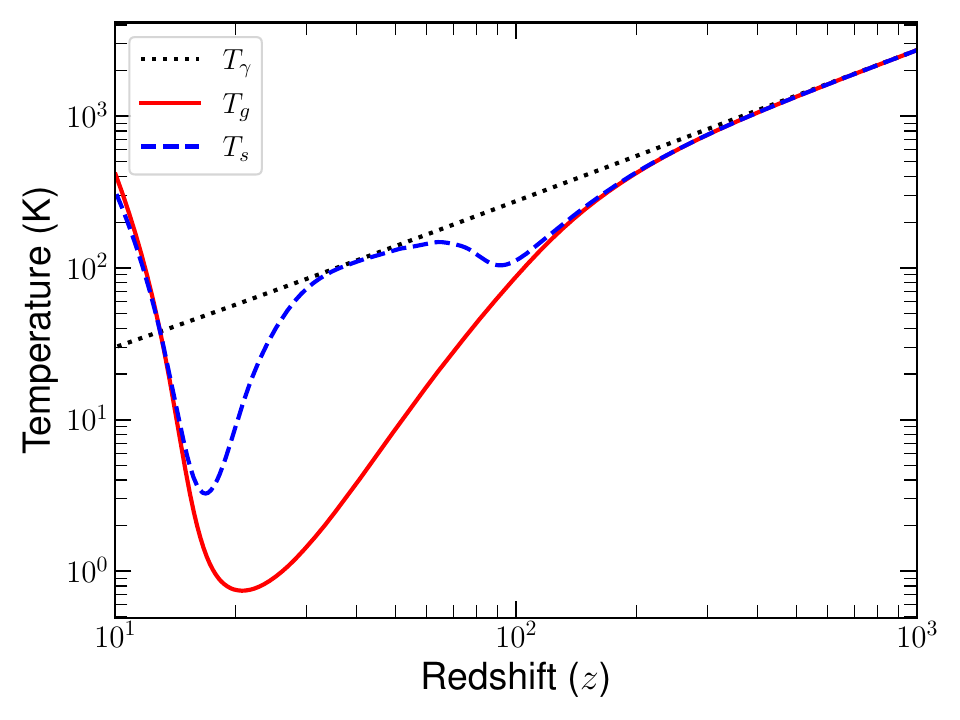}
    \caption[In the top panel, the evolution of the gas kinetic temperature, $T_g$, spin temperature, $T_s$, and CMBR temperature $T_{\gamma}$ with redshift are shown for considering the contribution of cosmic rays from Pop~III and Pop~II stars separately. In the bottom panel the same is shown in presence of the dark matter-baryon interaction.]{{{\bf Top panel:} The evolution of the gas kinetic temperature, $T_g$ with redshift is shown by black solid curve along with the spin temperature, $T_s$, and CMBR temperature $T_{\gamma}$ by black dashed and black dotted curves respectively. The blue solid and dashed curves represent the evolution in $T_g$ and $T_s$ respectively for considering the contribution of cosmic rays from Pop~II stars only and the red curves represent the same for considering Pop~III stars only.} {\bf Bottom panel:} The evolution of the gas kinetic temperature, $T_g$ with redshift is shown by red solid curve along with the spin temperature, $T_s$, and CMBR temperature $T_{\gamma}$ by blue dashed and black dotted curves respectively. The dark matter-baryon interaction is included here for excess cooling mechanism keeping in mind the EDGES deep absorption signal. The dark matter mass, and interaction cross-section $(m_{\chi}/{\rm Gev}, \sigma_{45}) = (0.1, 2)$ are considered for this particular plot.}
    \label{fig:Ts_Tg}
\end{figure*}

As soon as the first sources appear, Ly-$\alpha$ photons from these sources help the spin temperature $T_s$ to decouple from the CMBR temperature and start coupling with the gas temperature. This can be seen from Fig.~\ref{fig:Ts_Tg}, where we have plotted the CMBR temperature, spin temperature, and gas temperature starting from the recombination redshift $z = 1010$ {without DM-b interaction (top panel) and with DM-b interaction (bottom panel)}. It is clear from the figure that the decoupling begins at a redshift as early as $z \sim 30$ in our model. The spin temperature, $T_s$ then gradually approaches the IGM kinetic temperature, $T_g$, and gets fully coupled at redshift $z \approx 17$. Since the Pop~II stars dominate the Ly-$\alpha$ photon budget at redshift $z \lesssim 20$, the coupling is mainly determined by this population {as shown in the top panel of Fig.~\ref{fig:Ts_Tg}, where we plotted the spin temperatures by blue (Pop~II), red (Pop~III) and black (Pop~II + Pop~III) dashed curves}. We note that along with the standard adiabatic cooling of the IGM prior to the heating, we also consider dark matter-baryon interaction in our work. {The bottom panel of }Fig.~\ref{fig:Ts_Tg} shows result in presence of dark-matter baryon interaction for dark matter mass $m_{\chi}=0.1$ GeV and the interaction cross-section $\sigma_{45}=\frac{\sigma_0}{10^{-45} \, {\rm m^2}} =2$. This interaction  helps to cool the IGM  faster compared to the standard adiabatic cooling, as discussed in sec.~\ref{subsec:dark-matter}. This scenario is motivated by the EDGES observation that shows a strong absorption profile of {-0.5$~K$} that is elaborately discussed in sec.~\ref{sec:EDGES}. Redshifts of the decoupling of $T_s$ from the CMBR temperature and coupling to the IGM kinetic temperature can change to some extent depending on the Ly-$\alpha$ escape fraction into the IGM and redshift evolution of the IGM kinetic temperature. However, as we focus on the impact of cosmic rays on the heating of the IGM here, we defer this discussion to future works. 

\subsection{Impact of cosmic rays heating on IGM temperature}
One of the major aims of this work is to explore cosmic rays from first generation of stars as a source of IGM heating during the cosmic dawn. This section discusses our results on that.  As already discussed, we consider the collisional and ionization interaction as well as the magnetosonic interaction through which the energy gets transferred from cosmic rays to the IGM which leads to the increase in kinetic temperature of the IGM.
In Fig.~\ref{fig:Ts_Tg}, we have plotted the resulting IGM gas temperature by the red solid curve. 
Initially up to redshift $z \gtrsim 200$, the gas kinetic temperature $T_g$ and CMBR temperature $T_{\gamma}$ follow each other. This is enabled through the Compton scattering process. Afterward, $T_g$ cools faster than $T_{\gamma}$ due to adiabatic cooling and dark matter-baryon interaction that we consider. Fig.~\ref{fig:Ts_Tg} shows results for the dark matter mass of $m_{\chi} = 0.1$ GeV, and interaction cross-section of $\sigma_{45} = 2 $ that resulted in the sharp fall of IGM temperature. Further, note that, due to the collisional coupling hydrogen spin temperature follows the IGM temperature up to $z \gtrsim 100$. Afterward, the collisional coupling becomes weak due to the lower IGM density and temperature, and hence $T_s$ again starts to follow the CMBR temperature. As discussed before, the spin temperature again decoupled from the CMBR due to the presence of Ly-$\alpha$ photons around $z \sim 30$. Meanwhile, the IGM temperature reaches a minimum  at $z \sim 20$. By this time the first generation of stars has produced enough amount of cosmic rays that start heating the surrounding gas and increase the IGM temperature as can be seen from Fig.~\ref{fig:Ts_Tg}.  $T_g$ crosses the CMBR temperature at redshift $z \sim 12$ due to the cosmic ray heating. By this redshift, the Ly-$\alpha$ coupling has completely coupled the spin temperature to the gas temperature. The interplay between $T_s$ and $T_{\gamma}$ determines the 21-cm brightness temperature that we discuss in the next section.

\subsection{Global 21-cm signal}
\begin{figure}
    \centering
	\includegraphics[width=\columnwidth]{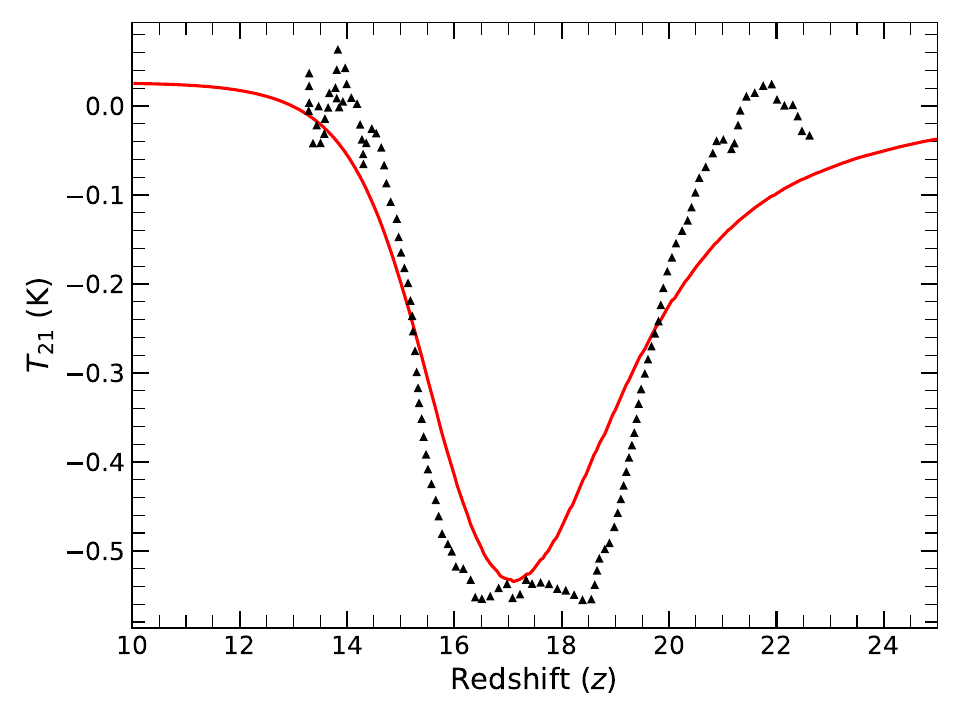}
    \caption[The brightness temperature, $T_{21}$, resulting from the temperatures shown in Fig.~\ref{fig:Ts_Tg} is represented here.]{The brightness temperature, $T_{21}$, resulting from the temperatures shown in the bottom panel of Fig.~\ref{fig:Ts_Tg} is represented here by red solid curve. This plot is generated considering Ly-$\alpha$ coupling and cosmic ray heating with efficiencies, $\epsilon_{\rm III} = 0.06$, and {$\epsilon_{\rm II}$ and/or $Q_{\rm II} = 0.15$} for cosmic rays produced in Pop~III and Pop~II stars respectively. The dark matter-baryon interaction parameters, $(m_{\chi}/{\rm Gev}, \sigma_{45})$ are same as in Fig. \ref{fig:Ts_Tg}. For reference, the measured $T_{21}$ signal by the EDGES is shown here by black triangles.}
    \label{fig:T21}
\end{figure}
The EDGES observation suggests a mean brightness temperature, $T_{21}$ between $-0.3$~K to $-1.0$~K. As already discussed, this strong absorption can be explained by considering a colder IGM resulting from dark matter-baryon interaction as adopted in our model. Thus, we will first focus our result in the light of EDGES detection. Later, we will also discuss the effect of cosmic ray heating on the global 21-cm signal in the absence of any dark matter-baryon interaction.

In the Fig.~\ref{fig:T21}, we show the estimated global 21-cm signal using $T_g$ and $T_s$ as discussed in the previous sections by solid red line. 
For comparison, we have also plotted the measured profile of $T_{21}$ with black triangles using publicly available data of \textcite{EDGES18}.
The absorption feature starting at $z \sim 20$ arises from the coupling of $T_s$ with $T_g$ by Ly-$\alpha$ photons produced by the early generation of Pop~III stars. The deeper absorption of $-0.5$~K is resulting from the dark matter-baryon interaction that cools the IGM temperature up to $\sim 1$~K by $z \sim 17$. Afterward, the heating due to cosmic ray protons generated from early Pop~III and Pop~II stars increases the IGM temperature sharply. By $z \sim 14$, the IGM temperature exceeds the CMBR temperature due to this cosmic ray heating. This causes the sharp disappearance of the absorption feature in the 21-cm signal as detected by the EDGES collaboration. {Thus we can say that the rising arm of the absorption signal matches reasonably well with the cosmic ray heating along with the dark matter-baryon interaction. }

We note that, for this particular profile, a very small amount of energy gets transferred from the cosmic rays to IGM. For instance, we have used $\epsilon_{\rm III} = 0.06$, and {$\epsilon_{\rm II}$ and/or $Q_{\rm II} = 0.15$} for Pop~III and Pop~II stars. 
{In passing we note that, even though we focus our results in light of EDGES detection, there are controversies about the detected signal by EDGES. For example, a recent work by SARAS3 collaboration \autocite{Saurabh_2021} has claimed a null detection of the same signal by a different experimental setup. However, cosmic ray heating is a more generic physics that one should consider while modeling the cosmic 21-cm signal. The exotic dark matter-baryon interaction was/is considered only to explain the unusually strong absorption profile detected by EDGES group. Hence, in order to highlight the effect of cosmic ray heating on 21-cm signal, and keeping in mind the controversy regarding the detected signal, we describe the effect of cosmic ray heating on 21-cm signal without considering the exotic dark matter-baryon interaction, in the next section.}

\subsection{21-cm signal without dark matter-baryon interaction}

\begin{figure}
    \centering
	\includegraphics[width=\columnwidth]{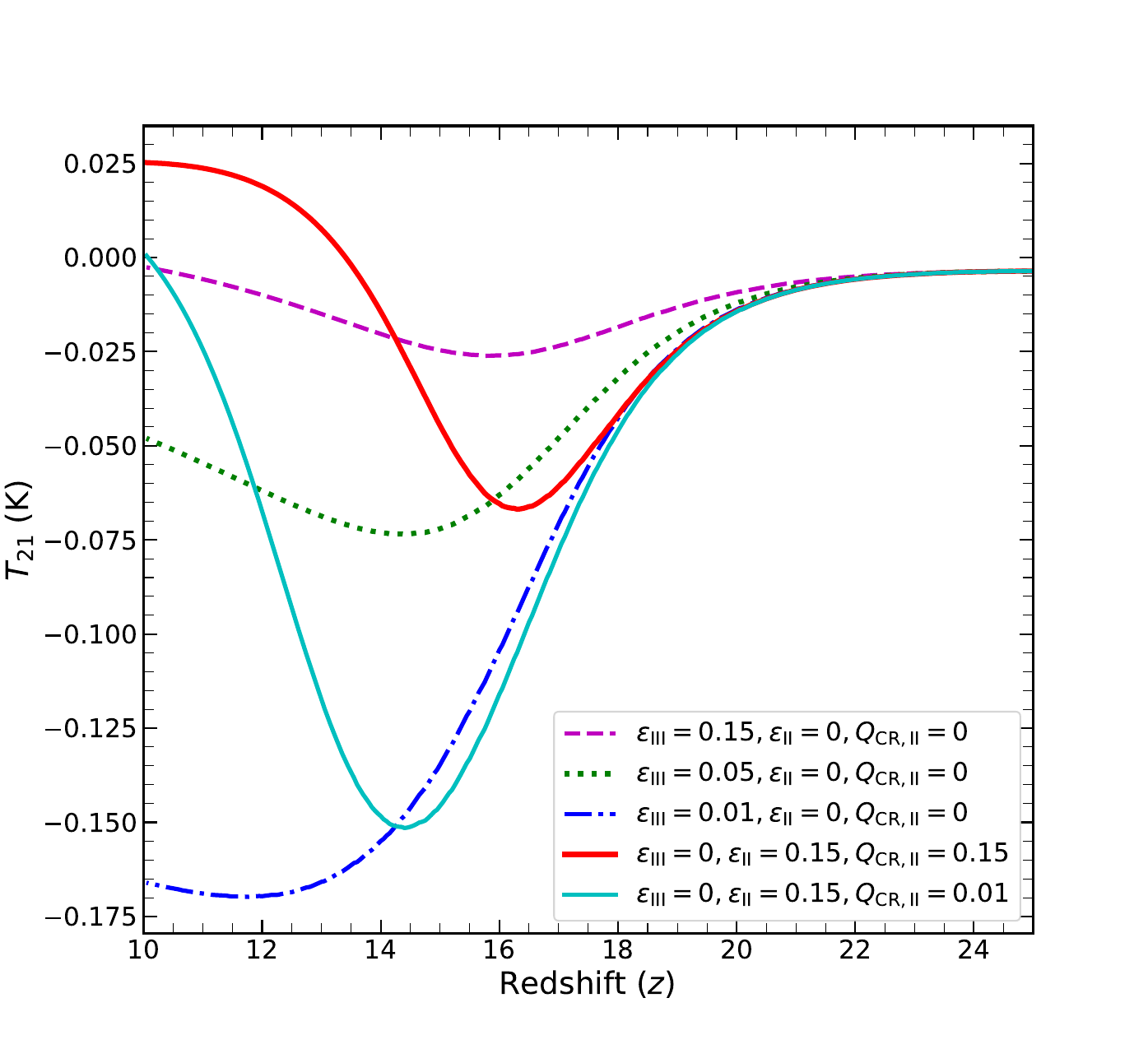}
    \caption[Variations in the global HI 21-cm signal $T_{21}$ due to the various efficiencies of cosmic ray heating without considering any exotic mechanism such as dark matter-baryon interaction.]{Variations in the global HI 21-cm signal $T_{21}$ due to the various efficiencies of cosmic ray heating are shown here. We kept the spectral index of cosmic ray spectra, $q$  fixed for this plot. All the evolutions are done here without considering any exotic mechanism such as dark matter-baryon interaction.}
    \label{fig:T21_noDM}
\end{figure}

In order to understand the contribution of cosmic ray heating from Pop~III and Pop~II stars separately, we vary the efficiency parameters $\epsilon_{\rm III}$, $\epsilon_{\rm II}$ {and the amount of energy transferring from cosmic rays to the IGM, $Q_{\rm CR, II}$} keeping the cosmic ray spectral index $q$ fixed. The resulting brightness temperature, $T_{21}$ as obtained for various $\epsilon_{\rm II}$, $\epsilon_{\rm III}$ {and $Q_{\rm CR,II}$} is shown in Fig.~\ref{fig:T21_noDM}. It is clear from the figure that in all such models, the absorption profile starts at $z \sim 20$ due to the Ly-$\alpha$ coupling. This depends only on the star formation rates of Pop~III and Pop~II stars and the resulting Ly-$\alpha$ flux. It is independent of cosmic ray heating as long as the gas temperature remains below the CMBR temperature. However, the depth and width of the absorption spectra highly depend on the heating due to cosmic rays. For example, if the Pop~III SNe are more efficient in accelerating cosmic rays i.e. $\epsilon_{\rm III} = 0.15$, we get a very shallow absorption profile of $\sim -0.025$~{K} as can be seen by magenta dashed curve in Fig.~\ref{fig:T21_noDM}. If the efficiency is even higher, it is likely to wash out any possible 21-cm absorption profile. On the other hand, reducing the efficiency would increase the absorption depth as well as the duration of the absorption as can be seen from the green dotted and blue dash-dotted curves where $\epsilon_{\rm III} = 0.05$ \& $0.01$ respectively. Note that, in these three models, we didn't take any contribution of cosmic ray heating by Pop~II stars. Thus we can say that a very small amount of contribution in heating by cosmic rays generated from Pop~III stars ($\epsilon_{\rm III} = 0.01$), would reduce the absorption depth compared to the prediction of the same by any model where no cosmic ray heating is considered. Thus accurately determining the global 21-cm signal would constrain the contribution of cosmic ray heating which in turn can put constraints on the nature of Pop~III stars during the cosmic dawn.

Finally, we show the contribution of cosmic ray heating only from Pop~II stars by red solid curve in Fig.~\ref{fig:T21_noDM}, where we assumed $\epsilon_{\rm III} = 0$, $\epsilon_{\rm II} = 0.15$ and $Q_{\rm CR,II} = 0.15$. It is clear from the figure that Pop~II heating is also efficient in reducing the absorption depth as well as the duration of the absorption profile. In this case, we get a maximum absorption of $-0.07$~{K} at $z\sim 16$ which ends by $z\sim 13.5$.
Even for a very small efficiency of $Q_{\rm CR,II} = 0.01$, we get an absorption depth of {$\sim -0.15$~K} as can be seen from cyan solid curve in Fig.~\ref{fig:T21_noDM}. This also increases the duration of the absorption signal. Note that, the impact of $\epsilon_{\rm II}$ is the same as that of $Q_{\rm CR,II}$. Thus even the heating by cosmic rays generated from the Pop~II stars is an important physics that one should consider while modeling the cosmic 21-cm signal. We also note that Pop~III stars have a considerable impact on the global 21-cm signal at redshift  $z\gtrsim 16$ in all models we consider. 
{However, $T_g$, $T_s$, and the resulting absorption profile depend significantly on the assumed parameters of cosmic ray heating that we discuss in the next section.}

\subsection{Variation of model parameters}

\begin{figure*}
    \centerline{\includegraphics[width=0.5\columnwidth]{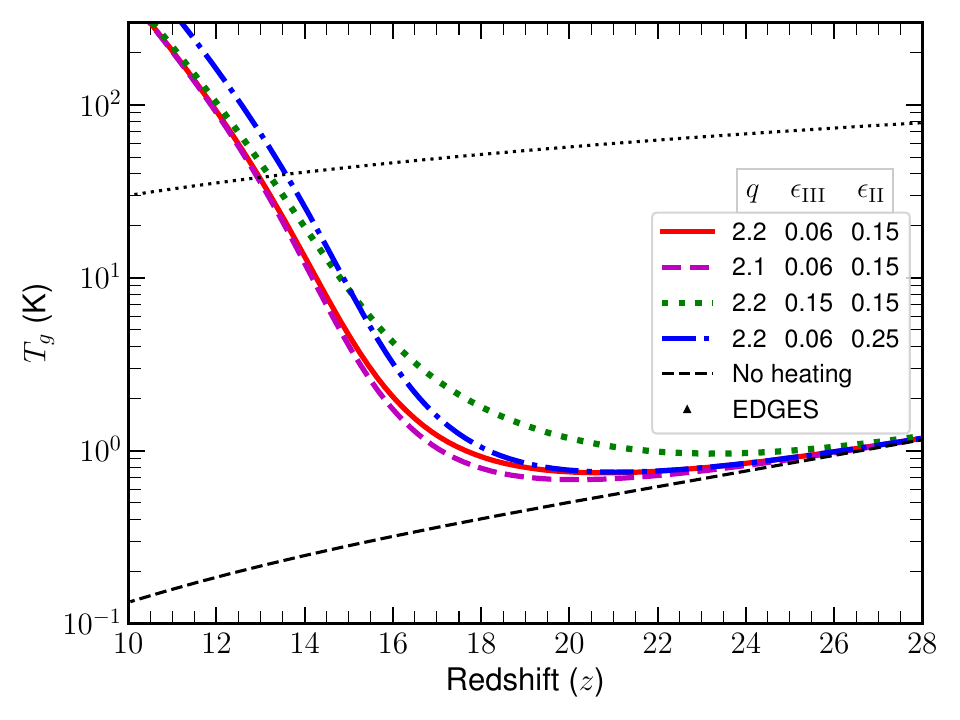}\includegraphics[width=0.5\columnwidth]{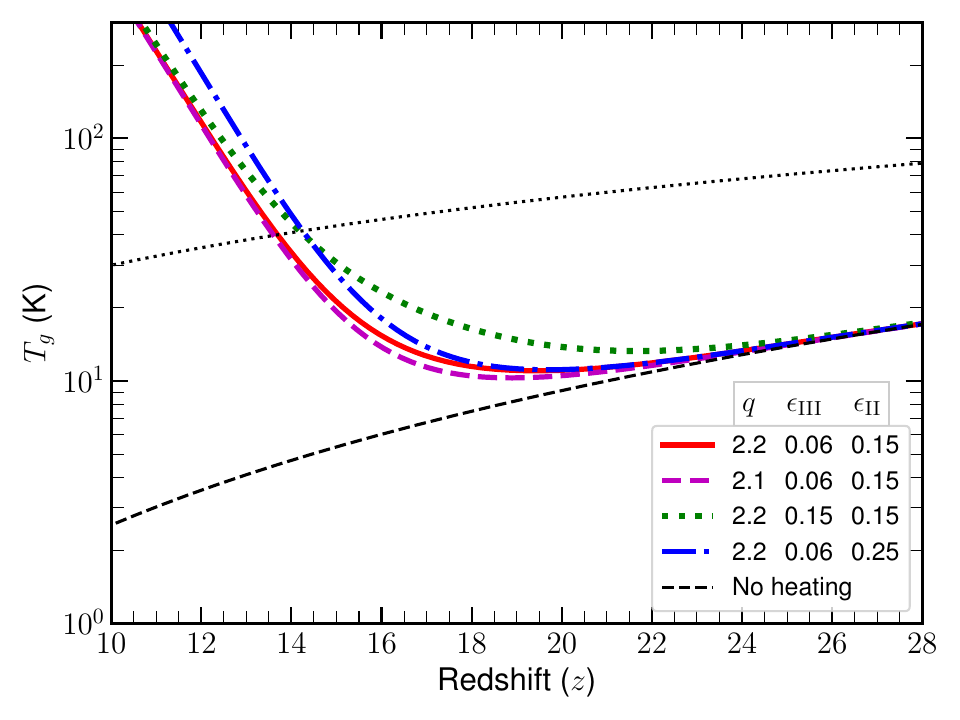}}
    \centerline{\includegraphics[width=0.5\columnwidth]{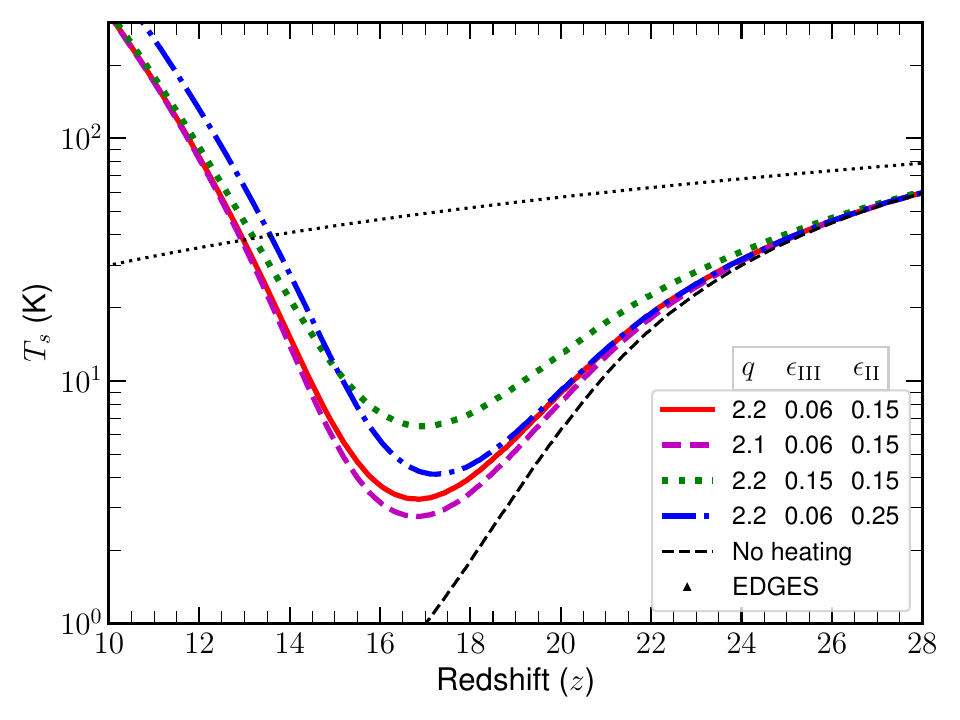}\includegraphics[width=0.5\columnwidth]{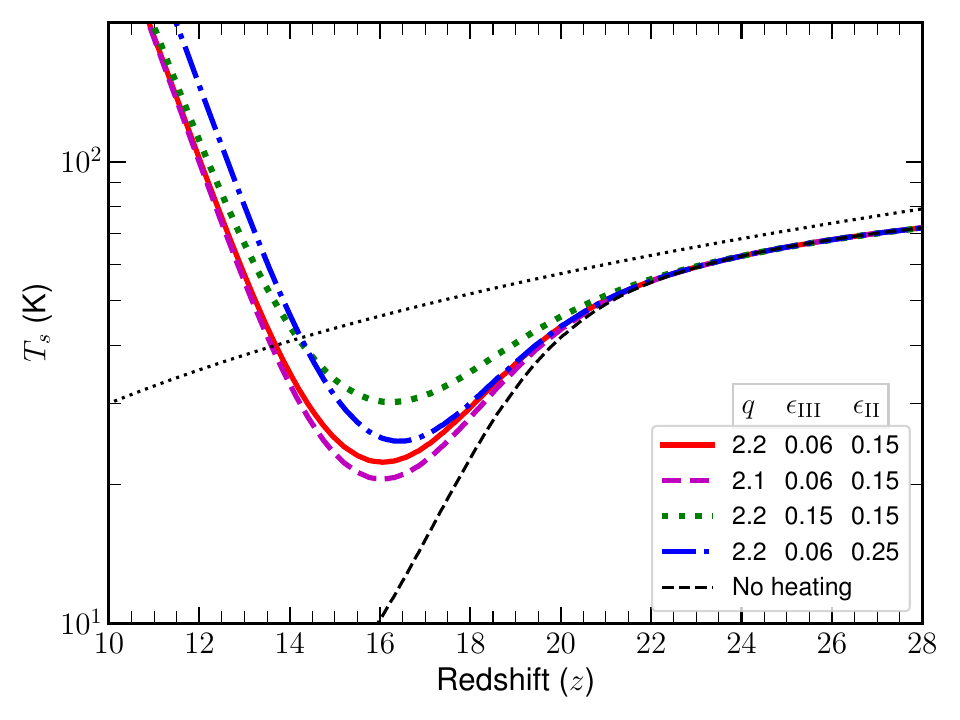}}
    \centerline{\includegraphics[width=0.5\columnwidth]{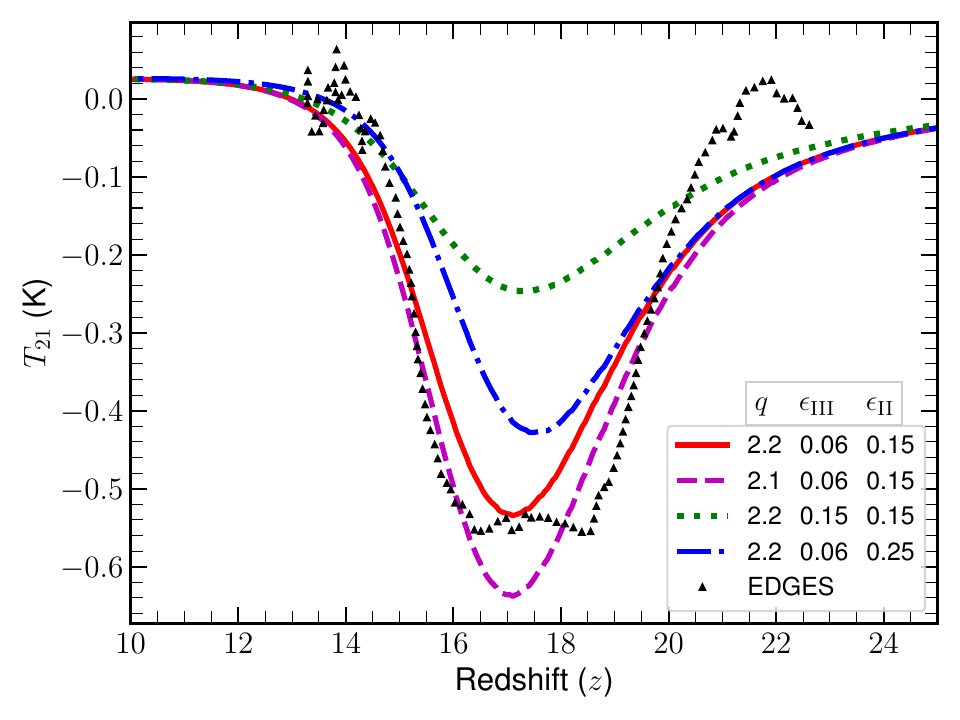}\includegraphics[width=0.5\columnwidth]{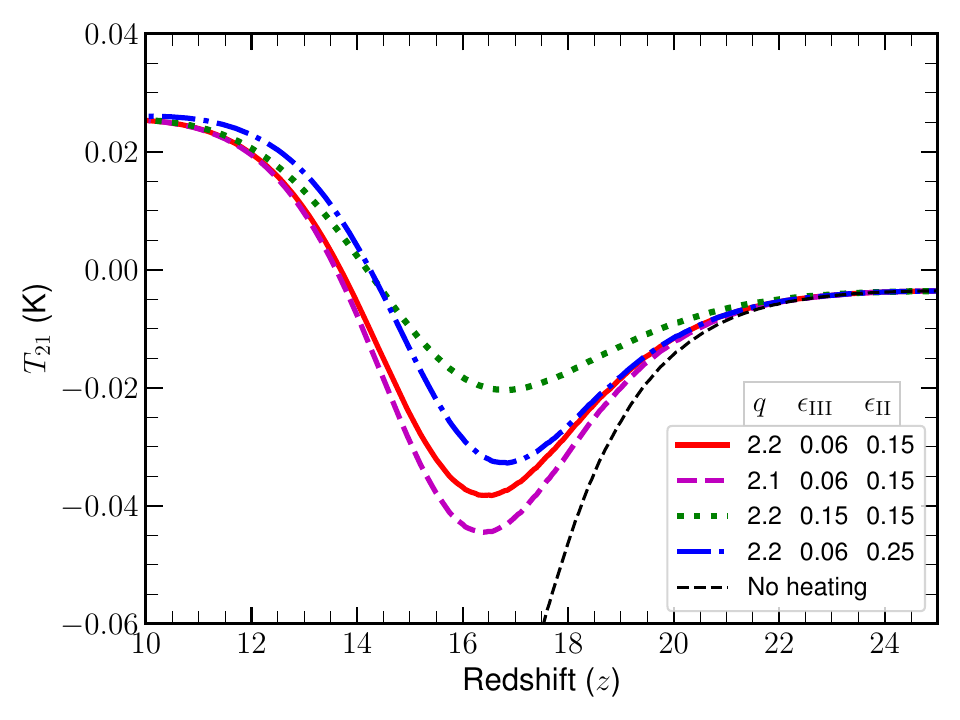}}
    \caption[The impact of cosmic ray heating on $T_g$, $T_s$ and resulting $T_{21}$ (upper to lower) are shown for different parameters describing the efficiencies of cosmic ray heating with (left panel) and without (right panel) dark-matter baryon interaction.]{{\bf Left panel:} The impact of cosmic ray heating on $T_g$, $T_s$ and resulting $T_{21}$ (upper to lower) are shown for different parameters describing the efficiencies of cosmic ray heating. The inset values represent the spectral index of cosmic ray spectra, $q$, the efficiencies of heating due to cosmic rays generated from Pop~III and Pop~II stars, $\epsilon_{\rm III}$ and $\epsilon_{\rm II}$ respectively. {We keep the other parameter, $Q_{\rm CR,II}=\epsilon_{\rm II}$ for this figure}. In this panel, the evolution of temperatures is done in the extra cooling scenario due to dark matter-baryon interaction in light of EDGES observation. The black dotted curves in the upper and middle panels represent the CMBR temperature, $T_{\gamma}$, whereas the black dashed curves are the temperatures where no heating source is present.  {\bf Right panel:} Same as left panel but in absence of any dark matter-baryon interaction.}
    \label{fig:temp}
\end{figure*}   

In this section, we show the changes in gas temperature, spin temperature, and resulting 21-cm brightness temperature by varying our model parameters such as $q$, $\epsilon_{\rm III}$, and $\epsilon_{\rm II}$ that govern the energy deposition by cosmic ray particles into the IGM from both Pop~III and Pop~II stars. {Due to the fact that the $\epsilon_{\rm II}$ and $Q_{\rm CR, II}$ alter the temperatures similarly, we are keeping $Q_{\rm CR, II}$ as same as $\epsilon_{\rm II}$. Further note that there are a few parameters of our model like IMF of Pop~III stars, supernova energetic, etc that are ill-constrained and can alter our results. However, all these uncertainties affect similar ways in the final temperature history of IGM and thus variation of the efficiency parameter $\epsilon_{\rm III}$ can encompass variation in any of the above-mentioned parameters.} The left panels of Fig.~\ref{fig:temp} show results in presence of the dark matter-baryon interaction with dark matter mass $m_{\chi}=0.1$ GeV and the interaction cross-section $\sigma_{45}=2$. The right panel shows results in the standard scenario without the dark matter-baryon interaction. In the left top, middle and bottom panels of Fig.~\ref{fig:temp}, $T_g$, $T_s$ and $T_{21}$ are plotted respectively with the model parameters shown in the legends. We have also shown our results for the default parameters discussed above by red solid curves.
Further in the top and middle panels, the black dotted curves represent the CMBR temperature. The black dashed curve in the top panel is the gas temperature where no cosmic ray heating is considered and the corresponding spin temperature is shown in the middle panel by black dashed curve for reference. 

The variation w.r.t the cosmic ray spectral index $q$ is shown by magenta lines where we consider a shallower slope of $q=2.1$. Given the same energy input, the number of cosmic ray particles below $30$ MeV decreases for a smaller $q$. These are the particles that take part in the collisional heating and hence the flatter slope makes the cosmic ray heating less efficient as can be seen from the figure. This further increases the absorption depth in 21-cm signal as can be seen from bottom panel.

As already discussed in the previous sections, the cosmic ray heating is directly proportional to the efficiency parameters such as, $\epsilon_{\rm III}$ and $\epsilon_{\rm II}$.
For example, it can be followed from green dotted curve that if we increase the efficiency of cosmic rays originating from Pop~III stars, $\epsilon_{\rm III}$ from $6\%$ to $15\%$, the dominance of Pop~III stars increases significantly and the IGM temperature increases faster compared to our default model. This leads to an increase in the hydrogen spin temperature resulting in a much shallower 21-cm absorption signal of $\sim -0.2$~{K}. Similarly, if we increase the cosmic ray efficiency coming from Pop~II stars, i.e from $\epsilon_{\rm II} = 15 \%$ to $\epsilon_{\rm II} = 25 \%$, the contribution to the heating by cosmic rays from Pop~II stars increases making $T_g$ to rise faster from $z \sim 17$. This, in turn, increases $T_s$ which leads to a lesser depth of $\sim -0.4$~{K} $T_{21}$ signal. It can be seen from blue dash-dotted curves in the left panel of Fig.~\ref{fig:temp}. Thus it is clear that any higher efficiency of cosmic ray heating would be ruled out by the EDGES absorption signal. Therefore any 21-cm signal from redshift $z \sim 14-20$ can be used to constrain early star formation and the corresponding cosmic ray heating efficiencies.

In the right panels of Fig.~\ref{fig:temp}, we show our results without considering any dark matter-baryon interaction that is neglecting the Eq.~\ref{eq:dQ_b}. All other parameters are the same as in the left panels of the figure. As expected, in general, the IGM temperature is much higher having a minimum value of $\sim 10$~K by $z \sim 17$ for our default parameter set (solid red curves). This resulted in a spin temperature of $\sim 20$~K due to the Ly-$\alpha$ coupling. Afterward, the IGM temperature and hence $T_s$ increases due to the rapid cosmic ray heating. Finally, $T_s$ crosses the CMBR temperature by $z \sim 14$ which leads to a weaker absorption profile of maximum {$\sim -0.04$~K} as can be seen from bottom right panel. The exact position of the absorption maxima depends on the interplay of Ly-$\alpha$ coupling and the onset of cosmic ray heating. For example, more efficient cosmic ray heating as shown by the green dotted line resulted in an absorption depth of {$\sim -0.02$~K} at an earlier redshift. Therefore, we conclude that cosmic ray heating is likely to play an important role in shaping the global 21-cm signal. In particular, it is likely to reduce the depth of any absorption signal if at all present.
In passing we note that, the Pop~II stars are likely to heat the IGM by magnetosonic heating which is sensitive to the IGM magnetic field. Thus the global 21-cm signal can, in principle, be used to constrain the magnetic field during the cosmic dawn \autocite{Minoda19, Bera_2020}.


{
\subsection{Comparison with X-ray heating}
\begin{figure}
    \centering
	\includegraphics[width=0.7\columnwidth]{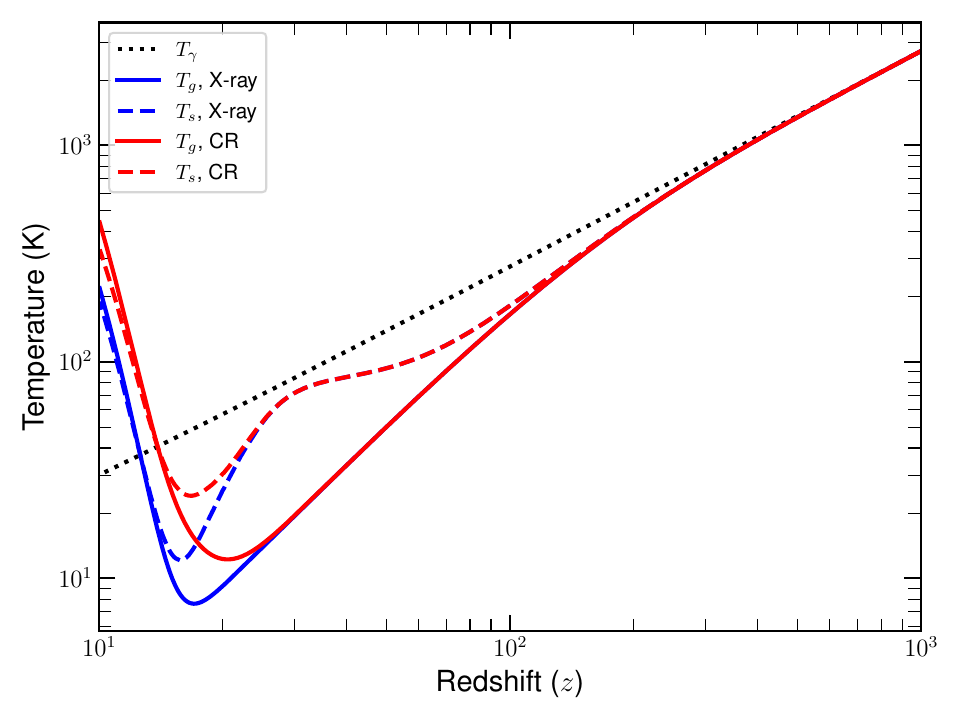}
	\includegraphics[width=0.7\columnwidth]{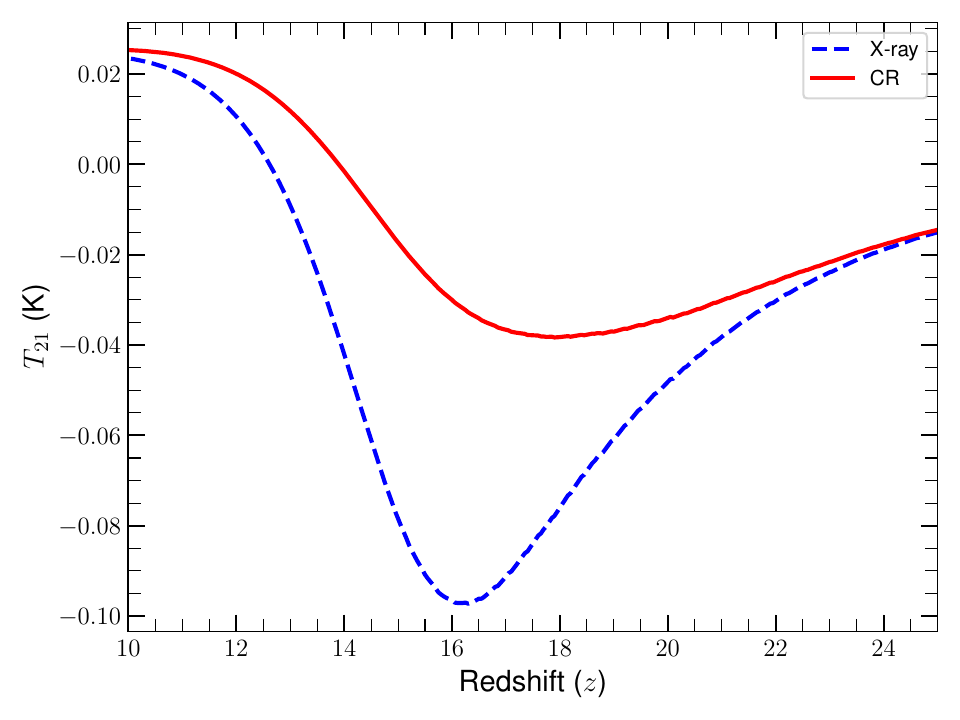}
    \caption[The comparison in the temperature evolution due to cosmic ray and X-ray heating.]{{The temperature evolution shown in red is due to the presence of cosmic ray heating for the parameters $\epsilon_{\rm III}=0.1$, $\epsilon_{\rm II}$ and/or $Q_{\rm CR,II}=0.15$, while the temperatures shown in blue are due the X-ray heating for the parameters $f_{\rm Xh,III}=0.001$, $f_{\rm Xh,II}=0.2$. The resulting brightness temperatures, $T_{21}$ are shown in the lower panel due to X-ray heating (blue dashed) and cosmic ray heating (red solid).}}
    \label{fig:Temp_CR_Xray}
\end{figure}
There are other processes that are likely to be responsible for heating of the intergalactic medium such as, X-ray heating \autocite{pritchard2007, baek09, ghara15, fialkov14b}, Ly-$\alpha$ \autocite{Chuzhoy2006, ghara2020, mittal21}, CMB \autocite{Venumadhav2018}, etc. At present there is no evidence showing the domination of any heating mechanism over others. All possible heating mechanisms have their own free parameters which are not well-constrained. Thus only when more observational evidences are available, detailed modeling of 21-cm signal including all possible sources of heating is effective and that is beyond the scope of the current work. However, to demonstrate the importance of cosmic ray heating, we compare our results due to cosmic ray heating with X-ray heating without the presence of any exotic physics such as dark matter-baryon interaction.
In the upper panel of Fig.~\ref{fig:Temp_CR_Xray}, we have shown the comparison of gas (solid) and spin (dashed) temperatures for cosmic ray heating (red) and X-ray heating (blue) for the same star formation rate densities and the Ly-$\alpha$ coupling. In order to include the X-ray heating in our model we followed Eq.~14 given in \textcite{atri2021}. In case of cosmic ray heating, we have taken the efficiencies as $\epsilon_{\rm III}=0.1$ and $\epsilon_{\rm II}$ and/or $Q_{\rm CR,II}=0.15$ for this particular plot. On the other hand, for X-rays the contribution to the heating from Pop~III stars and Pop~II stars are $f_{\rm Xh,III}=0.001$ and $f_{\rm Xh,II}=0.2$ which is motivated by \textcite{atri2021} and \textcite{furlanetto06}. We varied the heating rate due to X-rays coming from Pop~III stars from $f_{\rm Xh,III}=0.001$ to $f_{\rm Xh,III}=0.1$ and the changes in temperatures are negligible as stated in \textcite{atri2021}. It is clear from the blue and red curves that the contribution of cosmic rays and X-rays to the heating are quite similar. Although a small change in $T_s$ can make a big difference in the resulting brightness temperature that is shown in the lower panel of Fig.~\ref{fig:Temp_CR_Xray}. For the set of parameters that we used here, the cosmic ray heating is more efficient as compared to the X-ray heating for the same star formation history, though depending upon the parameters the magnitude of heating can be changed. Thus we conclude that the cosmic ray heating is comparable to X-ray heating during cosmic dawn and has similar effect to shape the global 21-cm signal.
}

\section{Summary}
\label{main_sec:summary}

We present a semi-analytical model of global HI 21-cm signal from cosmic dawn focusing on the heating by cosmic rays generated from the early generation of stars. In our model, we take into account both metal-free Pop~III and metal-enriched Pop~II stars. We consider the meta-galactic Lyman-Werner feedback on the Pop~III stars along with supernova feedback in Pop~II galaxies in order to calculate the star formations. The global HI 21-cm signal is calculated self-consistently taking into account the Ly-$\alpha$ coupling. The temperature evolution of IGM has been calculated using cosmic ray heating along with the dark matter-baryon interaction and adiabatic cooling. We consider the ionization and collisional heating by low-energy cosmic ray protons generating and escaping from Pop~III halos.
Besides accounting for the amount of cosmic ray energy in the IGM, our model captures the evolution of the cosmic ray spectrum, as they propagate through the IGM and lose their energy by ionizing and exciting neutral Hydrogen atoms, and due to the adiabatic expansion. We further consider the cosmic rays produced by the Pop~II stars which transfer energy to the intergalactic medium through the generation of magnetosonic Alfv\'en waves.
Here, we would like to mention that, all the quantities such as IGM kinetic/spin temperature, ionization fraction, Lyman-$\alpha$/Lyman-Werner background radiation, star formation rate, etc estimated here are globally averaged quantities. In principle, one should first simulate them on 3D cubes self-consistently at different redshifts and then perform the averaging. We must note that,  although estimating the global 21-cm signal directly from the global quantities is faster, it can introduce bias in the estimated signal.

{We find that the cosmic ray is an important source of IGM heating and shapes the global 21-cm signal during cosmic dawn which is in agreement with the existing literature.} The depth, duration and timing of the absorption signal are highly modulated by the cosmic ray heating that we consider here. In particular, the EDGES signal can be well explained by our model of cosmic ray heating along with the Ly-$\alpha$ coupling and dark matter-baryon interaction with a suitable choice of efficiency parameters.
In fact, the required efficiency parameters of cosmic ray heating are reasonably small like $\epsilon_{\rm III}=0.06$ \& $\epsilon_{\rm II}$ {and/or $Q_{\rm CR,II}=0.15$} to produce significant heating of the IGM and match the EDGES observed profile. Further, we explore the various efficiency parameters related to the cosmic ray heating and show that the brightness temperature highly depends on these parameters. In particular, highly efficient cosmic ray heating reduces $T_{21}$ by a significant amount. We also showed that the cosmic rays can highly impact the IGM temperature or in turns the 21-cm signal, in absence of any dark matter-baryon interaction, and could even potentially wash out the absorption signal during cosmic dawn. 
Thus cosmic rays need to be considered as a potential source of IGM heating along with widely explored sources of heating such as through soft X-rays during the cosmic dawn. 
We further argued that, since the cosmic 21-cm signal can be highly modulated by the heating due to cosmic rays produced by the early generation of stars, accurately determined 21-cm signal by experiments such the EDGES  \autocite[Experiment to Detect the Global Epoch of Reionization Signature,][]{EDGES18}, SARAS  \autocite[Shaped Antenna measurement of the background RAdio Spectrum,][]{raghunathan21}, LEDA  \autocite[Large-aperture Experiment to Detect the Dark Ages,][]{price18}, REACH  \autocite[Radio Experiment for the Analysis of Cosmic Hydrogen,][]{acedo19}, SKA (Square Kilometer Arrays), HERA (Hydrogen Epoch of Reionization Array),  etc.  could be used to probe the early cosmic ray heating and constrain the nature of these early generations of stars.  

Currently, there is only one observational evidence of global HI 21-cm signal by \citet{EDGES18} along with a null detection by \citet{Saurabh_2021} during cosmic dawn. However, there are several observational constraints during the reionization and post-reionization epochs.
Hence, in the following chapter, we combine the majority of the available observations from cosmic dawn and the epoch of reionization to explore the prospects of bridging the gap between these two epochs.

\chapter[Bridging the gap between CD \& EoR]{Bridging the gap between cosmic dawn and epoch of reionization\footnote{This chapter is adapted from the paper, ""Bridging the Gap between Cosmic Dawn and Reionization favors Faint Galaxies-dominated Models" by \citet{2022arXiv220914312B}.}}
\epigraph{\itshape  Not until the empirical resources are exhausted, need we pass on to the dreamy realms of speculation.}{-- Edwin Powell Hubble}
\label{chap:Bridging}
\startcontents[chapters]
\printmyminitoc{
As already mentioned, the measurements of the 21-cm global signal at $z>15$ by the EDGES collaboration \citep{2018Natur.555...67B} provide the constraints on cosmic dawn. On the other hand, currently, there are several observational global constraints on reionization at $z\gtrsim5$, including measurements of the volume-averaged neutral hydrogen fraction of the IGM
using different techniques such as the evolution in Lyman-$\alpha$ and Lyman-$\beta$ forests \citep{2001AJ....122.2833F, White_2003, 2015MNRAS.447..499M} and Lyman-$\alpha$ emitters \citep{Ouchi_2010, 2011ApJ...726...38Z, 10.1093/mnras/stu2089, 10.1093/mnras/stv1751,2019MNRAS.485.3947M}, the optical depth to Thomson scattering of cosmic microwave background (CMB) photons as measured by \citet{2020A&A...641A...6P}, and the ionizing emissivity constraints as compiled by \citet{2013MNRAS.436.1023B}.  
Inferred from the EDGES signal, \citet{2018MNRAS.480L..43M} have derived a constraint on the early star formation and shown that it is consistent with an extrapolation of UV measurements at lower redshifts ($4 < z < 9$). While there have been many theoretical \citep[see e.g.,][]{Hills_2018, Bradley2019, Singh2019, Sims2020} and observational works \citep[see e.g.,][]{2022NatAs...6..607S} suggesting that the measured profile by EDGES is not of astrophysical origin, it remains crucial to understand what implications such a detection might have on cosmic dawn and reionization. In particular, could additional constraints be placed on the role of faint and bright galaxies during these epochs if the measurement holds? In the light of the recent successful launch of the James Webb Space Telescope (JWST) and the growing observational efforts to detect the 21-cm signal during cosmic dawn, many high-redshift constraints are expected, and hence it is currently a scientific priority to ask the question: {\it What is required to bridge the gap between cosmic dawn and reionization?} 

We here use a semi-analytical approach coupled with a Markov Chain Monte Carlo (MCMC) sampler to explore the range of scenarios and models that are mostly consistent with the combined constraints from cosmic dawn and reionization. Since all considered constraints are globally-averaged quantities, a semi-analytical approach is sufficiently accurate while also being much faster than using reionization simulations where numerical limitations in the resolution, box size, sub-grid physics, and radiation transport prescriptions complicate the interpretability and feasibility of parameter explorations \citep[e.g.][]{2011MNRAS.411..955M, 2017MNRAS.468..122H, 2020NatRP...2...42V, 2021JCAP...02..042W}. To obtain insights on the role of faint and bright galaxies, we use a physically motivated source model for the ionization rate ($R_{\rm ion}$) that was derived from radiative transfer simulations \citep{2015MNRAS.447.2526F, 2016MNRAS.457.1550H}. This parameterization assumes a non-linear relation between the ionization rate and the halo mass $R_{\rm ion} \propto M_h^{C+1}$, where $C=0$ corresponds to a linear relation, as is typically assumed in many semi-numerical models of reionization via the efficiency parameter \citep{2011MNRAS.411..955M}. In addition to the $C$ parameter, we aim to constrain the most debated parameter in reionization models, namely, the escape fraction of ionizing photons $f_{\rm esc}$. Harnessing the MCMC framework, our aim is to explore the joint $C$--$f_{\rm esc}$ parameter space to determine the range of models that naturally reproduce the combined constraints from reionization and cosmic dawn. Finally, we aim to study the implications these constraints might have on galaxy evolution during these early formation epochs.

This chapter is organized as follows: we describe our empirical source model in section~\ref{sec:source} and calibrate it to cosmic dawn (EDGES) constraints in section~\ref{sec:mcd}. We then constrain our model to several reionization observables in section~\ref{sec:eor}. Further, we calibrate our source model jointly to both cosmic dawn and reionization in section~\ref{sec:cdEoR}. Finally, we summarise our findings in section~\ref{sec:conc}.
}

\begin{figure*}
    \centering
    \includegraphics[width=0.75\columnwidth]{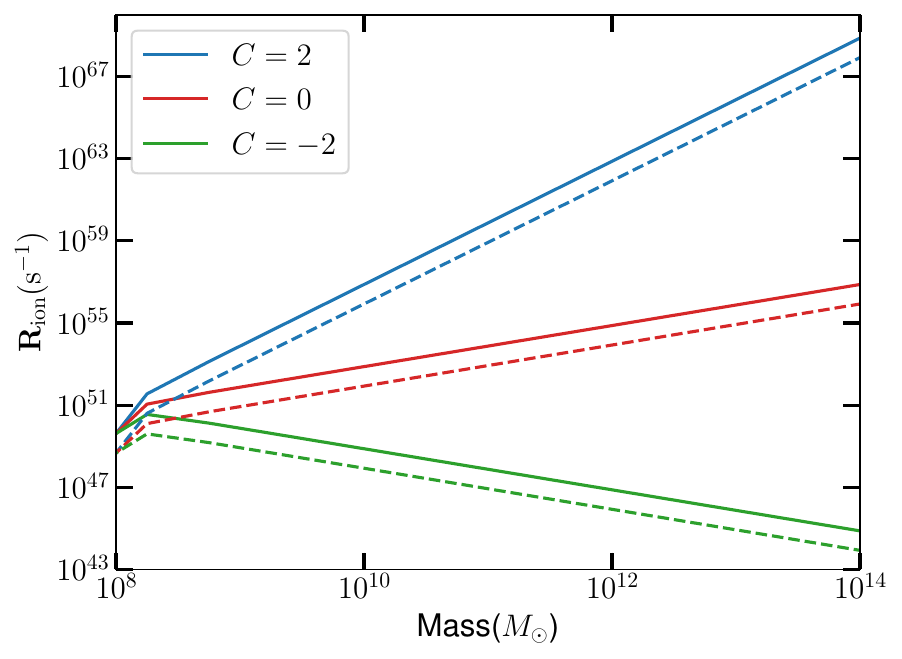}
    \includegraphics[width=0.75\columnwidth]{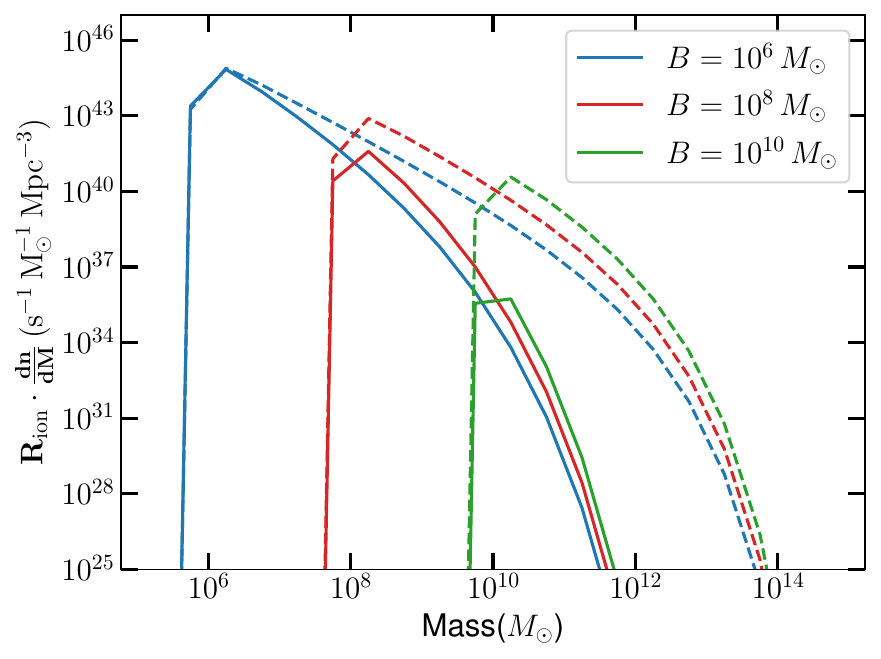}
    \caption[The ionization rates for different model parameters.]{{\bf Top panel:} The ionization rates for different $C$ values at $z\sim17$ (solid) and $z\sim6$ (dashed) while fixing other parameters to values of $A=10^{40}\,\Msun^{-1}\text{s}^{-1}$, $B=10^{8}\,\Msun$, and $D=2.28$ as derived from radiative transfer simulations \citep{2016MNRAS.457.1550H}. This shows that positive and negative $C$ correspond to higher ionization rate efficiencies from massive and low-mass halos, respectively. {\bf Bottom panel:} The ionization rates multiplied by the Sheth--Tormen halo mass function for different values of $B$, while keeping the other parameters fixed at $A=10^{40}\,\Msun^{-1}\text{s}^{-1}$, $C=-0.5$, and $D=2.28$. The $B$ parameter represents the minimum halo mass ($M_{h,\rm min}$) considered in the model.}
    \label{fig:Rion}
\end{figure*}

\section{Source model}
\label{sec:source}
In this section, we discuss the details of the source model considered in this work. To study the role of different populations of galaxies, we consider a physically motivated source model that was derived in \citet{2016MNRAS.457.1550H} from the radiative transfer simulations of reionization described in \citet{2018MNRAS.480.2628F}. In this model, the ionization rate $R_{\rm ion}$ is parameterized as a function of halo mass ($M_h$) and redshift ($z$). The mass dependence of $R_{\rm ion}$ is analogous to the Schechter function, which can be characterized as a power law on the bright end and an exponential cut-off on the faint end. The redshift dependence follows a simple power law.
This parameterization of $R_{\rm ion}$ accounts for the non-linear dependence on halo mass, which can be expressed as follows:
\begin{equation} \label{eq:Rion}
    \frac{R_{\rm ion}}{M_h} = A\,(1+z)^D \left(\frac{M_h}{B}\right)^C \exp\left[-\left(\frac{M_h}{B}\right)^{-3}\right] \, ,
\end{equation}
where $A$, $B$, $C$, and $D$ are free parameters, and their best-fit values were obtained by calibrating to radiative transfer simulations. We refer the reader to \citet{2016MNRAS.457.1550H} for more details about the derivation of this model. We now discuss the physical meaning of these free parameters. Parameter $A$ acts as an amplitude of $R_{\rm ion}$, which scales the ionizing emissivity over the entire halo mass range at a given redshift by the same amount. {Parameter $B$ determines the minimum halo mass, which can be thought of as the quenching mass scale due to feedback from star formation and photoionization heating.} Parameter $C$ quantifies the slope of the $R_{\rm ion}$--$M_h$ relation, which controls the contribution of different mass scales to the total emissivity. Lastly, parameter $D$ accounts for the redshift dependence of ionization rate for a given halo mass.

Having defined our source model $R_{\rm ion}$, we adopt the \citet{1999MNRAS.308..119S} halo mass function ($\frac{\text{d}n}{\text{d}M}$), which provides the number density of halos per unit halo mass, to compute the cosmic evolution of the global quantities governing reionization and comic dawn. In this work, we consider the halo mass range from $10^{5}$ to $10^{15}\,\Msun$ to account for the contribution from all source populations including the most faint halos.

The most interesting parameters of this source model are the $B$ and $C$ parameters since they allow us to draw conclusions about the role of different source populations during cosmic dawn and reionization. 
Therefore, we set the amplitude $A = 10^{40}\,\Msun^{-1}\text{s}^{-1}$ and redshift index $D = 2.28$, following the calibration to radiative transfer simulations \citep[for more details see][]{2016MNRAS.457.1550H}\footnote{We have varied the $A$ and $D$ parameters to reproduce several observations using MCMC and found similar values. Hence, fixing these parameters has a minimal impact on the presented results.} throughout. To illustrate how the source model parameters ($B$ and $C$) affect the $R_{\rm ion}$--$M_h$ relation, in Figure~\ref{fig:Rion} we show several $R_{\rm ion}$ models with different $C$ and $B$ values at cosmic dawn ($z\sim17$, solid) and reionization ($z\sim6$, dashed). As mentioned earlier, the $C$ parameter (top panel) scales the ionization rate ($R_{\rm ion}$) as a function of halo mass following $R_{\rm ion} \propto M_h^{C+1}$. In the top panel of Figure~\ref{fig:Rion}, $C=0$ (red curves) corresponds to a linear relation.
However, in the case of $C = 2$ (blue curves), the emissivity increases with halo mass and hence favors a relatively larger contribution from more massive halos than low-mass halos, and vice versa for the case of $C = -2$ (green curves), where reionization is dominated by low-mass halos. {While this depends on the photon escape fraction, under the assumption of a mass-independent escape fraction, 
the intrinsic $R_{\rm ion}$ increases by several orders of magnitude from low to high mass halos as seen in Figure~\ref{fig:Rion}.} In the bottom panel, we show the impact of varying the $B$ parameter, which sets the minimum halo mass scale. 

\begin{figure*}
    \centering
    \includegraphics[width=0.65\columnwidth]{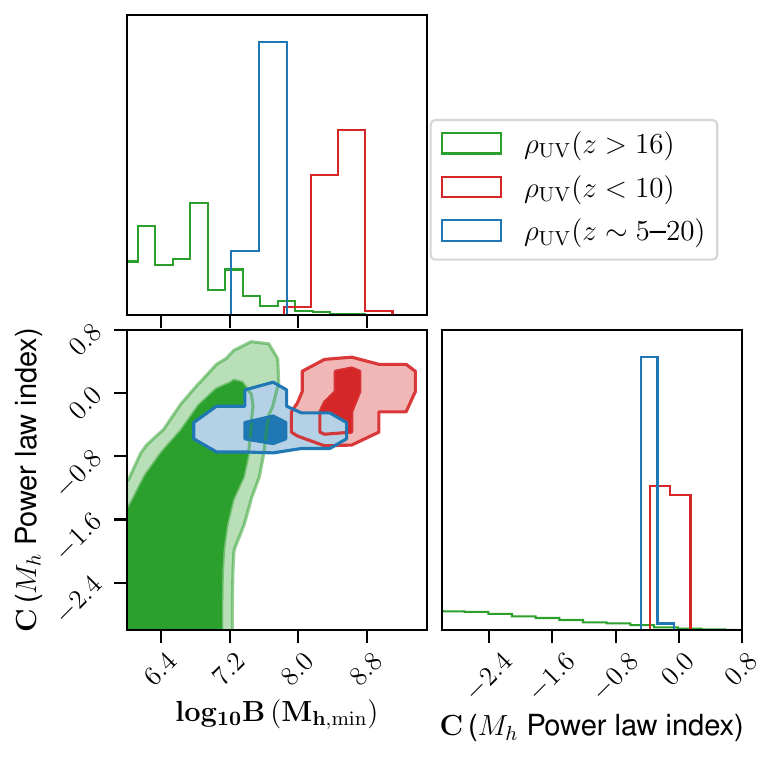}
    \includegraphics[width=0.65\columnwidth]{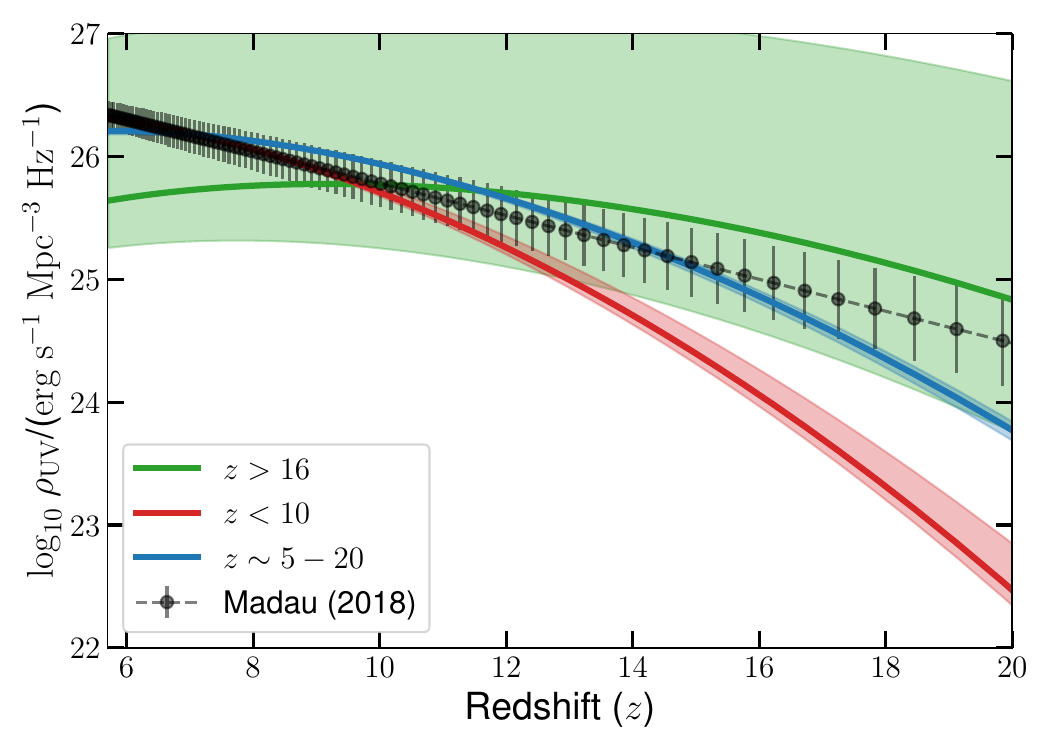}
    \caption[Several posterior distributions derived using the UV luminosity density \citep{2018MNRAS.480L..43M} including constraints for three different redshift ranges are shown in the top panel. Whereas, in the right panel, a comparison between several UV luminosity densities calculated with the inferred parameter values is given.]{{\bf Top panel:} Several posterior distributions derived using the UV luminosity density \citep{2018MNRAS.480L..43M} including constraints for three different redshift ranges, namely $z>16$ (representing cosmic dawn only, green), $z<10$ (during reionization only, red), and $z\sim5$--$20$ (cosmic dawn and reionization, by combining the full redshift range, blue). {\bf Bottom panel:} Comparison between several UV luminosity densities calculated from $R_{\rm ion}$ with the inferred parameter values using MCMC. This figure indicates that fitting to $z>16$ favors more negative $C$ values, which implies stronger contributions from low-mass (faint) galaxies.}
    \label{fig:madau_sfr}
\end{figure*}

\section{Source model calibration to cosmic dawn (EDGES) constraints}
\label{sec:mcd}
As previously mentioned, our goal is to find the possible range of models that can reproduce the existing observational constraints during cosmic dawn and reionization. To do so, we first calibrate our source model parameters ($C, B$) to the intrinsic UV luminosity density ($\rho_{\rm UV}$) inferred from the EDGES signal in the redshift range $z \sim 5$--$20$, where the function $\log_{10} [\rho_{\rm UV}/({\rm erg\,s^{-1}\,Mpc^{-3}\,Hz^{-1}})] = (26.30 \pm 0.12) + (-0.130 \pm 0.018)(z-6)$ is the best fit as compiled by \citet{2018MNRAS.480L..43M}.

Using our source model, we compute the UV luminosity density as follows. First, the ionization rate $R_{\rm ion}$ (Equation~\ref{eq:Rion}) can be converted to a star-formation rate (SFR) by accounting for a stellar metallicity-dependent parameterization of the ionizing photon flux $Q_{\rm ion}$ following:

\begin{equation} \label{eq:SFR}
    {\rm SFR} = R_{\rm ion}(M_h,z)/ Q_{\rm ion}(Z) \, .
\end{equation}
The sub-solar metallicity-dependent parameterization is provided by \citet{2011ApJ...743..169F}:
\begin{equation}
    \log Q_{\rm ion}(Z) = 0.639 (-\log Z)^{1/3} + 52.62 - 0.182 \, ,
\end{equation}
where $Q_{\rm ion}$ is in units of $\rm s^{-1} \, (\Msun\,yr^{-1})^{-1}$. This form of $Q_{\rm ion}$ is consistent with the equilibrium values measured by \citet{2003A&A...397..527S} assuming a \citet{2003PASP..115..763C} initial mass function (IMF). We also adopt the \citet{2017ApJ...840...39M} redshift evolution of the mass-weighted metallicity ($Z$) given by:
\begin{equation}
    \log\left(Z/\Zsun\right) = 0.153 - 0.074\,z^{1.34} \, .
\end{equation}
The SFR can then be written in terms of the UV luminosity density $L_{\rm UV}$ \citep{1998ARA&A..36..189K, 1998ApJ...498..106M} as follows:
\begin{equation}
    {\rm SFR} [{\rm \Msun\,yr^{-1}}] = 1.25 \times 10^{-28} L_{\rm UV} [{\rm ergs\,s^{-1}\,Hz^{-1}}] \, .
\label{eq:UVtoSFR}
\end{equation}
The global $\rho_{\rm UV}$ is obtained by integrating $\int L_{\rm UV}\frac{\text{d}n}{\text{d}m}\text{d}m$ over the entire mass range at different redshifts.

We now constrain our source model parameters ($B$ and $C$) to the \citet{2018MNRAS.480L..43M} UV luminosity evolution using {\sc emcee}, which is an affine-invariant ensemble sampler for MCMC \citep{2013PASP..125..306F}. We assume a flat prior in the range: $\log (B/\Msun) \in [5,10]$ and  $C \in [-5,5]$.
Our task is to find the range of models that minimizes the following multivariate $\chi^{2}$ distribution:
\begin{equation}
    \chi^{2} {\equiv} \sum_{z} \frac{(\rho_{\rm UV, model}(z) - \rho_{\rm UV,obs}(z))^{2}}{2\,(\sigma_{\rho_{\rm UV, obs}}(z))^{2}} \, .
    \label{eq:chi2}
\end{equation}

\begin{table}
    \centering
    \renewcommand{\arraystretch}{2.5}
    \caption{Best-fit values of photon escape fraction ($f_{\rm esc}$) and power dependence on halo mass ($C$). \\}
    \label{table:BandC}
    \begin{adjustbox}{width=0.8\textwidth}
    \small
    \begin{tabular}{||c||c|c|c||}
    \hline \hline
    Parameters & $z>16$ & $z<10$ & $z\sim5-20$ \\
    \hline \hline
    ${\bf \log_{10}B}$ & $6.34^{+0.69}_{-0.80}$ & $8.49^{+0.16}_{-0.17}$ & $7.67^{+0.07}_{-0.15}$ \\
    \hline
    ${\bf C}$ & $-6.43^{+3.78}_{-9.0}$ & $-0.11^{+0.04}_{-0.05}$ & $-0.34^{+0.03}_{-0.03}$ \\
    \hline 
    \end{tabular}
    \end{adjustbox}
    \vspace{1ex}
    
    {\raggedright \small{Note. -- The other model parameters are fixed to values obtained by fitting to radiative transfer simulations \citep[$A=10^{40}\,\Msun^{-1}\text{s}^{-1}$ and $D=2.28$, see][]{2016MNRAS.457.1550H}.} \par}
\end{table}

In the top panel of Figure~\ref{fig:madau_sfr} we show several posterior distributions of the parameters $B$ and $C$ by considering different epochs, namely, cosmic dawn ($z>16$, green), reionization ($z<10$, red), and the combination of cosmic dawn plus reionization ($z\sim 5$--$20$, blue). The dark and light shaded contours correspond to 1$\sigma$ and 2$\sigma$ levels, respectively. From the 1-dimensional probability distribution function (PDF) of the $B$ parameter (i.e. $M_{h,\rm min}$) it is clear that the minimum halo mass or the mass cut-off varies between $\sim 10^{6-8}\,\Msun$. It is also evident that calibrating the model to reproduce higher-$z$ $\rho_{\rm UV}$ constraints ($z > 16$) favors models with lower mass cut-offs ($\sim 10^{6}\,\Msun$) and vice versa. Likewise, a more negative $C$ is preferred to match with higher-$z$ constraints. As discussed before, the negative value of $C$ and the low value of $B$ both suggest that the low-mass (faint) halos play an important role in reproducing the extrapolated $\rho_{\rm UV}$ constraints. Overall, using $\rho_{\rm UV}$ measurements in the entire redshift range ($z\sim5$--$20$) provides much tighter posterior contours than constraining to either $z>16$ or $z<10$. This is mainly due to the number of redshift bins used in each case. Since the time duration in the redshift range of $z=16$--$20$ is shorter than that of $z=5$--$20$ or $z=5$--$10$, more redshift bins exist in the latter than the former, and hence more constraining power is expected. Each of the best-fit values with 1$\sigma$ errors is provided in Table~\ref{table:BandC} for the three scenarios. In the bottom panel of Figure~\ref{fig:madau_sfr}, we compare the $\rho_{\rm UV}$ evolution predictions from these three models with the \citet{2018MNRAS.480L..43M} $\rho_{\rm UV}$ evolution. Shaded areas show the 1$\sigma$ confidence levels which are obtained by translating the parameter 1$\sigma$ levels from Table~\ref{table:BandC} into constraints on $\rho_{\rm UV}$. All models are within the 1$\sigma$ level of the $\rho_{\rm UV}$ constraints over the relevant redshift ranges. This figure clearly demonstrates that a stronger contribution from low-mass halos is required to reproduce the $\rho_{\rm UV}$ constraints inferred from EDGES during cosmic dawn (green curve). 
This is in agreement with the picture that massive halos are rare and biased toward later formation times, whereas {low-mass halos are predominant at these early epochs}. 
This can be seen in the bottom panel of Figure~\ref{fig:Rion}, where the maximum halo mass during reionization ($z=6$, dashed) is about 2 orders of magnitude higher than during cosmic dawn ($z=17$, solid).

Since our aim is to bridge the gap between cosmic dawn and reionization, we will fix the $B$ (i.e. $M_{h,\rm min}$) parameter to the value obtained by calibrating the source model to the $\rho_{\rm UV}$ constraints over the entire redshift range ($z=5$--$20$). Nevertheless, we will keep the halo mass power-law index parameter ($C$) as a free parameter to explore its correlations with the escape fraction of ionizing photons ($f_{\rm esc}$), and to test whether similar 
values can be derived by adding reionization constraints, which we discuss next.

\begin{figure*}
    \centering
    \includegraphics[width=0.5\columnwidth]{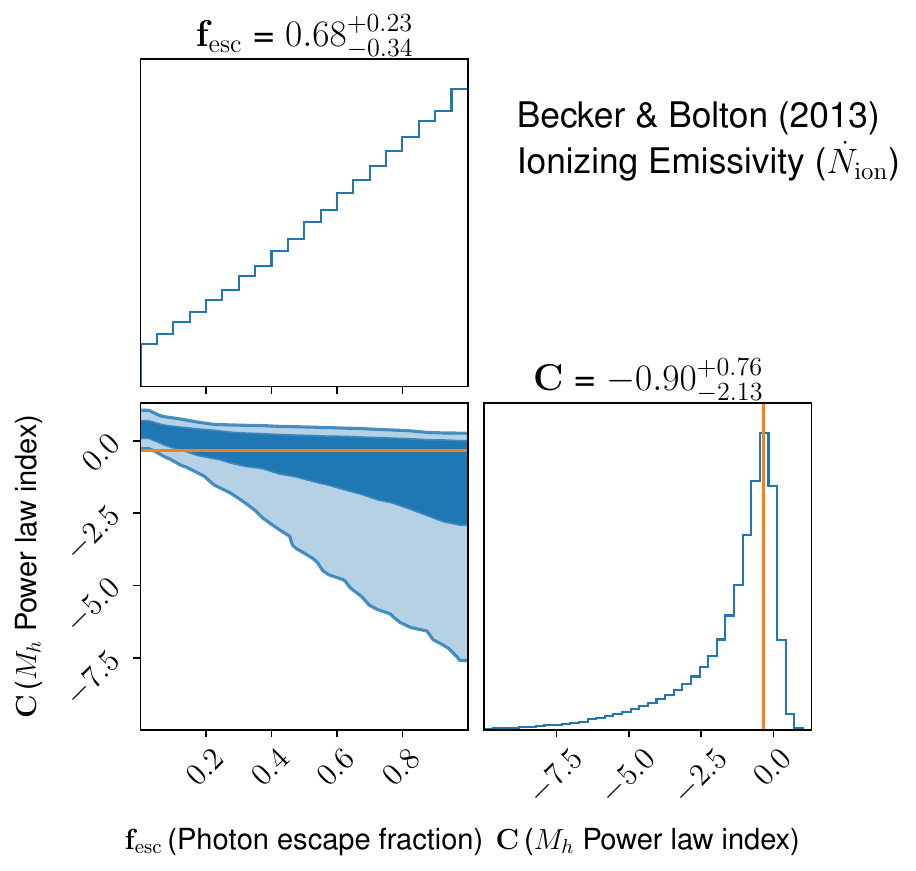}\includegraphics[width=0.5\columnwidth]{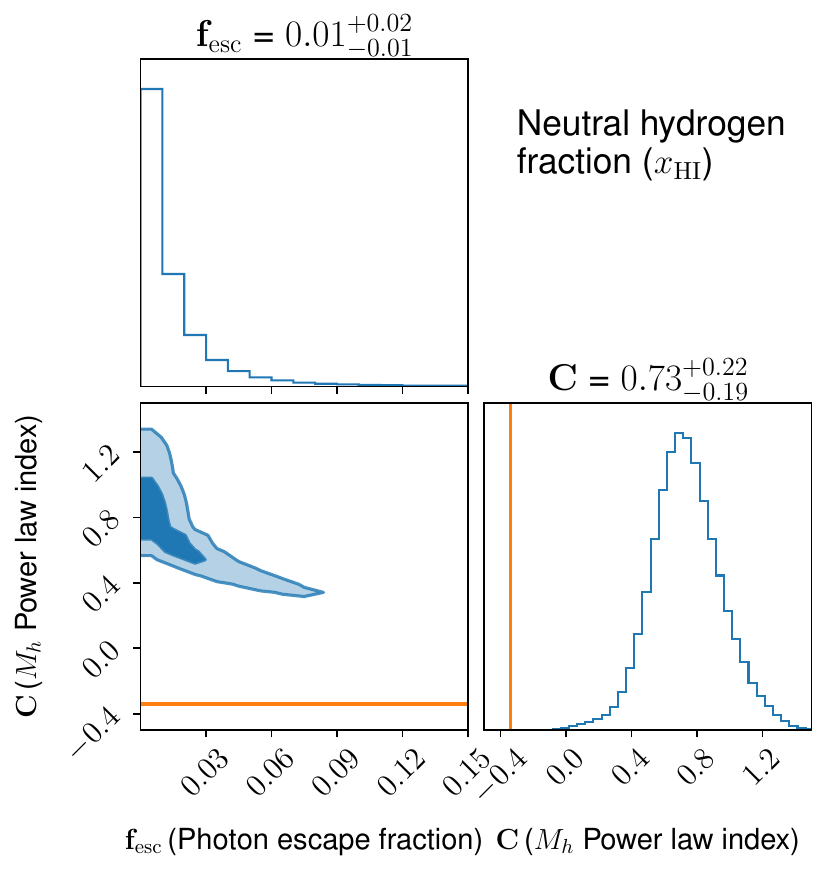}
    \includegraphics[width=0.5\columnwidth]{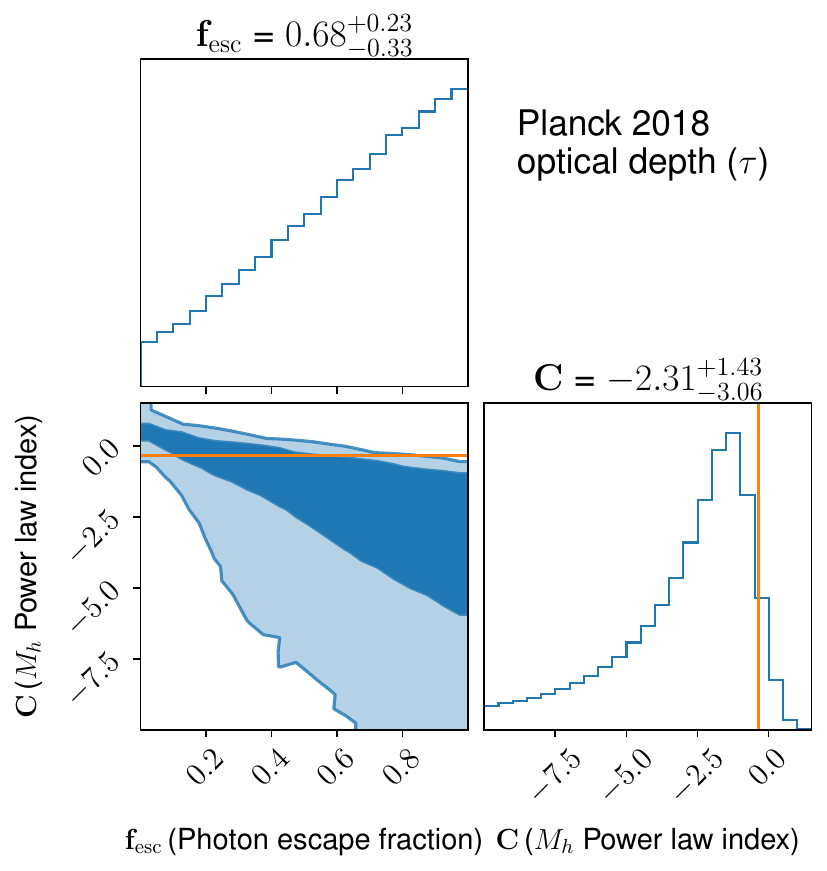}
    \caption[Posterior distributions for the escape fraction of ionizing photons $f_{\rm esc}$ and power-law index $C$ as constrained by the ionizing emissivity ($\dot{N}_{\rm ion}$; left), IGM neutral fraction ($x_{\rm HI}$; right), and Thomson optical depth ($\tau$; bottom).]{Posterior distributions for the escape fraction of ionizing photons $f_{\rm esc}$ and power-law index $C$ as constrained by the ionizing emissivity ($\dot{N}_{\rm ion}$; left), IGM neutral fraction ($x_{\rm HI}$; right), and Thomson optical depth ($\tau$; bottom). {The orange line in each plot denotes the value of $C$ parameter obtained in Table~\ref{table:BandC} for $z \sim 5$--$20$, for reference.} The emissivity and optical depth do not provide tight constraints on our parameters, due to the large uncertainties. The IGM neutral fraction measurements provide tighter constraints on our parameters and favor models with low $f_{\rm esc}$ and positive $C$ values, suggesting that reionization is driven by massive (bright) galaxies.}
    \label{fig:eor_con}
\end{figure*}

\section{Constraining the source model for key reionization observables}
\label{sec:eor}
We consider three key reionization observables to place constraints on our source model parameters,
namely the ionizing emissivity, IGM neutral fraction, and Thomson optical depth to the CMB.
In addition to the halo mass power-law index parameter ($C$) we also vary
the fraction of ionizing photons ($f_{\rm esc}$) that successfully escapes the remaining neutral hydrogen clumps and dust extinction in the interstellar medium (ISM) and circumgalactic medium (CGM) to contribute to the IGM ionization process. In this work, we assume a constant $f_{\rm esc}$ at all redshifts, since our $R_{\rm ion}$ function (Equation~\ref{eq:Rion}) already accounts for redshift and mass dependence.
We assume a flat prior in the range: $f_{\rm esc} \in [0,1]$ and  $C \in [-10,10]$, and combine different data using a multivariate Gaussian likelihood (similar to Equation~\ref{eq:chi2}).

\subsection{The ionizing emissivity, \texorpdfstring{$\dot{N}_{\rm ion}$}{Ndot}}
The integrated emission rate density of ionizing photons, or ionizing emissivity ($\dot N_{\rm ion}$), is a measure of the total number of ionizing photons per second per volume that escape from all ionizing sources to the IGM. In our case, the ionizing emissivity is related to the ionization rate ($R_{\rm ion}$) as follows:
\begin{equation} \label{eq:Nion}
    \dot{N}_{\rm ion} [{\rm s^{-1} Mpc^{-3}}] = f_{\rm esc}  \int R_{\rm ion}(M_h, z) \frac{\text{d}n}{\text{d}M_h} \text{d}M_h \, ,
\end{equation}
where $\frac{\text{d}n}{\text{d}M_h}$ is the \citet{1999MNRAS.308..119S} differential halo mass function, which gives the number density of halos in the mass range of $M$ and $M+\text{d}M$ per unit comoving volume.
We constrain our source model parameters ($C, f_{\rm esc}$) to the \citet{2013MNRAS.436.1023B} ionizing emissivity constraint at $z=4.75$ of $\log_{10} [\dot{N}_{\rm ion}/({\rm 10^{51}\,photons\,s^{-1}\,Mpc^{-3}})] = -0.014^{+0.454}_{-0.355}$.

In the top panel of Figure~\ref{fig:eor_con}, we show the $f_{\rm esc}$--$C$ joint posterior distribution constrained to match the ionizing emissivity measurement.
This shows that the $\dot{N}_{\rm ion}$ measurement alone cannot place a tight constraint on these parameters due to the large uncertainty. However, $\dot{N}_{\rm ion}$ data favors models with higher $f_{\rm esc}$ and more negative $C$, leading to a stronger contribution from low-mass halos. Similar parameter constraints were found in our earlier work on calibrating semi-numerical simulations to reionization observables \citep{2017MNRAS.468..122H}. {The best-fit value of $C$ parameter obtained in the previous section for $z \sim 5$--$20$ is also within the $1\sigma$ level as shown by orange horizontal line in Figure~\ref{fig:eor_con}.}

\subsection{The IGM neutral fraction, \texorpdfstring{$x_{\rm HI}$}{xHI}}
\label{sec:xHI}
We compute the reionization history from our models as follows. The rate of change in the ionized fraction of intergalactic hydrogen ($x_{\rm HII}$) is given by \citep{1999ApJ...514..648M},
\begin{equation}
    \frac{\text{d}x_{\rm HII}}{\text{d}t} = \frac{\dot N_{\rm ion}}{\bar n_{\rm H}} - \frac{x_{\rm HII}}{\bar t_{\rm rec}} \, .
\end{equation}
The first term describes the growth as a ratio between comoving ionizing emissivity ($\dot N_{\rm ion}$) and volume-averaged comoving number density of intergalactic hydrogen $\bar n_{\rm H}$, which is given by
\begin{equation}
    \bar n_{\rm H} = X \Omega_{b,0} \rho_{{\rm crit},0} / m_{\rm H} \, .
\end{equation} 
Here, $X$ is the cosmic hydrogen mass fraction (0.76), $\rho_{{\rm crit},0}$ is the present-day critical density, and $m_{\rm H}$ is the mass of a hydrogen atom. The second term models the sink of ionizing photons, where the recombination time-scale for the IGM is given by
\begin{equation}
    t_{\rm rec} = \left[ C_{\rm HII} \alpha_{\rm A} (1+Y/4X) {\bar n_{\rm H}} (1+z)^3 \right]^{-1} \, .
\end{equation}
Here, $C_{\rm HII}$ is the redshift-dependent clumping factor, which we adopt from \citet{2015MNRAS.451.1586P}. This clumping factor accounts for the overall density fluctuations in ionized medium, which boosts recombination rate by a factor up to $\sim 5$ near the end of reionization, predominantly contributed by ionized medium in the vicinity of halos. 
The $Y$ denotes the helium mass fraction (0.24) and $\alpha_{\rm A}$ is the case~A recombination coefficient ($4.2 \times 10^{-13}\,{\rm cm^3\,s^{-1}}$), corresponding to a temperature of $10^4\,{\rm K}$ \citep{Kaurov_2014}.
Having computed the reionization history, we now constrain our source model parameters to the IGM neutral fraction measurements given in Table~\ref{table:xHI_obs}.

\begin{table}
    \centering
    \renewcommand{\arraystretch}{3.0}
    \caption{IGM neutral hydrogen fraction measurements. \\}
    \label{table:xHI_obs}
    \begin{adjustbox}{width=1\textwidth}
    \small
    \begin{tabular}{||c||c|c|c||}
        \hline \hline
        Redshift($z$) & Constraints & Observables & References\\
        \hline \hline
        ${5.9}$ & $\leq 0.06\pm 0.05$ & Ly-$\alpha$ and Ly-$\beta$ forest dark fraction & \citet{2015MNRAS.447..499M} \\ \hline
        ${7.0}$ & $0.59^{+0.11}_{-0.15}$ & Ly-${\alpha}$ EW distribution & \citet{Mason_2018} \\
        \hline
        ${7.09}$ & $0.48 \pm 0.26$ &  & \citet{2018ApJ...864..142D} \\ 
        ${7.5}$ & $0.21^{+0.17}_{-0.19}$ & QSO damping wings & \citet{2019MNRAS.484.5094G} \\ 
        ${7.54}$ & $0.60^{+0.20}_{-0.23}$ &  & \citet{2018ApJ...864..142D} \\ \hline
        ${7.6}$ & $0.88^{+0.05}_{-0.10}$ & Lyman-break galaxies emitting Ly-$\alpha$ & \citet{2019ApJ...878...12H} \\
        \hline
    \end{tabular}
    \end{adjustbox}
\end{table}

In the right panel of Figure~\ref{fig:eor_con}, we show the parameter constraints given the above combination of IGM neutral fraction data. The $x_{\rm HI}$ data provides the tightest constraints,
with a clear tendency for the data to favor models with low $f_{\rm esc}$ and positive $C$ parameters. This implies that massive (bright) galaxies-dominated models are preferred. The low $f_{\rm esc}$ value (1\%) found here is a consequence of the non-linear relation between $R_{\rm ion}$ and $M_h^C$ through the $C$ parameter, where an anti-correlation between $f_{\rm esc}$ and $C$ is observed.
{However, we find that the resulting $C$ value is not within the range reported in Table~\ref{table:BandC}, hence we combine different constraints in the next section to understand the degeneracy between $f_{\rm esc}$ and $C$.}
Using the same source model in semi-numerical simulations of reionization, we have previously found that $f_{\rm esc}=4\%$ is sufficient to match the reionization observations \citep{2016MNRAS.457.1550H}, which is still within the 2$\sigma$ level of the current constraints based on updated observational data. We also find a lower $f_{\rm esc}$ value since our current analysis includes contributions from the wider halo mass range of $10^{5-15}\,\Msun$ as opposed to $10^{8-12}\,\Msun$ in \citet{2016MNRAS.457.1550H}. Such low $f_{\rm esc}$ values have been favored in several works \citep{gnedin2008escape, 2014MNRAS.442.2560W, 10.1093/mnras/stv1679,2019PhRvD..99b3518C,Rosdahl2022,Yeh2022}. 
{However, it is worth noting that in this work, we consider a constant $f_{\rm esc}$, and defer exploring the impact of assuming mass and/or redshift dependent $f_{\rm esc}$ to future works.}

\subsection{Thomson optical depth, \texorpdfstring{$\tau$}{tau}}
The optical depth is a measure of the scattering of CMB photons by free electrons produced by reionization. Given a reionization history, it is straightforward to obtain the Thomson scattering optical depth ($\tau$) as follows:
\begin{equation}
    \tau = \int_0^{\infty} \text{d}z \frac{c(1+z)^2}{H(z)}\, x_{\rm HII}(z)\, \sigma_T\, \bar{n}_{\rm H}\, (1+Y/4X) \, ,
\end{equation}
where $\sigma_T$ is the Thomson cross-section and $c$ is the speed of light. This optical depth can also be used to constrain the timing of reionization, where lower/higher $\tau$ corresponds to later/earlier reionization redshifts, respectively.

In the bottom panel of Figure~\ref{fig:eor_con}, we show the resulting posterior distribution of fitting to the measured value of $\tau = 0.054 \pm 0.007$ from \citet{2020A&A...641A...6P}.
Similar to the case with $\dot N_{\rm ion}$ (left panel), $\tau$ alone does not place tight constraints on our parameters due to the large uncertainty that is consistent with a broad range of reionization histories. The tendency to favor models with high $f_{\rm esc}$ and negative $C$ is also seen. However, allowing much more negative $C$, as compared to values derived from $\dot N_{\rm ion}$ data is suggesting that a stronger contribution from low-mass halos is needed to reproduce the \citet{2020A&A...641A...6P} $\tau$ in our model. {The constraints placed on the $C$ parameter using $\rho_{\rm UV}$ data for $z\sim5$--$20$ is consistent with constraints obtained by calibrating to \citet{2020A&A...641A...6P} $\tau$ at $1$--$2 \sigma$ level as shown by orange horizontal line.}

\begin{figure}
    \centering
    \includegraphics[width=0.95\textwidth]{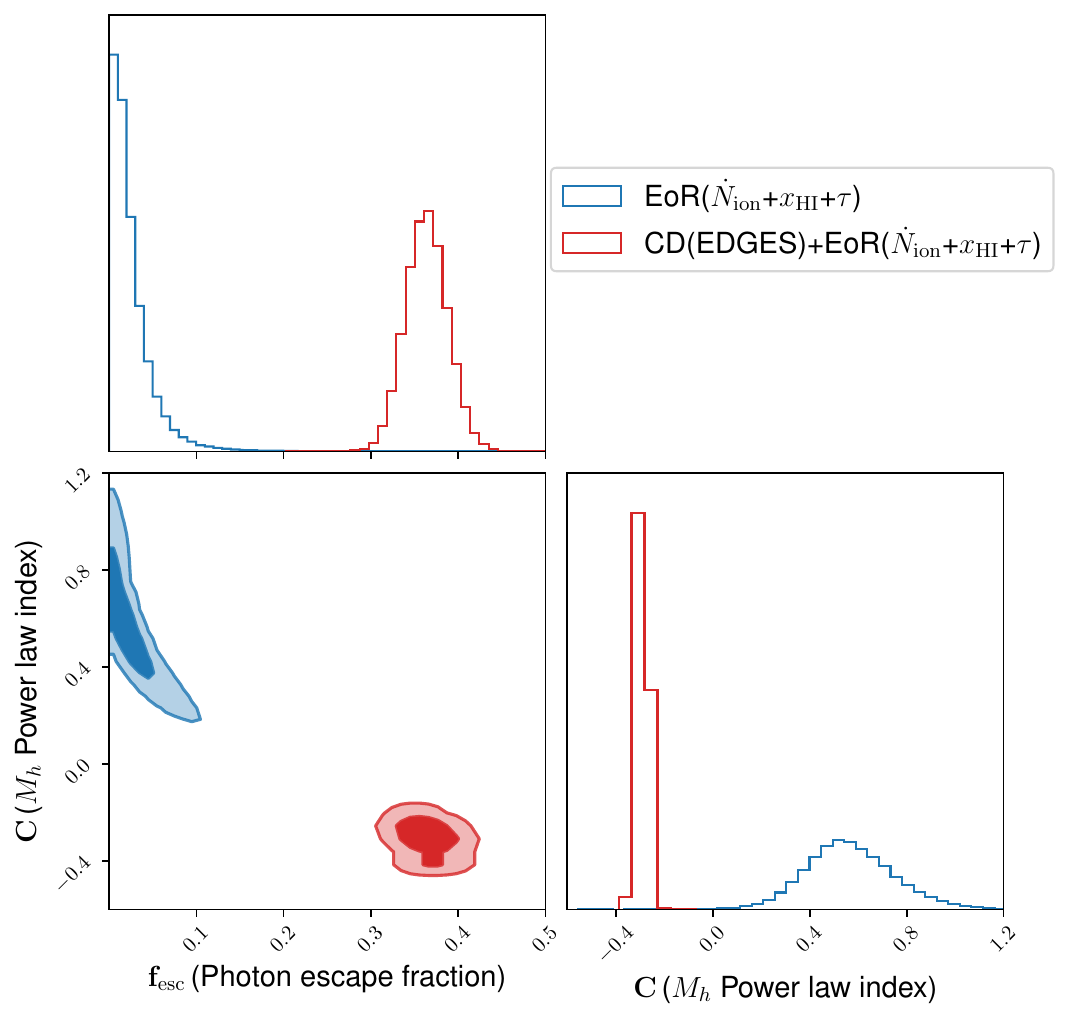}
    \caption[Comparison between the combined constraints from reionization, EoR ($\dot N_{\rm ion}, x_{\rm HI}$, $\tau$), and by adding cosmic dawn to reionization, CD ($\rho_{\rm UV}$ from EDGES) + EoR ($\dot N_{\rm ion}, x_{\rm HI}$, $\tau$).]{Comparison between the combined constraints from reionization, EoR ($\dot N_{\rm ion}, x_{\rm HI}$, $\tau$) in blue, and by adding cosmic dawn to reionization, CD ($\rho_{\rm UV}$ from EDGES) + EoR ($\dot N_{\rm ion}, x_{\rm HI}$, $\tau$) in red. The dark and light shaded contours correspond to the $1\sigma$ and $2\sigma$ levels, respectively. This demonstrates that adding cosmic dawn constraints shifts contours towards higher $f_{\rm esc}$ and negative $C$ parameters. This suggests that low-mass (faint) galaxies play a major role in bridging the gap between cosmic dawn and reionization.}
    \label{fig:cdEoR}
\end{figure}

\begin{figure}
    \centering
    \includegraphics[scale=0.67]{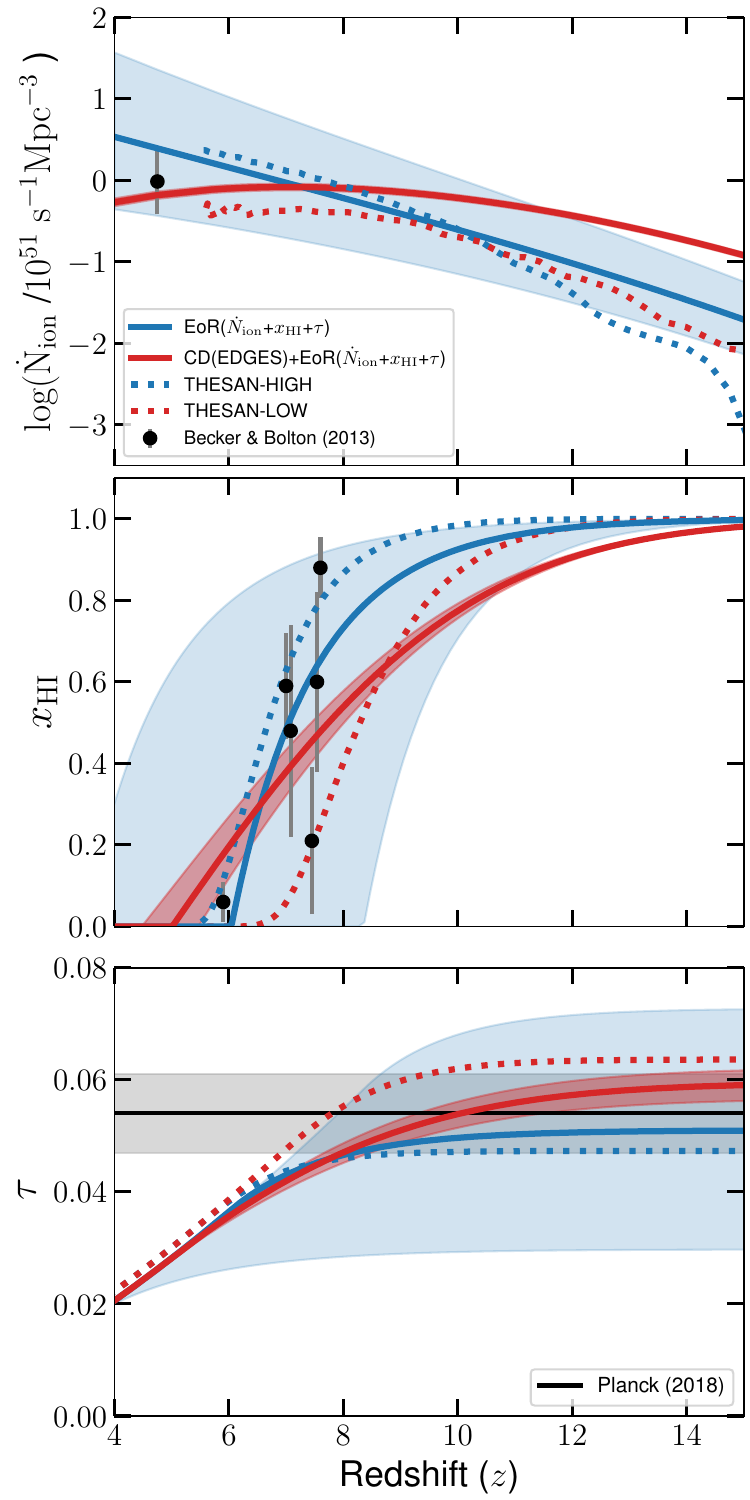}
    \caption[Redshift evolution of the ionizing emissivity, $\dot N_{\rm ion}$ (top), neutral hydrogen fraction, $x_{\rm HI}$ (middle), and Thomson optical depth, $\tau$ (bottom), as predicted by our models, EoR (solid blue) and CD+EoR.]{{\bf Upper:} Redshift evolution of the ionizing emissivity, $\dot N_{\rm ion}$, as predicted by our models, EoR (solid blue) and CD+EoR (solid red) along with the observational constraint by \citep{2013MNRAS.436.1023B}. Dotted blue and red curves show the \thesanhigh and \thesanlow simulations, respectively. {\bf Middle:} The neutral hydrogen fraction, $x_{\rm HI}$, as compared to
    data listed in Section~\ref{sec:xHI}.
    {\bf Lower:} The Thomson optical depth, $\tau$, as compared to the \citet{2020A&A...641A...6P} measurement (black line) with $1\sigma$ uncertainty (grey shaded region).
    Massive galaxies-dominated models (blue) favor a gradually increasing emissivity which in turn produces a later onset of reionization, more sudden and shorter duration, and lower optical depth. Faint galaxies-dominated models (red) favor a nearly flat emissivity evolution with the opposite characteristics.}
    \label{fig:pred}
\end{figure}

\section{Constraining the source model to reionization and cosmic dawn}
\label{sec:cdEoR}
\begin{table}
    \centering
    \renewcommand{\arraystretch}{3.0}
    \caption{Best-fit values of photon escape fraction ($f_{\rm esc}$) and power dependence on halo mass ($C$). \\}
    \label{table:cdEoR}
    \begin{adjustbox}{width=1\textwidth}
    \small
    \begin{tabular}{||c||c|c|c|c|c||}
        \hline \hline
        Parameters & \multicolumn{3}{|c|}{Constraints} & EoR & CD+ EoR  \\ [1ex]
        \hline 
        & $\dot N_{\rm ion}(z=4.75)$ & $x_{\rm HI}(z=5.9-7.6)$ & $\tau$ & $\dot N_{\rm ion}+x_{\rm HI}+\tau$ & $\rho_{\rm UV}+\dot N_{\rm ion}+x_{\rm HI}+\tau$ \\ [1ex]
        \hline \hline
        ${\bf f_{\rm esc}}$ & $0.68^{+0.23}_{-0.34}$  & $0.01^{+0.02}_{-0.01}$ & $0.68^{+0.22}_{-0.33}$ &  $0.02^{+0.02}_{-0.01}$ & $0.36^{+0.02}_{-0.02}$ \\ [1ex]
        \hline 
        ${\bf C}$ & $-0.90^{+0.76}_{-2.13}$ & $0.73^{+0.22}_{-0.19}$ & $-2.31^{+1.43}_{-3.06}$ & $0.56^{+0.19}_{-0.17}$ & $-0.29^{+0.02}_{-0.02}$\\ [1ex]
        \hline 
    \end{tabular}
    \end{adjustbox}
    \vspace{1ex}
    
    {\raggedright \small{Note. -- The $A$ and $D$ parameters are fixed to values obtained by calibration to radiative transfer simulations ($A=10^{40}\,\Msun^{-1}\text{s}^{-1}$, $D=2.28$, see \citealt{2016MNRAS.457.1550H}), and $B$ to the derived value from fitting to $\rho_{\rm UV}$ in the entire redshift range $z\sim5$--$20$ ($\log (B/\Msun) = 7.67$, see Table~\ref{table:BandC})}. \par}
\end{table}
We now turn our attention to the main goal of this work, which is to determine the range of models that can successfully reproduce reionization and cosmic dawn constraints. For reionization, we consider all measurements, from the previous section, which include the $\dot N_{\rm ion}$, $x_{\rm HI}$, and $\tau$ data. As mentioned earlier, the only existing constraint during cosmic dawn comes from the EDGES detection at $z \sim 17$, from which \citet{2018MNRAS.480L..43M} have been able to place constraints on the UV luminosity density $\rho_{\rm UV}$ (see \S\ref{sec:mcd}). We now constrain our source model to reionization only constraints ($\dot N_{\rm ion} + x_{\rm HI} + \tau$), and combined reionization and cosmic dawn constraints ($\dot N_{\rm ion} + x_{\rm HI} + \tau + \rho_{\rm UV}$). We refer to the former as EoR($\dot N_{\rm ion} + x_{\rm HI} + \tau$), and to the {latter} as CD(EDGES) + EoR($\dot N_{\rm ion} + x_{\rm HI} + \tau$). We compare the two resulting posterior distributions in Figure~\ref{fig:cdEoR}. Blue and red correspond to reionization only (EoR) and cosmic dawn plus reionization (CD+EoR), respectively. In both cases, combined observations place tight constraints on our parameters but in different parts of the parameter space. Comparing with Figure~\ref{fig:eor_con}, the reionization combined observations (blue, EoR) are mainly driven by the reionization history measurements ($x_{\rm HI}$). This shows that constraining to reionization observations favors models with low $f_{\rm esc}$ and positive $C$, indicating that massive galaxies play a major role to drive reionization. On the other hand, constraining to both reionization and cosmic dawn observations (red, CD+EoR) favors models with high $f_{\rm esc}$ and negative $C$, suggesting that low-mass galaxies play a dominant role to bridge the gap between these two epochs. This is represented by the diagonal shift from the blue to red contours once the cosmic dawn constraint ($\rho_{\rm UV}$ from EDGES) is added to the likelihood. All best-fit parameters and their 1$\sigma$ confidence intervals are listed in Table~\ref{table:cdEoR}.

In Figure~\ref{fig:pred}, we present predictions based on these best-fit parameters for the ionizing emissivity (top), reionization history (middle), and cumulative optical depth (bottom) from the EoR (blue) and CD+EoR (red) models. Shaded regions reflect the 1$\sigma$ uncertainty of the derived constraints (see Table~\ref{table:cdEoR}). In each panel, we add the relevant observations used in the MCMC analysis. We then compare our predictions with results from the \thesan project \citep{2022MNRAS.511.4005K, 2022MNRAS.512.4909G, 2022MNRAS.512.3243S}, which provides a suite of radiation-magneto-hydrodynamic simulations that self-consistently evolve reionization on large-scales ($L_{\text{box}} = 95.5\,\text{cMpc}$) and resolve the majority of ionizing sources responsible for it ($m_{\text{gas}} = 5.82 \times 10^5\Msun$) with galaxy formation physics based on the state-of-the-art IllustrisTNG model \citep{2014Natur.509..177V, 2014MNRAS.444.1518V, 2018MNRAS.480.5113M, 2018MNRAS.477.1206N, 2018MNRAS.475..624N, 2018MNRAS.475..676S, 2018MNRAS.475..648P}. Specifically, we compare with the \thesanhigh and \thesanlow simulation runs (dotted curves), where reionization is mainly driven by high-mass and low-mass galaxies, respectively.

In the upper panel of Figure~\ref{fig:pred}, we compare our predictions for the ionizing emissivity evolution with the \thesan runs \citep[based on the ray-tracing escape fraction calculations of][]{Yeh2022} and \citet{2013MNRAS.436.1023B} measurements. Our models (EoR, CD+EoR) and the different \thesan runs are consistent within the 1--2$\sigma$ levels of the observations. The emissivity evolution in the EoR model (blue) gradually increases towards decreasing redshifts as more massive halos form. This is a consequence of the positive $C$ value inferred by reionization observations, which leads to stronger contributions from massive halos. On the other hand, the CD+EoR model (red) infers a negative $C$ value, which gives more weight to the low-mass halos. This results in a higher $\dot{N}_{\rm ion}$ predicted by the CD+EoR model as compared to the EoR model predictions at high redshifts ($z \gtrsim 10$). As more massive halos form at lower redshifts ($z \lesssim 10$), the EoR model produces higher $\dot{N}_{\rm ion}$ than the CD+EoR model, where the emissivity in the latter starts to decrease due to the negative $C$ value. This leads to a flatter emissivity evolution in the CD+EoR model (red). Besides observations, our different models produce emissivity histories that are consistent with model predictions from the \thesan simulations. Our EoR and CD+EoR models show similar emissivity evolution to the \thesanhigh and \thesanlow, respectively since the models share similar assumptions.

In the middle panel of Figure~\ref{fig:pred}, we show a reionization history comparison between our models and the \thesan runs against observationally inferred measurements. As with the emissivity, we see similar reionization histories between EoR/CD+EoR models and \thesanhigh/\thesanlow simulations, respectively. Reionization begins earlier in faint galaxies-dominated models (CD+EoR, \thesanlow) than in bright galaxies-dominated models (EoR, \thesanhigh), since {faint (bright) galaxies are predominant (rare) at high redshift}. Reionization is also more gradual (longer duration) in faint galaxies-dominated models, whereas bright galaxies-dominated scenarios yield a more sudden (shorter duration) neutral-to-ionized transition of the Universe.
This earlier onset of reionization in the CD+EoR and \thesanlow models translates to a higher cumulative optical depth, as seen in the bottom panel of Figure~\ref{fig:pred}, that is consistent within the 1--2$\sigma$ levels of the \citet{2020A&A...641A...6P} measurements. On the other hand, the sudden (late) reionization history in the massive galaxies-dominated models (EoR and \thesanhigh) produces a lower optical depth that is also consistent with the 1$\sigma$ level of the data. In addition,
our EoR model is in agreement with the late reionization models presented in \citet{2019MNRAS.485L..24K}, which reproduce the large-scale opacity fluctuations measurements \citep{2018ApJ...863...92B}.

{To test whether our present models (EoR, EoR+CD) provide plausible predictions in terms of star formation, we calculate the star-formation efficiency, $f_{*}$ using Equations 1 and 4 in \citet{2016MNRAS.460..417S}. 
In case of EoR model, we find that it is $\sim 10\%$ at $M_{\text{h}} = 10^9\,\Msun$ and then increases with halo mass. On the contrary, the efficiency peaks at $1\%$ for our CD+EoR model and then decreases with increasing halo mass as this model favors a greater 
contribution from faint galaxies. The $f_{*}$ evolution at $z\sim8$ in \citet{2016MNRAS.460..417S} lies between the predicted $f_{*}$ in our models. We further compute the star-formation timescales ($t_{\text{SF}}$) in our models using the approximate relation, $\rm SFR = f_* \frac{\Omega_b}{\Omega_m} \frac{M_h}{t_{\rm SF}}$. We find that it increases/decreases with $M_{\text{h}}$ in the EoR/EoR+CD models, respectively.
For both of our models, the average $t_{\text{SF}}$ is about $\sim 1\,\text{Gyr}$, consistent with commonly assumed values in traditional star formation models~\citep[e.g.][]{2003MNRAS.339..289S}.
It is worth mentioning that our models are calibrated to the globally averaged quantities, and hence we do not expect our models to reproduce detailed instantaneous observables such as the star formation rate functions (SFRF) or UV luminosity functions (UVLF) at different redshifts. For instance, we find that our EoR (CD+EoR) model over (under) -produces the SFRF at $z\sim$ 7 and 8 as compared to \citet{2012ApJ...756...14S} star formation rate functions. However, our CD+EoR model which is calibrated to all observables is broadly in good agreement with the \citet{2014ARA&A..52..415M} cosmic star formation densities at low redshifts.}
As demonstrated by the CD+EoR model, it is entirely possible to reconcile the different observational constraints from the cosmic dawn to post-reionization using faint galaxies-dominated models of reionization, without the need to invoke new physics/exotic sources to explain the potential detection by EDGES.

\section{Summary}
\label{sec:conc}
We have presented a detailed analysis to explore the conditions by which the controversial EDGES detection \citep{2018Natur.555...67B} is consistent with reionization and post-reionization measurements including the ionizing emissivity ($\dot{N}_{\rm ion}$), IGM neutral hydrogen fraction ($x_{\rm HI}$), and Thomson optical depth ($\tau$) measurements. To account for the EDGES detection during cosmic dawn, we use the inferred constraint on the UV luminosity density ($\rho_{\rm UV}$) following \citet{2018MNRAS.480L..43M}, which is consistent with a simple extrapolation of deep Hubble Space Telescope (HST) observations of $4 < z < 9$ galaxies.

Using a semi-analytical framework of reionization with a physically-motivated source model \citep{2016MNRAS.457.1550H} coupled with MCMC likelihood sampling, our key findings are as follows:
\begin{itemize}
    \item Calibrating our source model $R_{\rm ion}$ to cosmic dawn constraints ($\rho_{\rm UV}$ from EDGES) favors models with negative $C$ values. This indicates that a stronger contribution from low-mass (faint) galaxies is required to reproduce the inferred $\rho_{\rm UV}$ constraints from EDGES (see Figure~\ref{fig:SFR} and Table~\ref{table:BandC}).
    
    \item Constraining our source model $R_{\rm ion}$ to different reionization observables is mainly driven by the IGM neutral fraction measurements where models with low $f_{\rm esc}$ and positive $C$ values are favored. This suggests that massive (bright) galaxies play the major role to drive reionization (see Figures~\ref{fig:eor_con} and \ref{fig:cdEoR} and Table~\ref{table:cdEoR}).
    
    \item Further constraining our source model $R_{\rm ion}$ to both reionization ($\dot{N}_{\rm ion}+x_{\rm HI}+\tau$) and the cosmic dawn constraint ($\rho_{\rm UV}$ from EDGES) favors models with high $f_{\rm esc}$ and negative $C$ values. This implies that low-mass (faint) galaxies would play a crucial role to bridge the gap between cosmic dawn and reionization (see Figure~\ref{fig:cdEoR} and Table~\ref{table:cdEoR}).
    
    \item Massive (bright) galaxies-dominated models produce an increasing emissivity that results in a later onset of reionization, more sudden and shorter reionization duration, and lower optical depth (see Figure~\ref{fig:pred}). Low-mass (faint) galaxies-dominated models result in a flatter emissivity evolution with the opposite reionization history characteristics.
\end{itemize}

It is worth mentioning several limitations to this work. First, the inferred $\rho_{\rm UV}$ values from EDGES compiled by~\citet{2018MNRAS.480L..43M} do not take into account the depth nor shape of the detected profile, but rather the constraints are imposed by the required Wouthuysen-Field coupling strength on UV radiation backgrounds at the detection redshift. As mentioned in~\citet{2018MNRAS.480L..43M}, accounting for the amplitude might require new physics, which in turn might alter our finding. We leave accounting for the whole properties of the detected profile to future works. Second, we use a linear relationship between SFR and $L_{\rm UV}$ in the entire redshift range ($z=5$--$20$), that has been derived from low redshift. This SFR--$L_{\rm UV}$ relation might not be linear at high redshift and assuming a different form might change our results quantitatively. Third, due to the high degeneracies between the $R_{\rm ion}$ parameters, we have fixed the amplitude $A$ and redshift dependence $D$ to values found in \citet{2016MNRAS.457.1550H} by calibrating to radiative transfer simulations of reionization. We have checked the results when $A$ is included as a free parameter, and found relatively the same value obtained in \citet{2016MNRAS.457.1550H}. The parameter $D$ might not be the same at different redshifts due to evolution in galactic feedback. However, we have obtained our fit from radiative transfer simulations using data from $z=6$--$12$, and it has been shown that the observed SFR function parameters show a weak dependence on redshift in \citet{2012ApJ...756...14S}. Hence, we do not expect a qualitative change in our results if $D$ is varied. Fourth, the quantitative results might be different if the $A$, $B$, and $D$ parameters were fixed to different values. Nevertheless, the qualitative result (high/low redshift data prefers models with more negative/positive $C$ and higher/lower $f_{\rm esc}$, respectively) would be similar. {Fifth, our models have been adjusted to match the globally averaged quantities. As a result, we observe that the EoR and CD+EoR model overproduces and underproduces the observables, such as the star formation rate functions or UV luminosity functions respectively. We leave to future works to perform a detailed analysis to calibrate our models to all global and local quantities, including the SFRF and UVLF at different redshifts using the recently high redshift data by JWST and the previous HST low redshift data.}

{It is worth noting that calibrating our models to a different $\tau$ value, $\tau = 0.0627^{+0.0050}_{-0.0058}$, as recently determined by \citet{2021MNRAS.507.1072D}, and to the updated measurements of $\dot N_{\rm ion}$ by \citet{2021MNRAS.508.1853B}, does not alter our findings.}
In summary, our results demonstrate that it is entirely possible to reproduce both cosmic dawn and reionization constraints with faint galaxies-dominated models without requiring new physics or exotic sources {as our models with deduced parameters align well with the star formation efficiencies and time scales that have been reported in previous literature.} Our results shed additional light on the roles of faint and bright galaxies during cosmic dawn and reionization, which can be tested by upcoming JWST surveys.

\chapter{Conclusions \& Discussion}

\epigraph{\itshape We are not at the end but at the beginning of a new physics. But whatever we find, there will always be new horizons continually awaiting us.}{-- Michio Kaku}

\label{chap:conclusion}
\startcontents[chapters]
This thesis work is dedicated to investigating the impacts of various physical processes on the cosmological HI 21-cm signal, with particular attention to the global HI 21-cm signal. 
Our study starts with the modeling of the first generation of stars and galaxies and goes on to explore the evolution and energy deposition by magnetic fields and cosmic rays into the intergalactic medium. The research also pays attention to the observational constraints obtained by the EDGES.
Moreover, our work also explores the reionization constraints and how they can be combined with the cosmic dawn constraints to highlight the parameter space related to different physical processes that are constrained by different observations together.

In chapter~\ref{chap:PMF}, we present the constraints on the primordial magnetic field using the global HI 21-cm absorption signal in the `colder IGM' background. Constraining the primordial magnetic field is very important as it can shed light on its origin and evolution. We show that the constraints in the colder IGM scenario are different from constraints obtained in the standard scenario. The reason behind it is the colder IGM that enhances the Hydrogen recombination rate which, in turn, reduces the residual free electron fraction during the dark ages and cosmic dawn.
In addition, the coupling between the ionized and neutral component which has a direct impact on the IGM heating also get suppressed in the `colder IGM' scenario. Moreover, the heating rate due to the Compton process, which depends on the IGM kinetic temperature and the residual free electron fraction, too gets affected when the background IGM temperature is lower. Together all these effects enhance the IGM heating rate due to the primordial magnetic field through ambipolar diffusion and decaying turbulence. Consequently, a small amount of magnetic field would be enough to keep the IGM temperature at a level that can produce global 21 cm signal consistent with the EDGES observation. On the contrary, significantly more magnetic energy would be transferred to the IGM due to the enhanced heating rate which would affect the redshift evolution of the primordial magnetic field itself and the heating at later reshifts. In the light of EDGES data, and considering the DM-baryon interaction we find that the upper bound on the primordial magnetic field can be as high as $\sim 0.4\,\rm nG$ which is ruled out in the standard model. However, $B_0 \gtrsim 1 \, {\mathrm{nG}}$ may not be allowed as this could only be explained with a very highly efficient cooling mechanism that requires dark-matter mass (interaction cross-section) to be quite low (high) than the current predictions.
As the 21-cm absorption signal also depends on the mass of the DM particles and the interaction cross-section, we are also able to put limits on the dark matter mass and cross-section using EDGES low band measurements. We see that the allowed region of dark-matter mass decreases while the interaction cross-section increases with the increase in magnetic field. Furthermore, due to the decaying nature of the magnetic field at lower redshifts, we find that the heating due to the primordial magnetic field becomes inefficient in explaining the sharp rise in the 21-cm profile observed by EDGES.

Hence, in chapter~\ref{chap:CR}, we explore another possible mechanism that can heat the IGM substantially. One such candidate is cosmic ray protons that are generated in the termination shocks of supernova explosions.
It is well known that the SNe are the main sources of high-energy cosmic rays. Given the expected top-heavy IMF of Pop~III stars, they are all likely to explode as a SNe that can accelerate copious amounts of cosmic rays. Due to the smaller sizes of the host galaxies as well as the higher energetics of the Pop~III SNe, the cosmic rays are likely to be generated outside the virial radius of the halo and can escape to the IGM easily. 
While traversing through the IGM, low energy protons (with energy $\lesssim 30$ MeV) interact with the neutral hydrogen and free $e^-$. In case of collisions with free $e^-$, the entire energy loss by these particles becomes the thermal energy of the IGM, whereas, the interaction between cosmic ray protons and neutral hydrogen results in primary and secondary ionizations, and finally contributes to the heating. We find that cosmic rays resulting from Pop~III stars can potentially alter the thermal state of the IGM, and hence the spin temperature, $T_s$ during cosmic dawn.
Contrary to Pop~III halos, the low energy protons generated from the SNe exploding in massive atomic cooling halos, get confined within the galaxy itself. However, the high-energy protons are likely to escape from Pop~II galaxies like our Galaxy and can contribute to the heating. In this case, if a sufficient magnetic field is present in the IGM, these high-energy cosmic ray particles gyrate along the magnetic field lines and excite magnetosonic Alfv\`en waves. When these waves get damped, the energy gets transferred to the thermal gas, and can potentially change the temperature of the IGM.
%

Our model accounts for both metal-free Pop III and metal-enriched Pop II stars, and considers Lyman-Werner feedback on Pop III star formation, and supernova, AGN and radiative feedbacks in Pop II galaxies. The temperature evolution of the intergalactic medium, or in turn, global HI 21-cm signal is calculated self-consistently, taking into account Lyman-$\alpha$ coupling, cosmic ray heating, dark matter-baryon interaction, and adiabatic cooling. We find that cosmic rays are an important source of IGM heating and shape the global 21-cm signal during cosmic dawn. We show that the EDGES signal can be explained by our model, with reasonable efficiency parameters of cosmic ray heating. We also find that even a small fraction of cosmic ray protons is capable of heating the IGM that could wash out the absorption in 21-cm signal. Our results suggest that accurately determined 21-cm signals could be used to probe early cosmic ray heating and constrain the nature of early-generation stars.

Furthermore, the reionization constraints provide insights into the process by which the first galaxies reionized the universe, while the cosmic dawn constraints provide information on the earliest stages of the universe. By combining these measurements, we identified the parameter space that is constrained by the joint observations in chapter~\ref{chap:Bridging}, which can be further verified by ongoing and upcoming surveys such as the JWST and the SKA. In particular, we explore the conditions under which
the EDGES detection is consistent with other reionization and post-reionization observations, including ionizing emissivity, IGM neutral hydrogen fraction, and Thomson scattering optical depth. The analysis is based on a semi-analytical framework of reionization with a physically-motivated ionizing source model and an MCMC likelihood sampling to calibrate the source model to different observables. The key findings from this work suggest that low-mass (faint) galaxies play a crucial role in reproducing the inferred UV luminosity density constraint from EDGES, while massive (bright) galaxies dominate the reionization process. This analysis indicates that low-mass galaxies would bridge the gap between cosmic dawn and reionization. 
This work also shows that low-mass galaxies-dominated models result in a flatter emissivity evolution, resulting in a later onset of reionization, more sudden and shorter reionization duration, and lower optical depth and vice versa. This study acknowledges several limitations, such as the assumption of a linear relationship between $\rm SFR-L_{UV}$ and the fixed values of some model parameters. Nevertheless, the results demonstrate that it is possible to reproduce both cosmic dawn and reionization constraints with faint galaxies-dominated models without requiring new physics or exotic sources. The study sheds light on the roles of faint and bright galaxies during cosmic dawn and reionization and provides avenues for further investigation using upcoming JWST surveys.

In conclusion, this thesis work focuses on investigating the impacts of various physical processes on the cosmological HI 21-cm signal, with particular attention to the global HI 21-cm signal. The study explores the modeling of the first generation of stars and galaxies, the impact of magnetic fields and cosmic rays on the HI 21-cm signal, and the observational constraints obtained by EDGES. The study also combines the reionization constraints with the cosmic dawn constraints to identify the parameter space that is constrained by joint observations. The insights gained from this research can help us to better understand the evolution of the universe and the processes that shaped it. The ongoing and upcoming surveys such as the JWST and the SKA are expected to provide further insights into the early universe and help to refine our understanding of the HI 21-cm signal.

\chapter{Future outlook}
\epigraph{\itshape  A ship is always safe at the shore - but that is {\rm not} what it is built for.}{-- Albert Einstein}
\label{chap:future}

\startcontents[chapters]
In this thesis work, we have investigated some heating mechanisms in detail that affect the global HI 21-cm signal at different epochs. However, we must acknowledge that there are several other processes that are likely to play a role in the heating of the intergalactic medium, including X-rays \citep{pritchard2007, baek09, mesinger13, ghara14, Pacucci2014, Fialkov2014, fialkov14b, Arpan2017, Ma2021}, Ly-$\alpha$ \citep{Madau1997, Chuzhoy2006, ghara2020, Ciardi2010, Reis2021, mittal21}, and the cosmic microwave background \citep{Venumadhav2018}. Despite extensive research, there is currently no conclusive evidence indicating which heating mechanism dominates over the others. Each possible heating mechanism has its own set of free parameters that are not well-constrained.

One of the heating mechanisms that is thought to have a significant impact on the IGM's heating during the cosmic dawn and the epoch of reionization is heating through the X-ray background. During these periods, the first galaxies and quasars started emitting X-rays, which ionized the gas around them and heated the surrounding IGM. The X-ray background is also believed to play a role in heating the IGM during the later stages of reionization.
The heating mechanism through the Ly-$\alpha$ background is also thought to be important during the later stages of reionization when the IGM is partially ionized. At this stage, the ionizing radiation from the first galaxies is thought to create a Ly-$\alpha$ background, which heats the IGM. The Ly-$\alpha$ background is also believed to play a role in heating the IGM during the cosmic dawn and the early stages of reionization.
The CMB can also heat the IGM by Compton scattering off free electrons, and it is believed to be an important heating mechanism during the cosmic dawn and the early stages of reionization.

Despite the progress made in understanding the heating mechanisms of IGM, there are still many unanswered questions. For instance, we do not know the exact nature of the sources responsible for the X-ray and Ly-$\alpha$ backgrounds. Additionally, the free parameters associated with each heating mechanism are not well-constrained. Hence, there is a need for more observational and theoretical studies to shed light on these issues.

Therefore, the aim of our future work is to conduct a comprehensive study of the various heating processes that occur in the intergalactic medium. This is necessary to determine if the effects of different heating processes and their sources can be distinguished from the time evaluation of the HI 21-cm signal. If we find that the global signal is inadequate for this purpose, we will further investigate the fluctuations in the 21-cm power spectrum. As more observational data becomes available through future surveys, we will conduct detailed modeling of the 21-cm signal, taking into account all possible sources of heating. This will enable us to better understand the physical processes that govern the evolution of the intergalactic medium and the formation of structures in the universe.

\backmatter

\fancyhead[RO]{\leftmark}
\fancyhead[LE]{\leftmark}

\printbibliography[heading=bibintoc,title={Bibliography}]

\appendix
\chapter{Appendix}

\label{chap:appendix}
As given in equation~\ref{eq:dndM}, the differential halo mass function can be written as,
\begin{equation}
    \frac{dn(M, z)}{dM} = \frac{\bar \rho_0}{M} f(\sigma) \left| \frac{d \ln \sigma}{dM} \right| .
    \label{eq:n(M)}
\end{equation}
The RMS linear overdensity of the density field $\sigma^2 (M) = \frac{b^2(z)}{2\pi^2} \int_0^{\infty} k^2 P(k) W^2(k;M) dk$ or, $\sigma (z, M) = b(z) \sigma_0$, where, $W(k;M)$ is the Fourier Transform of the real space top-hat filter, $P(k)$ is the linear power spectrum of the density fluctuations (at $z=0$) and $b(z)$ is the growth factor normalized to unity at $z=0$ \citep{Peebles_1993} and can be written in terms of cosmological parameters as \citep{2002tagc.book.....P},
\begin{equation}
    b(z) = \left[ \frac{\Omega_m + 0.4545 \Omega_{\Lambda}}{\Omega_m (1+z)^3 + 0.4545 \Omega_{\Lambda}} \right]^{1/3} .
\end{equation}
The mass function, $f(\sigma)$ is defined as the fraction of mass in collapsed halos per unit interval in mass variable $ln \sigma^{-1}$. The Sheth-Tormen mass function can be expressed as, 
\begin{equation}
    f_{ST} (\sigma) = A \sqrt{\frac{2a}{\pi}} \frac{\delta_c}{\sigma} \left[ 1 + \left( \frac{\sigma^2}{a \delta_c^2} \right)^p \right] \exp\left( -\frac{a \delta_c^2}{2 \sigma^2} \right) .
    \label{eq:f_ST}
\end{equation}
Now using equation~\ref{eq:f_ST}, equation~\ref{eq:n(M)} can be written as,
\begin{equation}
    \frac{dn}{dM} = \frac{\bar \rho_0}{M} A \sqrt{\frac{2a}{\pi}} \frac{\delta_c}{b(z) \sigma_0} \left[ 1 + \left( \frac{b^2(z) \sigma^2_0}{a \delta_c^2} \right)^p \right] \exp\left( -\frac{a \delta_c^2}{2 b^2(z) \sigma^2_0} \right) \left| \frac{1}{\sigma_0} \frac{d \sigma_0}{dM} \right| .
\end{equation}
So, the redshift derivative of the differential halo mass function is given by, 
\begin{dmath}
    \frac{d^2n}{dz dM} = \frac{\bar \rho_0}{M} A \sqrt{\frac{2a}{\pi}} \frac{\delta_c}{b(z) \sigma_0} \exp\left( -\frac{a \delta_c^2}{2 b^2(z) \sigma^2_0} \right) \left| \frac{1}{\sigma_0} \frac{d \sigma_0}{dM} \right| \left\{ \left[ 1+ \left( \frac{b^2(z) \sigma^2_0}{a \delta_c^2} \right)^p \right]  \times \left(-\frac{\dot b(z)}{b(z)} \right) + \left( \frac{b^2(z) \sigma^2_0}{a \delta_c^2} \right)^p \times 2p \frac{\dot b(z)}{b(z)} + \left[ 1+\left( \frac{b^2(z) \sigma^2_0}{a \delta_c^2} \right)^p \right] \left(\frac{a \delta_c^2}{b^2(z) \sigma_0^2} \right) \frac{\dot b(z)}{b(z)} \right\} \\
    = \frac{\bar \rho_0}{M} A \sqrt{\frac{2a}{\pi}} \frac{\delta_c}{\sigma_0} \exp\left( -\frac{a \delta_c^2}{2 b^2(z) \sigma^2_0} \right) \left| \frac{1}{\sigma_0} \frac{d \sigma_0}{dM} \right| \frac{\dot b(z)}{b^2(z)} \times \left\{ 2p \left( \frac{b^2(z) \sigma^2_0}{a \delta_c^2} \right)^p + \left[ 1+\left( \frac{b^2(z) \sigma^2_0}{a \delta_c^2} \right)^p \right] \times \left[ \left(\frac{a \delta_c^2}{b^2(z) \sigma_0^2} \right) -1 \right] \right\}
\end{dmath}
This gives the redshift derivative of the differential halo mass function discussed in Chapter~\ref{sec:SFRD_th}.

\end{document}